\documentclass[12pt,letterpaper,nohyper]{JHEP3}

\usepackage{amsmath,amssymb,amsfonts,theorem,epsfig,xspace,cite}

\title{Lectures\footnote{Lectures delivered at the
RTN Winter School on Strings, Supergravity and Gauge theories,
(CERN, January 16-20, 2006), the 11-th APCTP/KIAS String Winter
School (Pohang, Feb 8-15 2005) and the Winter School on the 
Attractor Mechanism (Frascati, March 20-24, 2006). Updated, Feb. 2007} 
\ on Black Holes, Topological Strings\\
 and Quantum  Attractors (2.0)}

\preprint{\hepth{0607227}\\LPTENS-07-04}

\author{Boris Pioline\\
--
Laboratoire de Physique Th\'eorique et Hautes Energies\footnote{Unit\'e mixte
de recherche du CNRS UMR 7589}, \\
Universit\'e Pierre et Marie Curie - Paris 6,\\
4 place Jussieu, F-75252 Paris cedex 05 \\ 
-- Laboratoire de Physique Th\'eorique de l'Ecole Normale
Sup\'erieure\footnote{Unit\'e mixte
de recherche du CNRS UMR 8549}\\
24 rue Lhomond, F-75231 Paris cedex 05\\
-- E-mail: {\tt pioline@lpthe.jussieu.fr} }

\abstract{In these lecture notes, we review some recent developments on the
relation between the macroscopic entropy of four-dimensional 
BPS black holes and the microscopic
counting of states, beyond the thermodynamical, large charge limit.
After a brief overview of charged black holes in supergravity
and string theory, we give an extensive introduction to special 
and very special geometry, attractor flows
and topological string theory, including holomorphic anomalies.
We then expose the Ooguri-Strominger-Vafa (OSV) conjecture which
relates microscopic degeneracies to the topological string amplitude,
and review precision tests of this formula on ``small'' black holes.
Finally, motivated by a holographic interpretation of the OSV
conjecture, we give a systematic approach to the radial quantization 
of BPS black holes (i.e. quantum attractors). This suggests 
the existence of a one-parameter generalization of the topological string 
amplitude, and provides a general framework for constructing automorphic 
partition functions for black hole degeneracies in theories with sufficient
degree of symmetry.
}

\renewcommand{\Im}{{\rm Im}}
\renewcommand{\Re}{{\rm Re}}
\newcommand{\pa}{\partial}
\newcommand{\CN}{{\cal N}}
\newcommand{\nn}{\nonumber}
\newcommand{\cc}{\mbox{c.c.}}
\newcommand{\IR}{\mathbb{R}}
\newcommand{\IQ}{\mathbb{Q}}
\newcommand{\IA}{\mathbb{A}}
\newcommand{\IC}{\mathbb{C}}
\newcommand{\IP}{\mathbb{P}}
\newcommand{\IZ}{\mathbb{Z}}
\newcommand{\IH}{\mathbb{H}}
\newcommand{\IO}{\mathbb{O}}
\newcommand{\Tr}{\mbox{Tr}}
\newcommand{\Aut}{\mbox{Aut}}
\newcommand{\Str}{\mbox{Str}}
\newcommand{\Conf}{\mbox{Conf}}
\newcommand{\QConf}{\mbox{QConf}}
\newcommand{\Pf}{\mbox{Pf}}

\newcommand{\tzeta}{\tilde\zeta}

\newcommand{\eps}{\epsilon}
\newcommand{\BesselI}[3]{\hat I_{#1}\left( #2 \pi \sqrt{ #3 }\right)}

\newcommand{\ri}{{\rm i}}
\newcommand{\rj}{{\rm j}}
\newcommand{\rk}{{\rm k}}
\newcommand{\rl}{{\rm l}}
\newcommand{\half}{\frac{1}{2}}

\newcommand{\cV}{\mathcal{V}}

\newcommand{\cC}{\mathcal{C}}
\newcommand{\cS}{\mathcal{S}}

\newcommand{\cK}{\mathcal{K}}
\newcommand{\cM}{\mathcal{M}}
\newcommand{\cN}{\mathcal{N}}

\newcommand{\hk}{{hyperk\"ahler}\xspace}
\newcommand{\qk}{{quaternionic-K\"ahler}\xspace}

\newcommand{\CP}{{\mathbb CP}}

\newcommand{\Z}{{\mathbb Z}}
\newcommand{\cZ}{\mathcal{Z}}

\newcommand{\abs}[1]{\lvert#1\rvert}

\newcommand{\txi}{\tilde\xi}
\newcommand{\inprod}[1]{\langle#1\rangle}

\newcommand{\I}{\mathrm{i}}

\def\bea{\begin{eqnarray}}
\def\eea{\end{eqnarray}}
\def\be{\begin{equation}}
\def\ee{\end{equation}}
\def\ba{\begin{align}}
\def\ea{\end{align}}
\def\bse{\begin{subequations}}
\def\ese{\end{subequations}}

\theoremstyle{plain}
\newtheorem{exo}{Exercise}
\begin{document}

\numberwithin{equation}{section}

\section{Introduction}
Once upon a time regarded as unphysical solutions of General
Relativity, black holes now occupy the central stage. In astrophysics,
there is mounting evidence of stellar size and supermassive black holes
in binary systems and in galactic centers (see e.g. \cite{Hughes:2005wj}).
In theoretical particle physics, black holes are believed to 
dominate the high energy behavior of
quantum gravity (e.g. \cite{Banks:2003vp}). 
Moreover, the Bekenstein-Hawking entropy
of black holes is one of the very few clues in our hands about the
nature of quantum gravity: just as the macroscopic thermodynamical
properties of perfect gases hinted at their microscopic atomistic structure, 
the classical thermodynamical properties of black holes suggest the
existence of quantized micro-states, whose dynamics should account 
for the macroscopic production of entropy. 

One of the great successes of string theory is to have made this idea precise,
at least for a certain class of black holes which admittedly 
are rather remote from reality: supersymmetric, charged black holes 
can indeed be viewed as 
bound states of D-branes and other extended objects, whose microscopic
``open-string'' fluctuations account for the macroscopic Bekenstein-Hawking 
entropy \cite{Strominger:1996sh}. 
In a more modern language, the macroscopic gravitational dynamics
is holographically 
encoded in microscopic gauge theoretical degrees of freedom living
at the conformal boundary of the near-horizon region. Irrespective of
the language used, the agreement is quantitatively exact in 
the ``thermodynamical'' limit of large charge, where the counting of 
the degrees of freedom requires only a gross understanding of their dynamics.

While the prospects of carrying this quantitative agreement over to
more realistic black holes remain distant, it is interesting to
investigate whether the already remarkable agreement found 
for supersymmetric extremal black holes can be pushed beyond 
the thermodynamical limit. Indeed, this regime in principle allows to probe
quantum-gravity corrections to the low energy Einstein-Maxwell Lagrangian, 
while testing our description of the microscopic degrees of 
freedom in greater detail. 


The aim of these lectures is to describe some recent developments in this
direction, in the context of BPS black holes in $\CN\geq 2$ supergravity.

In Section \ref{ern}, we give an overview of extremal 
Reissner-Nordstr\"om black holes, recall their embedding
in string theory and the subsequent microscopic derivation of their
entropy at 
leading order, and briefly discuss an early proposal to relate
the exact microscopic degeneracies to Fourier coefficients of a
certain modular form.

In Section \ref{spegeo}, we recall the essentials of special geometry,
and describe the ``attractor flow'', which 
governs the radial evolution of the scalar fields and determines the
horizon geometry in terms of asymptotic charges. We illustrate these
results in the context of ``very special supergravities'', an interesting
class of toy models whose symmetries properties allow to get very
explicit results.

In Section \ref{topprim},
we give a self-contained introduction to topological string theory,
which allows to compute an infinite set of higher-derivative ``F-term''
corrections in the low energy Lagrangian. We emphasize the wave function
interpretation of the holomorphic anomaly, which underlies much of
the subsequent developments.

In Section \ref{highder}, we discuss the effects of these ``F-term'' 
corrections on the macroscopic entropy, and formulate the 
Ooguri-Strominger-Vafa (OSV) conjecture \cite{Ooguri:2004zv},
which relates these macroscopic corrections to the micro-canonical counting.

In Section \ref{smallbh}, based on \cite{Dabholkar:2005by,Dabholkar:2005dt}, 
we submit this conjecture to a precision test, in the context 
of ``small black holes'': these are dual to perturbative
heterotic states, and can therefore be counted exactly using standard 
conformal field theory techniques. 

Finally, in Section \ref{qatt}, motivated by a holographic interpretation
of the OSV conjecture put forward by Ooguri, Vafa and 
Verlinde \cite{Ooguri:2005vr}, we turn to the subject of ``quantum
attractor flows''. We give a systematic treatment of the radial 
quantization of BPS 
black holes, and compute the exact radial wave function for a black hole
with fixed electric and magnetic charges.  In the course of this discussion,
we find evidence for a one-parameter generalization
of the usual topological string amplitude, and provide a framework for
constructing automorphic partition functions for black hole degeneracies
in theories with a sufficient degree of symmetry, in the spirit (but not
the letter) of the genus-2 modular forms discussed in Section \ref{dvvform}.
This section is based on \cite{Pioline:2005vi,Gunaydin:2006bz,
Gunaydin:2005mx,Neitzke:2007ke}
and work in progress \cite{gnpw-in-progress,gnopw-in-progress}.

We have included a number of exercices, most of which are quite easy,
which are intended to illustrate, complement or extend the discussion
in the main text. The dedicated student might learn more from 
solving the exercices than from pondering over the text.

\section{Extremal Black Holes in String Theory\label{ern}}
In this section, we give a general overview of extremal black holes
in Einstein-Maxwell theory, comment on their embedding in string theory, 
and outline their microscopic description as bound states of D-branes.
We also review an early conjecture that relates the exact microscopic 
degeneracies of BPS black holes to Fourier coefficients of
a certain modular form. We occasionally make use of notions that
will be explained in later Sections. For a general introduction to black
hole thermodynamics, the reader may consult e.g. \cite{Townsend:1997ku,
Damour:2004kw}.

\subsection{Black Hole Thermodynamics}
Our starting point is the Einstein-Maxwell Lagrangian for gravity
and a massless Abelian gauge field in 3+1 dimensions,
\be
\label{einmax}
S = \int d^4x\   \frac{1}{16\pi G} \left[ ~\sqrt{-g}~R 
-\frac{1}{4} F \wedge \star F \right]
\ee
Assuming staticity and spherical symmetry, the only solution with
electric and magnetic charges $q$ and $p$ is the Reissner-Nordstr\"om
black hole
\be
\label{rn}
ds^2 = - f(\rho)~dt^2 + f^{-1}(\rho)~d\rho^2 + \rho^2~
d\Omega^2 \ ,\quad
F = p~ \sin\theta d\theta \wedge d\phi + q ~ \frac{dt \wedge d\rho}{\rho^2}
\ee
where $d\Omega^2 = d\theta^2 + \sin^2\theta\ d\phi^2$ is the metric
on the two-sphere, and $f(\rho)$ is given 
in terms of the ADM mass $M$ and the charges $(p,q)$ by
\be
\label{frn}
f(\rho) = 1 - \frac{2~GM}{\rho} + \frac{p^2+q^2}{\rho^2}
\ee
For most of what follows, we set the Newton constant $G=1$. 
The Schwarzschild black hole is recovered in the
neutral case $p=q=0$.

The solution \eqref{rn} has a curvature singularity at $r=0$, 
with diverging curvature invariant $R_{\mu\nu} R^{\mu\nu} \sim 
4 (p^2+q^2)^2/ \rho^8$. When $M^2<p^2+q^2$, this is a naked singularity
and the solution must be deemed unphysical. When $M^2>p^2+q^2$
however, there are two horizons at the zeros of $f(\rho)$,
\be
\label{rhopm}
\rho_\pm= M \pm \sqrt{M^2-p^2-q^2}
\ee
which prevent the singularity to have any physical 
consequences for an observer at infinity, see the Penrose diagram
on Figure 1. We shall denote by I, II, III
the regions outside the horizon, between the two horizons and inside the
inner horizon, respectively. Since the time-like 
component of the metric changes sign twice between regions I and III, 
the singularity at $\rho=0$ is time-like, and may be imputed to 
the existence of a physical source at $\rho=0$. This is
unlike the Schwarzschild black hole, whose space-like singularity 
at $\rho=0$ raises more serious concerns.

\begin{figure}
\centerline{\hfill\includegraphics[height=9cm]{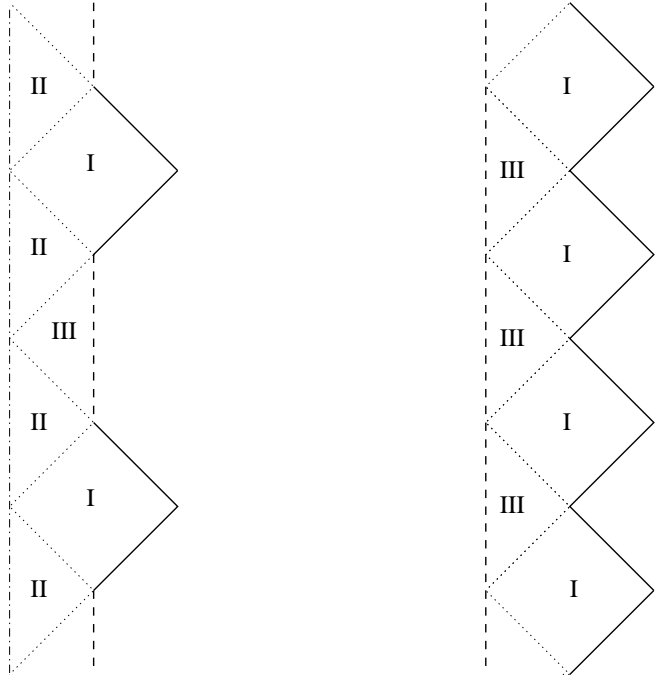}\hfill}
\caption{Penrose diagram of the non-extremal (left) and extremal (right)
Reissner-Nordstr\"om black holes. Dotted lines denote event
horizons, dashed lines represent time-like singularities. The 
diagram on the left should be doubled along the dashed-dotted
line.\label{penro}}
\end{figure}

Near the outer horizon, one may approximate
\be
f(\rho) = \frac{(\rho-\rho_+)(\rho-\rho_-)}{\rho^2} \sim
\frac{(\rho_+ - \rho_-)}{\rho_+^2} r
\ee
where $\rho=\rho_+ + r$, and the line element \eqref{rn} by
\be
\label{nhne}
ds^2 \sim \left[ 
- \frac{(\rho_+ - \rho_-)}{\rho_+^2} r ~ dt^2 
+ \frac{\rho_+^2}{(\rho_+ - \rho_-)} \frac{dr^2}{r} \right]
+ \rho_+^2 ~ d\Omega_2^2
\ee
Defining $t = 2\rho_+^2 \tau / (\rho_+-\rho_-)$ and $r = \eta^2$,
the first term is recognized as Rindler space while the second term 
is a two-sphere of fixed radius,
\be
ds^2 = \frac{4\rho_+^2}{\rho_+-\rho_-} ( - \eta^2 d\tau^2 + d\eta^2  )
+ \rho_+^2 ~ d\Omega_2^2\ .
\ee
Rindler space describes the patch of Minkowski space accessible
to an observer ${\cal O}$ with constant acceleration $\kappa$. As spontaneous 
pair production takes place in the vacuum, ${\cal O}$
may observe only one member of that pair, while its correlated partner 
falls outside of ${\cal O}$'s horizon. Hawking and Unruh have
shown that, as a result,   ${\cal O}$ detects a thermal spectrum of particles  
at temperature $T=\kappa/(2\pi)$, where $\kappa$ is the acceleration,
or ``surface gravity'' at the horizon \cite{Hawking:1974sw,Unruh:1976db}. 
Equivalently,
smoothness of the Wick-rotated geometry $\tau\to i\tau$ requires that 
$\tau$ be identified modulo $2\pi i$. In terms of the inertial time $t$
at infinity, this requires $t\sim t+i\beta$ where $\beta$ is 
the inverse temperature
\be
\label{ht}
\beta = \frac{1}{T} = \frac{4\pi \rho_+^2}{\rho_+ - \rho_-}
\ee
Given an energy $M$ and a temperature $T$, it is natural to define
the ``Bekenstein-Hawking'' entropy $S_{BH}$ such that 
$dS_{BH}/dM=1/T$ at fixed charges. 

\begin{exo}
By integrating \eqref{ht}, show that the entropy
of a Reissner-Nordstr\"om black hole is equal to
\be
S_{BH} =  \pi \left( M + \sqrt{M^2 - p^2 -q^2} \right)^2 = \pi \rho_+^2
\ee
\end{exo}

Remarkably, the result is, up to a factor $1/(4G)$, just equal to the
area of the horizon:
\be
\label{sbh}
S_{BH} = \frac{A}{4G}
\ee
This is a manifestation the following general statements, 
known as the ``laws of black hole thermodynamics'' (see
e.g. \cite{Damour:2004kw,Wald:2001} and references therein):
\begin{itemize}
\item[0)] The temperature $T=\kappa/(2\pi)$ is uniform on the horizon;
\item[I)] Under quasi-static changes, $dM = (T/4G) dA + \phi dq+\chi dp$;
\item[II)] The horizon area always increases with time.
\end{itemize}
These statements rely purely on an analysis of the classical 
solutions to the action \eqref{einmax}, and their singularities.
The modifications needed to preserve the validity of these laws in the 
presence of corrections to the action \eqref{einmax}
will be discussed in Section \ref{bhw}. 

The analogy of 0),I),II) with the usual laws of thermodynamics strongly
suggests that it should be possible to identify the Bekenstein-Hawking
entropy with the logarithm of the number of micro-states which lead to
the same macroscopic black hole,
\be
\label{som}
S_{BH} = \log \Omega (M, p ,q)
\ee
where we set the Boltzmann constant to 1. In writing this equation, 
we took advantage of the ``no hair'' theorem which
asserts that the black hole geometry, after transients, 
is completely specified by the charges measured at infinity. 

Making sense of \eqref{som} microscopically
requires quantizing gravity, which for us means using string theory.
As yet, progress on this issue has mostly been restricted the case of 
extremal (or near-extremal) black holes, to which we turn now.

\subsection{Extremal Reissner-Nordstr\"om Black Holes}
In the discussion below \eqref{frn}, we left out one special case, namely
$M^2=p^2+q^2$. When this happens, the inner and outer horizons
coalesce in a single degenerate horizon at $r=\sqrt{p^2+q^2}$, 
where the scale factor vanishes quadratically:
\be
f(\rho) = \left( 1 - \frac{\sqrt{p^2+q^2}}{\rho} \right)^2
\sim \frac{r^2}{p^2+q^2}
\ee
Such black holes are called ``extremal'', for reasons that will become
clear below.
In this case, defining $r=(p^2+q^2)/z$, we can rewrite the near-horizon
geometry as
\be
\label{ads2s2}
ds \sim (p^2+q^2)~ \left[ \frac{-dt^2+dz^2}{z^2} + d\Omega^2 \right]
\ee
which is now recognized as the product of two-dimensional Anti-de Sitter
space $AdS_2$ times a two sphere. In contrast to \eqref{nhne}, this is
now a bona-fide solution of \eqref{einmax}. The appearance of the $AdS_2$
factor raises the hope that such ``extremal'' black holes have an holographic 
description, although holography in $AdS_2$ is far less understood than in 
higher dimensions (see \cite{Strominger:1998yg} for an early discussion).

An important consequence of $f(r)$ vanishing quadratically is that the
Hawking temperature \eqref{ht} is zero, so that the black hole no
longer radiates: this is as it should, since otherwise its mass would
go below the bound 
\be
\label{bpsb} M^2\geq p^2+q^2\ ,
\ee
producing a naked singularity. Black holes saturating this bound can be 
viewed as the stable endpoint of Hawking evaporation\footnote{
The evaporation end-point of neutral black holes is far less understood,
and in particular leads to the celebrated ``information paradox''.}, assuming
that all charged particles are massive. Moreover, 
the Bekenstein entropy remains finite
\be
\label{bhbps}
S_{BH} = \pi (p^2+q^2)
\ee
and becomes large in the limit of large charge. 
This is not unlike
the large degeneracy of the
lowest Landau level in condensed matter physics.

\subsection{Embedding in String Theory}
String theory compactified to four dimensions typically involves
many more fields than those appearing in the Einstein-Maxwell Lagrangian
\eqref{einmax}. Restricting to compactifications which preserve $\CN\geq 2$
supersymmetry in four dimensions, there are typically many Abelian
gauge fields and massless scalars (or ``moduli''), 
together with their fermionic partners,
and the gauge couplings in general have a complicated dependence on the 
scalar fields. As a result, the static, spherically symmetric solutions 
are much more complicated, involving in particular a non-trivial radial 
dependence of the scalar fields. The first smooth solutions were constructed
in the context of the heterotic string compactified on $T^6$ in 
\cite{Cvetic:1995uj}, and the general solution was obtained in
\cite{Cvetic:1995kv} using spectrum-generating symmetries. 
Charged solutions exhibit the same causal structure as the
Reissner-Nordstr\"om black hole, and become extremal when a certain
``BPS'' bound, analogous to \eqref{bpsb} is saturated. 

In fact, in the context of supergravity with $\CN\geq 2$ extended
supersymmetry, the BPS bound is a 
consequence of unitarity in a sector with non-vanishing central charge 
$Z=\sqrt{p^2+q^2}$, see \eqref{bpsbound} below. 
The saturation of the bound implies
that the black hole preserves some fraction of the supersymmetry of the
vacuum. Since the corresponding representation of the supersymmetry
algebra has smaller dimension that the generic one, such states are 
absolutely stable (unless they can pair up with 
an other extremal state with the same energy) \cite{Witten:1978mh}. 
They can be followed as the coupling is varied, which is part of the reason 
for their successful description in string theory. 

Another peculiarity of extremal black holes in supergravity is that 
the radial profile of the scalars simplifies: specifically, the 
values of the scalar fields at the horizon become independent of the 
values at infinity, and depend only on the electric and
magnetic charges. Moreover, the horizon area itself 
becomes a function of the charges only\footnote{Although it no longer takes 
the simple quadratic form \eqref{bhbps}, at tree-level it is still an 
homogeneous function of degree 2 in the charges.}. 
This is a consequence of the ``attractor mechanism'',
which we will discuss at length in Sections \ref{spegeo} and 
\ref{qatt}. This fits in nicely with the fact that the number
of quantum states of a system is expected to be invariant under
adiabatic perturbations \cite{Larsen:1995ss}. More practically,
it implies that a rough combinatorial, weak coupling counting
of the micro-states may be sufficient to reproduce the macroscopic
entropy.

As a side comment, it should be pointed out that even in supersymmetric 
theories, extremal black holes can exist which break all supersymmetries. 
In this case, the electromagnetic charges differ from the 
central charge, and the extremality bound is subject to quantum
corrections. In this case, there may exist non-perturbative decay processes 
whereby an extremal black hole may break into smaller ones. The
subject of non-supersymmetric extremal black holes has become
of much interest recently, see e.g. \cite{Tripathy:2005qp,Goldstein:2005hq,
Kallosh:2005ax,Kallosh:2006bt,Sahoo:2006rp,Arkani-Hamed:2006dz,Kaura:2006mv}.

\begin{exo}
Show that if black hole of mass and charge $(M,Q)$ breaks up into
two black holes of mass and charge $(M_1,Q_1)$ and $(M_2,Q_2)$, 
then at least one of $M_1/Q_1$ and $M_2/Q_2$ must be smaller than
$M/Q$. Conclude that quantum corrections should decrease
the ratio $M/Q$ \cite{Arkani-Hamed:2006dz,Kats:2006xp}.
\end{exo}

\subsection{Black Hole Counting via D-branes \label{stromv}}
The ability of string theory to account microscopically for the 
Bekenstein-Hawking entropy of BPS black holes \eqref{bhbps} is one
of its most concrete successes. Since this subject is well covered 
in many reviews, we will only outline the argument, referring e.g. to
\cite{Maldacena:1996ky,Peet:2000hn,David:2002wn,Mathur:2005ai} 
for more details and references.

The main strategy, pioneered by Strominger and Vafa \cite{Strominger:1996sh}, 
is to represent the black hole as a bound state of
solitons in string theory, and vary the coupling so that the degrees
of freedom of these solitons become weakly coupled. The BPS
property ensures that the number of micro-states will be
conserved under this operation.

Consider for exemple
1/8 BPS black holes in Type II string theory on $T^6$, or 1/4 BPS black 
holes on $K3 \times T^2$ \cite{Maldacena:1996gb}. 
Both cases can treated simultaneously 
by writing the compact 6-manifold as $X = Y \times S_1 \times S_1'$,
where $Y= T^4$ or $K3$. Now consider a configuration of $Q_6$ D6-branes
wrapped on $X$, $Q_2$ D2-branes wrapped on $S_1 \times S_1'$, 
$Q_5$ NS5-branes wrapped on $Y \times S_1$, carrying $N$ units of
momentum along $S_1$. The resulting configuration is localized in 
the four non-compact directions and supersymmetric, 
hence should be represented as a BPS black
hole in $\CN=8$ or $\CN=4$ supergravity\footnote{As usual in AdS/CFT
correspondence, the closed string description is valid at large value
of the t'Hooft coupling $g_s Q$, where $Q$ is any of the D-brane
charges.}. Its macroscopic entropy can be computed by
studying the flow of the moduli
with the above choice of charges, leading in either case to
(Eq. \eqref{sbhn8} below)
\be
\label{sbh4c}
S_{BH} = 2\pi \sqrt{Q_2~Q_5~Q_6~N}
\ee
The micro-states correspond to open strings
attached to the D2 and D6 branes, in the background of the 
NS5-branes. In the limit where $Y \times S_1'$ is very small,
they may be described by a two-dimensional field theory extending
along the time and $S_1$ direction.  
In the absence of the NS5-branes, the open strings are described 
at low energy by 
$U(Q_2)\times U(Q_6)$ gauge bosons together with bi-fundamental matter,
which is known to flow to a CFT with central charge $c=6 Q_2 Q_6$
in the infrared (see \cite{David:2002wn} for a detailed analysis of
this point). In the presence of the NS5-branes, localized at $Q_5$
points along $S_1'$, the D2-branes generally  break at the points
where they intersect the NS5-branes. This effectively leads to $Q_5 Q_2$  
independent D2-branes, hence a CFT with  central charge 
$c_{\rm eff}=6 Q_2 Q_5 Q_6$. The extremal micro-states
correspond to the right-moving ground states of that field theory, with
$N$ units of left-moving momentum along $S_1$.
By the Ramanujan-Hardy formula (Eq. \eqref{cardyform} below),
also known as the Cardy formula in the physics literature, 
the number of states carrying $N$ units of momentum
grows exponentially as
\be
\label{cardy256}
\Omega(Q_2,Q_5,Q_6,N) \sim 
\exp\left[ 2\pi \sqrt{\frac{c_{\rm eff}}{6} N} \,\right]
\sim \exp\left[ 2\pi \sqrt{Q_2~Q_5~Q_6~N} \,\right]
\ee
in precise agreement with the macroscopic answer \eqref{sbh4c}. 

While quantitatively successful, this argument has some obvious shortcomings.
The degrees of freedom of the NS5-branes have been totally neglected, and the
D2-branes stretching between each of the NS5-branes were treated independently.
A somewhat more tractable configuration can be obtained by T-dualizing 
along $S_1'$, leading to a bound state of D1-D5 branes in the gravitational
background of Kaluza-Klein monopoles \cite{Johnson:1996ga}. 
The latter are purely gravitational
solutions with orbifold singularities, so in principle can be treated 
by worldsheet techniques. 

Key to this reasoning was the ability to lift the 4-dimensional black hole to
a 5-dimensional black string, whose ground-state dynamics can be described 
by a two-dimensional ``black string CFT'', such that Cardy's formula
is applicable.
This indicates
how to generalize the above argument to 1/2-BPS black holes in $\CN=2$ 
supergravity: any configuration of
D0,D4 branes with vanishing D6-brane charge in type IIA string theory
compactified on a Calabi-Yau threefold $X$ can be lifted in M-theory
to a single M5-brane wrapped  around a 
general divisor (i.e. complex codimension one submanifold) $P$,
with $N$ (the D0-brane charge) units of momentum along the M-theory 
direction \cite{Maldacena:1997de}.
The reduction of the (0,2) tensor multiplet
on the M5-brane worldvolume along the divisor $P$ leads to a 
(0,4) SCFT in 1+1 dimensions, whose left-moving central charge can be computed
with some technical assumptions on $P$:
\be
c_L = 6 \,C(P) + c_2\cdot P
\ee
Here, $C(P)$ is the self-intersection of $P$, while $c_2$ is 
the second Chern class of $X$. Using again Cardy's formula, this leads to
\be
\label{msw}
\Omega(P,N) \sim \exp\left[ 2\pi \sqrt{N \, 
\left( C(P)+ \frac16 c_2\cdot P \right)} \,\right]
\ee
To leading order, this reproduces the macroscopic computation
in $\CN=2$ supergravity, T-dual to \eqref{cardy256},
\be
\label{msw2}
S_{BH} = 2\pi \sqrt{ Q_0 \ C(Q_4) } 
\ee
We shall return to formula \eqref{msw} in Section 6 (Exercise \ref{exomsw}), 
and show that the subleading
contribution proportional to $c_2$ agrees
with the macroscopic computation, provided one incorporates higher-derivative
$R^2$ corrections.

\subsection{Counting $\CN=4$ Dyons via Automorphic Forms \label{dvvform}}
While the agreement between the macroscopic entropy and microscopic 
counting at leading order is already quite spectacular, it is interesting to 
try and understand the corrections to the large charge limit. Ideally,
one would like to be able to compute the exact microscopic degeneracies
for arbitrary values of the charges. 
Here, we shall recall an interesting conjecture, due
to Verlinde, Verlinde and Dijkgraaf (DVV), which purportedly
relates the exact degeneracies of $1/4$-BPS states in $\CN=4$ string
theory, to Fourier coefficients of a certain automorphic 
form \cite{Dijkgraaf:1996it}.
This conjecture has been the subject of much recent work, which
we will not be able to pay justice to in this review. However, 
it plays an important 
inspirational role for some other conjectures relating black hole
degeneracies and automorphic forms, that we will develop in Section \ref{qatt}.

Consider the heterotic string compactified on $T^6$, or equivalently the
type II string on $K3\times T^2$. The moduli space factorizes into
 \be 
\label{modspace}
\frac{Sl(2,\IR)}{U(1)} \times \frac{SO(6,n_v,\IR)}{SO(6)\times
SO(n_v)} 
\ee 
with $n_v=22$.
The first factor is the complex scalar
in the $\CN=4$ gravitational multiplet, and corresponds to
the heterotic axio-dilaton $S$, or equivalently 
to the complexified K\"ahler modulus of $T^2$ on the type II side.
Points in
\eqref{modspace} related by an action of the duality group
$\Gamma=Sl(2,\IZ)\times SO(6,22,\IZ)$ are conjectured to be 
equivalent under non-perturbative dualities.

The Bekenstein-Hawking entropy for $1/4$-BPS black holes is given by
\cite{Cvetic:1995bj}
\be
\label{sqrpp}
S_{BH} = \pi \sqrt{(\vec q_e\cdot \vec q_e)(\vec q_m\cdot \vec q_m) - (\vec q_e\cdot \vec q_m)^2}
\ee
where $\vec q_e$ and $\vec q_m$ are the electric and magnetic charges
in the natural heterotic polarization. $(\vec q_m,\vec q_e)$ 
transform as a doublet of $SO(6,n_v)$ 
vectors under $Sl(2)$. Equation \eqref{sqrpp} is manifestly
invariant under the continuous group $Sl(2,\IR)\times SO(6,22,\IR)$,
a fortiori under its discrete subgroup $\Gamma$.

DVV proposed that the exact degeneracies should be given by the Fourier
coefficients of the inverse of $\Phi_{10}$, the unique cusp form 
of $Sp(4,\IZ)$ with modular weight 10:
\be
\label{dvv}
\Omega(\vec q_e,\vec q_m) \stackrel{?}{=} 
\int_{\gamma} d\tau~
\frac{1}{\Phi_{10}(\tau)} e^{-i \left( \rho \vec q_m^2 
+ \sigma \vec q_e^2 + 2\nu \vec q_e \cdot \vec q_m \right)}
\ee
Here, $\tau=\begin{pmatrix} \rho & \nu \\
\nu & \sigma \end{pmatrix}$ parameterizes Siegel's upper half plane
$Sp(4,\IR)/U(2)$
and $\gamma$ is the contour
$0\leq \rho,\sigma\leq 2\pi, ,0\leq \nu\leq \pi$. 
One may think of $\tau$ as the 
period matrix of an auxiliary genus 2 Riemann surface, with modular group
$Sp(4,\IZ)$. The cusp form $\Phi_{10}$ has an infinite product representation
\be
\Phi_{10}(\tau)=
e^{i(\rho+\sigma+\nu)} 
\prod_{(k,l,m)>0}
\left( 1- e^{i(k\rho+l\sigma+m)} \right)^{c(4kl-m^2)}
\ee
where $c(k)$ are the Fourier coefficients of the elliptic genus of $K3$,
\be
\label{K3ell}
\chi_{K3}(\rho,\nu) =
\sum_{h\geq 0,m\in\IZ} c(4h-m^2) e^{2\pi i (h \rho+mz)}
=
 24 \left( \frac{\theta_3(\rho,z)}{\theta_3(\rho)}
                           \right)^2
     -2 \frac{ \left( \theta_4^4(\rho)-\theta_2^4(\rho) \right)
          \theta_1^2(\rho,z)}{\eta^6(\rho) }\ .
\ee
This shows that  the Fourier coefficients obtained in this fashion 
are (in general non-positive) integers.

The r.h.s. of \eqref{dvv} is manifestly invariant under continuous
rotations in $SO(6,22,\IR)$, hence under its discrete subgroup
 $SO(6,22,\IZ)$. The invariance under $Sl(2,\IZ)$ is more subtle,
and uses the embedding of $Sl(2,\IZ)$ inside $Sp(4,\IZ)$; using
the modular invariance of $\Phi_{10}$,
\bea
\Phi_{10}[(A\tau+B)(C\tau+D)^{-1}]=[\det(C\tau+D)]^{10}\ \Phi_{10}(\tau)\ ,
\eea
one can cancel the action
of $Sl(2,\IZ)$ by a change of contour $\gamma\to \gamma'$, and
deform $\gamma'$ back to $\gamma$ while avoiding singularities.

As a consistency check on this conjecture, one can extract the 
large charge behavior of $\Omega(\vec q_e,\vec q_m)$ by computing the
contour integral in \eqref{dvv} by residues, and obtain agreement
with \eqref{sqrpp} \cite{Dijkgraaf:1996it}. 

\begin{exo} By picking the residue at the 
divisor $D=\rho\sigma+\nu-\nu^2\sim 0$ and using
$\Phi_{10}\sim D^2\ \eta^{24}(\rho')\eta^{24}(\sigma')/\det^{12}(\tau)$
where $\rho'=-\frac{\sigma}{\rho\sigma-\nu^2}, 
\sigma'=-\frac{\rho}{\rho\sigma-\nu^2}$, reproduce the
leading charge behavior \eqref{sqrpp}. You may
seek help from \cite{Dijkgraaf:1996it,Cardoso:2004xf}. 
\end{exo}

A recent ``proof'' of the DVV conjecture has recently been given
by lifting 4D black holes with unit D6-brane charge to 5D, and 
using the Strominger-Vafa relation between degeneracies of
5D black hole and the elliptic genus of the Hilbert scheme
(or symmetric orbifold) ${\rm Hilb}(K3)$ \cite{Shih:2005uc}. We will
return to this 4D/5D lift in Section \ref{veryspe}. The conjecture
has also been generalized to other  $\CN=4$ ``CHL'' models with different
values of $n_v$ in \eqref{modspace} \cite{Jatkar:2005bh,David:2006ji,
Dabholkar:2006xa}. More recently, the $Sp(4,\IZ)$ symmetry has been motivated 
by representing
$1/4$-BPS dyons as string networks on $T^2$, which lift to 
M2-branes with genus 2 topology \cite{Gaiotto:2005hc}. Despite
this suggestive interpretation, it is fair to say that the origin of $Sp(4)$  
remains rather mysterious. In Section \ref{qatt},
we will formulate a similar conjecture, which relies on the
3-dimensional U-duality group $SO(8,24,\IZ)$ obtained by reduction
on a thermal circle, rather than $Sp(4)$.

\section{Special Geometry And Black Hole Attractors \label{spegeo}}
In this section, we expose the formalism of special geometry, which
governs the couplings of vector multiplets in $\CN=2$, $D=4$ supergravity.
We then derive the attractor flow equations, governing the radial evolution
of the scalars in spherically BPS geometries. Finally, we illustrate these
these constructions in the context of ``very special'' supergravity theories, 
which are simple toy models of $\CN=2$ supergravity with symmetric moduli
spaces. We follow the notations of \cite{Ceresole:1995ca}, which gives
a good overview of the essentials of special geometry. Useful
reviews of the attractor mechanism include \cite{Fre:1997jk,Moore:1998pn,
Mohaupt:2000mj}.

\subsection{$\CN=2$ SUGRA and Special Geometry}
A general ``ungauged'' $\CN=2$ supergravity theory in 4 dimensions 
may be obtained by combining massless supersymmetric multiplets
with spin less or equal to 2:
\begin{itemize}
\item[i)] The gravity multiplet, containing the graviton $g_{\mu\nu}$, 
two gravitini $\psi^\alpha_\mu$  and one Abelian gauge field known
as the graviphoton;
 \item[ii)] $n_V$ vector multiplets, each consisting of one Abelian gauge 
field $A_\mu$, two gaugini $\lambda^\alpha$ and one complex scalar.
The complex scalars $z_i$ take values in a {\it projective special K\"ahler
manifold} ${\cal M}_V$ of real dimension $2n_V$.
\item[iii)] $n_H$ hypermultiplets, each consisting of two complex scalars
and two hyperinis $\psi,\tilde\psi$. The scalars take values
in a {\it quaternionic-K\"ahler space} ${\cal M_H}$ of real dimension $4n_H$.
\end{itemize}
Tensor multiplets are also possible, and can be dualized 
into hypermultiplets with special isometries.
At two-derivative order, vector multiplets and hypermultiplets interact
only gravitationally\footnote{This is no longer true in ``gauged'' 
supergravities, where some of the hypermultiplets become charged
under the vectors.}. We will concentrate on the gravitational and
vector multiplet sectors, which control the physics of charged BPS 
black holes. Nevertheless, we will encounter hypermultiplet
moduli spaces in Section \ref{cmap}, when reducing the solutions 
to three dimensions.

The couplings of the vector multiplets, including the geometry
of the scalar manifold ${\cal M}_V$, are conveniently described by
means of a $Sp(2n_V+2)$ principal bundle ${\cal E}$ over ${\cal M}_V$,
and its associated bundle ${\cal E}_V$ 
in the vector representation of $Sp(2n_V+2)$.
The origin of the symplectic symmetry lies in electric-magnetic duality, 
which mixes the $n_V$ vectors ${\cal A}_\mu$
and the graviphoton ${\cal A}_\mu$ together with their magnetic duals.
Denoting a section $\Omega$ by its coordinates $(X^I,F_I)$, the
antisymmetric product
\be
\langle \Omega, \Omega' \rangle = X^I F_I^{'} - X^{'I} F_I 
\ee
endows the fibers with a phase space structure, derived from the 
symplectic form $\langle d\Omega, d\Omega \rangle=dX^I\wedge dF_I$.

The geometry of the scalar manifold ${\cal M}_V$ is completely determined
by a choice of a holomorphic section $\Omega(z)=(X^I(z),F_I(z))$ taking 
value in a Lagrangian cone, i.e. a dilation invariant subspace 
such that $dX^I\wedge dF_I=0$. At generic points, 
one may express $F_I$ in terms of their canonical conjugate $X^I$ via
a characteristic function $F(X^I)$ known as the {\it prepotential}:
\be
F_I = \frac{\pa F}{\pa X^I}\ ,\qquad   F(X^I) = \frac12 X^I F_I\ .
\ee
The second relation reflects the homogeneity of the Lagrangian,
and implies that $F$ is an homogeneous function of degree 2 in the $X^I$.
At generic points, the sections $X^I$ ($I=0\dots n_V)$ may
be chosen as {\it projective} holomorphic coordinates on ${\cal M}_V$ --
equivalently, the $n_V$ ratios $z^i=X^i/X^0$ ($i=1\dots n_V)$ may be
taken as the holomorphic coordinates; these are known as (projective)
special coordinates. Note however that a choice of $F$ breaks manifest
symplectic invariance, so special coordinates may not always be the
most convenient ones. 

\begin{exo} Show that a symplectic transformation $(X^I,F_I)\to(F_I,-X^I)$,
turns the prepotential into its Legendre transform. \label{exoleg}
\end{exo}

Once the holomorphic section $\Omega(z)$ is given, the metric on 
${\cal M}_{V}$ is obtained from the K\"ahler potential
\be
\label{kahl}
\cK(z^i,\bar z^{i}) = - \log K(X,\bar X)\ ,\quad
K(X,\bar X)= i \left( \bar X^I F_I - X^I \bar F_I\right) 
\ee
This leads to a well-defined metric $g_{i\bar j} =\pa_i \pa_{\bar j} {\cK}$,
since under a holomorphic 
rescaling $\Omega\to e^{f(z^i)}\Omega$, $\cK\to \cK-f(z)-\bar f(\bar z)$
changes by a K\"ahler transformation. Equivalently, $\Omega$ should
be viewed as a section of ${\cal E}_V\otimes {\cal L}$ where ${\cal L}$
is the Hodge bundle over ${\cal M}_{V}$, namely a line bundle
whose curvature is equal to the K\"ahler form; its connection
one-form is just $Q=(\pa_i \cK dz^i - \pa_{\bar i} \cK dz^{\bar i})/(2i)$.
The rescaled section 
$\tilde\Omega=e^{\cK/2}\Omega$ is then normalized to 1, and transforms by 
a phase under holomorphic rescalings of $\Omega$. For later purposes, 
it will be convenient to introduce the derived section 
$U_i = D_i \tilde \Omega = (f_i^I, h_{iI})$ where
\bea
\label{deffh}
f_i^I &=& e^{\cK/2} D_i X^I = e^{\cK/2} (\pa_i X^I + \pa_i \cK~X^I) \\
h_{iI} &=& e^{\cK/2} D_i F_I = e^{\cK/2} (\pa_i F_I + \pa_i \cK~F_I) 
\eea
The metric may thus be reexpressed as 
\be
g_{i\bar j} = -i \langle U_i, \bar U_{\bar j} \rangle 
= i \left(f_i^I \bar h_{\bar j I} - h_{iI} \bar f_{\bar j}^I \right)
\ee
After some algebra, one may show that the Riemann tensor on ${\cal M}_V$
takes the form
\be
\label{rcc}
R_{i\bar j k \bar l}= g_{i\bar j} g_{k \bar l} + 
g_{i\bar l} g_{k \bar j} - e^{2\cK} C_{ikm} \bar C_{\bar j\bar l \bar n}
g^{m\bar n}
\ee
where $C_{ijk}$ is a holomorphic, totally symmetric tensor\footnote{We
follow the standard notation in the topological string literature, 
which differs from \cite{Ceresole:1995ca} by a factor of $e^\cK$.}
\be
C_{ijk} = e^{-\cK}  ~ \langle D_i U_j, U_k \rangle 
\ee
The foregoing formalism was in fact geared to produce a solution of
Equation \eqref{rcc}, which embodies the constraint of supersymmetry,
and may be taken as the definition of a 
projective special K\"ahler manifold.

The kinetic terms of the $n_V+1$ Abelian 
gauge fields (including the graviphoton)
may also be obtained from the holomorphic section $\Omega$ as
\be
\label{gaugek}
\begin{split}
{\cal L}_{\rm Maxwell}
=& -\Im{\cal N}_{IJ} ~{\cal F}^I \wedge \star {\cal F}^J + 
\Re{\cal N}_{IJ} ~{\cal F}^I \wedge {\cal F}^J \\
=& \Im \left[ \bar{\cal N}_{IJ} ~{\cal F}^{I-} \wedge \star ~ {\cal F}^{J-} \right] 
+ \mbox{total der.}
\end{split}
\ee
where ${\cal F}^{I-}=({\cal F}^I-i \star {\cal F}^I)/2$, $I=0\dots n_V$  
is the anti-self dual part of the field-strength, and 
${\cal N}_{IJ}$ is defined by the relations
\be
\label{defn}
F_I = {\cal N}_{IJ} X^J\ ,\quad
h_{iI} = \bar{\cal N}_{IJ} f_i^J
\ee
In term of the prepotential $F$ and its Hessian $\tau_{IJ}=\pa_I\pa_J F$,
\be
\label{nijtau}
{\cal N}_{IJ} = \bar\tau_{IJ} + 2i \frac{ (\Im\tau\cdot X)_I ~
(\Im\tau\cdot X)_J}{X\cdot\Im\tau\cdot X} 
\ee
While $\Im\tau_{IJ}$ has indefinite signature $(1,n_V)$, 
$\Im{\cal N}_{IJ}$ is a negative definite matrix, as required for the
positive definiteness of the gauge kinetic terms in \eqref{gaugek}.

\begin{exo} For later use, prove the relations
\be
\label{idn}
\cK  = - \log \left[ -2  X^I [\Im{\cal N}]_{IJ} \bar X^J \right]\ ,\quad
f^I_i [\Im{\cal N}]_{IJ} X^J=0
\ee
\end{exo}

In order to study the invariance of \eqref{gaugek} 
under electric-magnetic duality, it is useful to introduce the
dual vector
\be
{\cal G}_{I;\mu\nu} = 
\frac12 \frac{\pa {\cal L}_{\rm Maxwell}}{\pa{\cal F}^{I;\mu\nu}}
= [\Re{\cal N}]_{IJ} ~{\cal F}^J
+[\Im{\cal N}]_{IJ} ~\star {\cal F}^I
\ee
Under symplectic transformations, ${\cal N}$ transforms
as a ``period matrix'' ${\cal N}\to (C+D {\cal N})(A+B{\cal N})^{-1}$,
while the field strengths $({\cal F}^{I-},{\cal G}_{I}^-= \bar{\cal N}_{IJ} 
{\cal F}^{J-}_{\mu\nu})$
transform as a  symplectic vector, leaving \eqref{gaugek}
invariant. The electric and
magnetic charges $(p^I,q_I)$ are measured by the integral on a 
2-sphere at spatial infinity of  $({\cal F}^{I-},{\cal G}_{I}^-)$, 
and transform
as a symplectic vector too.

One linear combination of the $n_V+1$ field-strengths, the graviphoton
\be
\label{tgrav}
T_{\mu\nu}^- = -2 i ~ e^{\cK/2} ~ X^I~ [\Im{\cal N}]_{IJ} {\cal F}^{J-}_{\mu\nu}
= e^{\cK/2} (X^I {\cal G}_I^- - F_I {\cal F}^{I -})
\ee
plays a distinguished r\^ole, as its associated charge 
measured at infinity
\be
\label{centchar}
Z = e^{\cK/2}~ \left( q_I X^I - p^I F_I \right) \equiv e^{\cK/2} W(X)
\ee
appears as the central charge in $\CN=2$ supersymmetry algebra,
\be
\left\{ Q_{\alpha}^i, \bar Q_{\dot\alpha j} \right\}
= P_\mu \sigma^\mu_{\alpha\dot\alpha} \delta^i_j\ ,\quad
\left\{ Q_{\alpha}^i, Q_{\beta}^j \right\} = Z \epsilon^{ij}
\epsilon_{\alpha\beta}
\ee
In particular, there is a Bogomolony-Prasad-Sommerfeld (BPS) bound on the mass
\be
\label{bpsbound}
M^2 \geq |Z|^2~ m_P^2
\ee
where $m_P$ is the (duality invariant) 4-dimensional Planck scale,
which is saturated when the state preserves 4 supersymmetries out
of the 8 supersymmetries of the vacuum.

\subsection{$\CN=2$ SUGRA and String Theory}
There are several ways to obtain $\CN=2$ supergravities in 4 dimensions 
from string theory. 
Type IIB string compactified on a Calabi-Yau three-fold $Y$ leads 
to $\CN=2$ supergravity with $n_V=h_{2,1}(Y)$ vector multiplets and
$n_H=h_{1,1}(Y)+1$ hypermultiplets. The scalars in ${\cal M}_V$ parameterize
the complex structure of the Calabi-Yau metric on $Y$. The associated
vector fields are the reduction of the 10D Ramond-Ramond 4-form on the 
various 3-cycles in $H_3(Y,\IR)$. The holomorphic section $\Omega$
is then given by
the periods of the holomorphic 3-form $\Omega$ (abusing the notation)
on a symplectic basis $(A^I,B_I)$ of $H_3(Y,\IR)$:
\be
X^I = \int_{A^I} \Omega\ ,\quad F_I = \int_{B^I} \Omega
\ee
The K\"ahler potential on the moduli of complex structures is just
\be
\cK = -\log \left[ i \int_Y \Omega \wedge \bar\Omega \right]
\ee
which agrees with \eqref{kahl} by Riemann's bilinear identity. As
we shall see later, it is determined purely at tree-level, and
can be computed purely in field theory. The 
central charge of a state with electric-magnetic charges $p^I,q_I$
may be rewritten as
\be
Z = \frac{\int_\gamma \Omega}{\sqrt{i \int_Y \Omega \wedge \bar\Omega}}
\ee
where $\gamma=q_I A^I - p^I B_I$, and is recognized as the mass of a D3-brane
wrapped on a special Lagrangian 3-cycle $\gamma\in H_3(Y,\IZ)$.

On the other hand, the scalars in 
${\cal M}_H$ parameterize the complexified K\"ahler structure of $Y$, 
the fluxes (or more appropriately, Wilson lines) of the 
Ramond-Ramond two-forms along $H_{\rm even}(Y,\IR)$, as well as 
the axio-dilaton. The axio-dilaton, zero
and six-form RR potentials form a ``universal hypermultiplet''
sector inside ${\cal M}_H$. In contrast to the vector-multiplet
metric, the hyper-multiplet metric receives one-loop and 
non-perturbative corrections
from Euclidean D-branes and NS-branes wrapped on $H_{\rm even}(Y)$. 

The situation in type IIA string compactified on a Calabi-Yau three-fold
$\tilde Y$ is reversed: the vector-multiplet moduli space describes the
complexified K\"ahler structure of $\tilde Y$, while the hypermultiplet
moduli space describes its complex structure, together with the Wilson
lines of the Ramond-Ramond forms along $H_{\rm odd}(\tilde Y)$
and the axio-dilaton. As in IIB,
the vector-multiplet moduli space is determined at tree-level only,
but receives $\alpha'$ corrections. Letting
$J=B_{NS}+i \omega_K$ be the complexified K\"ahler form,
$\gamma^A$ be a basis of $H_{1,1}(\tilde Y,\IZ)$ and $\gamma_A$ the
dual basis of $H_{2,2}(\tilde Y,\IZ)$,
the holomorphic section $\Omega$ (not to be confused with the
holomorphic three-form on $\tilde Y$) is determined projectively
by the special coordinates
\be
X^A/ X^0 = \int_{\gamma^A} J\ ,\quad F_A/X^0 = \int_{\gamma_A} J\wedge J
\ee
In the limit of large volume, the K\"ahler potential (in the
gauge $X^0=1$) is given by the volume in string units,
\be
\cK= -\log \int_{\tilde Y} J \wedge J \wedge J
\ee
originating from the cubic prepotential 
\be
\label{f00iia}
F = -\frac16 C_{ABC} \frac{X^A X^B X^C}{X^0}
+ \dots
\ee
Here, $C_{ABC}$ are the intersection numbers of the 4-cycles $\gamma_{A,B,C}$.
At finite volume, there are corrections to \eqref{f00iia} from
worldsheet instantons wrapping effective curves in 
$H_2^+(\tilde Y,\IZ)$, to which we will return 
in Section \ref{gwgkdt}. The 
central charge following from \eqref{f00iia} is
\be
Z = e^{\cK/2} X^0 \left( q_0 + q_A \int_{\gamma^A} J - p^A
\int_{\gamma_A} J \wedge J - p^0 \int_{\tilde Y} J\wedge J \wedge J \right)
\ee
so that $q_0,q_A,p^A,p^0$ can be identified as 
the D0,D2,D4 and D6 brane charge, respectively. 

While \eqref{f00iia} expresses the complete prepotential in terms of the
geometry of $\tilde Y$, the most practical way of computing it is to
use mirror symmetry, which relates type IIA compactified on $\tilde Y$
to type IIB compactified on $Y$, where $(Y,\tilde Y)$ form a ``mirror pair'';
this implies in particular that $h_{1,1}(Y) = h_{2,1}(\tilde Y)$ and
$h_{1,1}(\tilde Y)  = h_{2,1}(Y)$ (see \cite{MR2003030} for a review).

On the other hand, the tree-level metric 
on the hypermultiplet moduli space ${\cal M}_H$ in type IIA compactified
on $\tilde Y$ may be obtained from the vector-multiplet metric ${\cal M}_V$
in type IIB compactified on the {\it same} Calabi-Yau $\tilde Y$, by
compactifying on a circle $S^1$ to 3 dimensions, T-dualizing along $S^1$
and decompactifying back to 4 dimensions.
We shall return to this ``c-map'' procedure in Section \ref{cmap}.

Finally, another way to obtain $\CN=2$ supergravity in 4 dimensions is
to compactify the heterotic string on $K3 \times T^2$. Since the
heterotic axio-dilaton is now a vector-multiplet, ${\cal M}_V$ now
receives loop and instanton corrections, while ${\cal M}_H$ is
determined purely at tree-level (albeit with $\alpha'$ corrections).

\subsection{Attractor Flows and Bekenstein-Hawking Entropy \label{afbh}}
We now turn to static, spherically symmetric BPS black hole solutions of $\CN=2$ 
supergravity. The assumed isometries lead to the metric ansatz
\be
\label{statan}
ds^2 = -e^{2U}dt^2 + e^{-2U} ( dr^2 + r^2 d\Omega_2^2 )
\ee
where $d\Omega_2^2=d\theta^2+\sin^2\theta ~d\phi^2$ is the round metric on
$S^2$, and $U$ depends on $r$ only. We took advantage of the BPS
property to restrict to flat 3D spatial slices\footnote{This condition
may be relaxed if one allows for a non-trivial profile of the
hypermultiplets \cite{Huebscher:2006mr}.}. 
Moreover, the scalars $z^i$
in the vector multiplet moduli space are taken to depend on $r$ only.
The gauge fields are uniquely determined by
the equations of motion and Bianchi identities:
\be
{\cal F}^{I-} = \frac12 \left[ p^I - i [\Im{\cal N}]^{IJ}
\left( q_J - [\Re{\cal N}]_{JK} p^K \right) \right]\cdot
\left[ \sin\theta ~d\theta \wedge d\phi 
- i \frac{e^{2U}}{r^2} dt\wedge dr \right]
\ee
where $(p^I,q_I)$ are the magnetic and electric charges, and
$[\Im{\cal N}]^{IJ}= [ \Im{\cal N}]^{-1}_{IJ}$.  

Assuming that the solution preserves half of the 8 supersymmetries, the
gravitino and gaugino variations lead to a set of first-order 
equations \cite{Ferrara:1995ih,Ferrara:1996um,Ferrara:1997tw,Moore:1998pn}
\footnote{ We shall provide a full derivation of 
\eqref{att1},\eqref{att2} in Section \ref{qatt},
but for now we accept them and proceed with their consequences.}
\bea
r^2 \frac{dU}{dr} &=& |Z| ~e^{U} \label{att1}\\
r^2 \frac{dz^i}{dr} &=& 2 ~e^U~ g^{i\bar j} \pa_{\bar j} |Z| \label{att2}
\eea
where $Z$ is the central charge defined in \eqref{centchar}.
These equations govern the radial evolution of $U$ and $z^i(r)$,
and are usually referred to as ``attractor flow equations'', for reasons
which will become clear shortly. The boundary conditions are
such that $U(r\to \infty)\to 0$ at spatial infinity, while
the vector multiplet scalars $z^i$ go to their vacuum values 
$z^i_\infty$. The
black hole horizon is reached when the time component 
of the metric $g_{tt}=e^{2U}$ vanishes, i.e. at $U=-\infty$.

\begin{figure}
\centerline{\hfill\includegraphics[height=9cm]{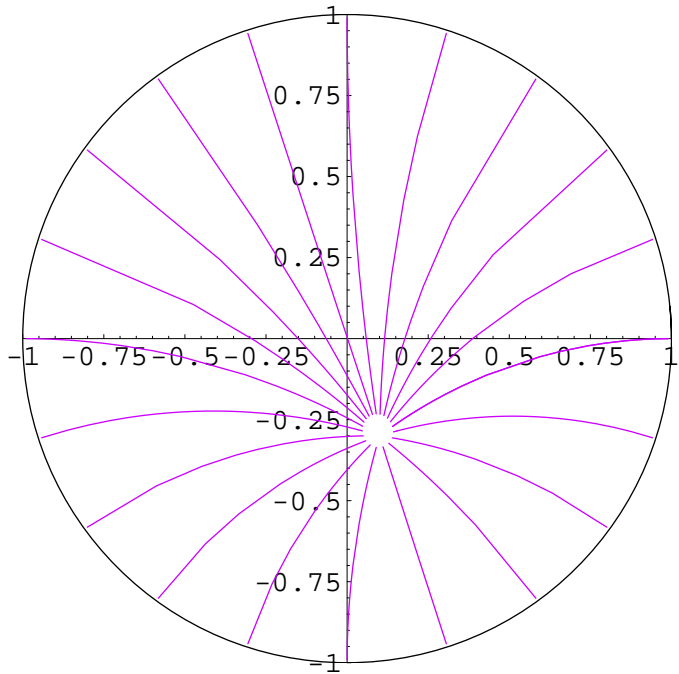}\hfill}
\caption{Radial flow for the Gaussian one-scalar model, for
charges $(p^0,p^1,q_1,q_0)=(4,1,1,2)$. All trajectories are
attracted to $z_*=X^1/X^0=(1-3i)/10$ at $r=0$.
\label{su22fig0}}
\end{figure}

Defining $\mu=e^{-U}$, 
so that $r^2 d\mu/dr=-|Z|$, the second
equation may be cast in the form of a gradient flow, or RG flow,
\be
\mu\frac{dz^i}{d\mu} = - g^{i\bar j} \pa_{\bar j} \log |Z|^2 \label{att3}
\ee
As a consequence, $|Z|$ decreases from spatial infinity, where $\mu=1$,
to the black hole horizon, when $\mu\to +\infty$. The scalars $z^i$
therefore settle to values $z^i_*(p,q)$ which minimize the
BPS mass $|Z|$; in particular, the vector multiplet scalars 
are ``attracted'' to a fixed value at 
the horizon, independent\footnote{In some cases, there can exist different
basins of attraction, leading to a discrete set of possible values
$z^i_*(p,q)$ for a given choice of charges. This is typically
connected with the ``split attractor flow'' phenomenon \cite{Denef:2000nb}.}
of the asymptotic values $z^i_{\infty}$, and determined 
only by the charges $(p^I,q_I)$. This attractor behavior is illustrated
in Figure \ref{su22fig0} for the case of the Gaussian one-scalar model with 
prepotential $F=-i[(X^0)^2-X^1)^2]/2$, whose moduli space corresponds to
the Poincar\'e disk $|z|<1$. It should be noted that 
the attractor behavior is in fact a consequence of extremality
rather than supersymmetry, as was first recognized in \cite{Ferrara:1997tw}.

We shall assume that the charges $(p^I,q_I)$  are chosen such that at the 
attractor point, $Z=Z_*\neq 0$, since otherwise the solution becomes singular.
Equation \eqref{att1} may be easily integrated near the horizon, 
\be
\mu = e^{-U} \sim |Z_*|/r
\ee
Defining $z=|Z_*|^2/r$, it is easy to see that the near-horizon metric
becomes $AdS_2\times S^2$, as in \eqref{ads2s2}, where the prefactor
$(p^2+q^2)$ is replaced by $|Z_*|^2$. The Bekenstein-Hawking
entropy is one quarter of the horizon area,
\be
\label{bhe}
S_{BH} = \frac{1}{4}\cdot 4\pi \lim_{r\to 0} e^{-2U} r^2 
= \pi |Z_*|^2
\ee
This is a function of the electric and magnetic charges only, by
virtue of the attractor mechanism, except for possible discrete labels
(or ``area codes'') corresponding to different basins of attraction.

We shall now put these results in a more manageable form, by making use
of some special geometry identities discussed in Section \ref{spegeo}.
First, using the derived section $U_i=(f_i^I,h_{iI})$ defined in \eqref{deffh}
and the property \eqref{defn}, one easily finds
\be
\label{dZ}
\pa_{i}Z = f_i^I(q_I - \bar {\cal N}_{IJ} p^J) - \frac12 Z \pa_i \cK\ ,\quad
\pa_{\bar i} Z = \frac12 Z \pa_{\bar i}\cK 
\ee
so that
\be
\label{dmodZ}
\frac{\partial_i |Z|}{|Z|}
=\frac12 \left( \frac{\partial_i Z}{Z} + \frac{\partial_i \bar Z}{\bar Z}
\right)
 = \frac{1}{Z} f_i^I 
\left( q_I - \bar {\cal N}_{IJ} p^J  \right)
\ee
This allows to rewrite \eqref{att2} as
\be
r^2 \frac{dz^i}{dr} = -\sqrt{\frac{Z}{\bar Z}} e^U g^{i\bar j}
\bar f^J_{\bar j} (q_I - {\cal N}_{IJ} p^J)
\ee
The stationary value of $z^i$ at the horizon is thus obtained by 
setting the rhs of this equation to zero, i.e.
\be
\label{statv}
f^J_{i} (q_I - \bar {\cal N}_{IJ} p^J) = 0
\ee
The rectangular matrix $f^I_i$ has a unique zero eigenvector, given by the
second equality in \eqref{idn}. Hence, \eqref{statv} implies 
\be
\label{qpalpha}
q_I - \bar{\cal N}_{IJ} p^J = C~ \Im{\cal N}_{IJ} X^J
\ee
Contracting either side with $\bar X^I$ and using the first
equation in \eqref{idn} allows to compute the value of $\alpha$,
\be
C = -2 {\bar Z}~e^{\cK/2}
\ee
Moreover, using again \eqref{defn},  one may rewrite \eqref{qpalpha} and
its complex conjugate, equivalently as two real equations
\be
\label{pcx}
p^I = \Im( C X^I)\ ,\quad q_I = \Im( C F_I) 
\ee
while the Bekenstein-Hawking entropy \eqref{bhe} is given by
\be
S_{BH} = \frac{\pi}{4} |C|^2 e^{-\cK(X,\bar X)} = 
\frac{i\pi}{4} |C|^2 \left( \bar X^I F_I - X^I \bar F_I \right)
\ee
Making use of the fact that near the horizon, $e^{-U}\sim |Z_*|/r$, 
it is convenient to rescale the holomorphic section $\Omega=(X^I,F_I)$ into 
\be
\begin{pmatrix} Y^I \\ G_I\end{pmatrix}
 = 2 i ~ r ~e^{\frac12\cK(X,\bar X) - U} \sqrt{\frac{\bar Z}{Z}} 
\begin{pmatrix} X^I \\ F_I\end{pmatrix}
\ee
in such a way that
\be
e^{-\cK(Y,\bar Y)} = 4 r^2 e^{-2U}\ ,\quad
\arg W(Y) =\pi/2
\ee
where we defined, in line with \eqref{kahl} and \eqref{centchar},
\be
\label{kywy}
K(Y,\bar Y)=\left[i\left( \bar Y^I G_I - Y^I \bar G_I 
\right) \right] = e^{-\cK(Y,\bar Y)}\ ,\quad W(Y)=q_I Y^I - p^I G_I
\ee
In this fashion, we have incorporated the geometric variable $U$ into the
symplectic section $(Y^I,G_I)$, and fixed the phase.
In this new ``gauge''\footnote{This is an abuse of language, since 
the scale factor is a priori not a holomorphic function of $z^i$.}, 
which amounts to setting $C\equiv i$,
\eqref{pcx} and \eqref{bhe} simplify into 
\bea
\label{stabeq}
\begin{pmatrix} p^I \\ q_I\end{pmatrix}
&=&\Re \begin{pmatrix} Y^I \\ G_I\end{pmatrix} \\
S_{BH}&=&\frac{\pi}{4} K(Y,\bar Y) 
=\frac{i\pi}{4} \left[ \bar Y^I G_I - Y^I \bar G_I \right] 
\label{stabeq2}
\eea
These equations, some times known as ``stabilization equations'', are
the most convenient way of summarizing the endpoint of the attractor 
mechanism, as will become apparent in the next subsection.

\subsection{Bekenstein-Hawking entropy and Legendre transform}
A key observation for later developments is that the Bekenstein-Hawking
entropy \eqref{stabeq} is simply related by Legendre transform\footnote{This
was first observed in \cite{Behrndt:1996jn}, and spelled out more clearly 
in \cite{Ooguri:2004zv}.} to
the tree-level prepotential $F$. To see this, note that the
first equation in \eqref{stabeq} is trivially solved by setting
$Y^I= p^I + i \phi^I$, where $\phi^I$ is real. The entropy is then 
rewritten as
\bea
S_{BH} &=&\frac{i\pi}{4} \left[ (Y^I-2i\phi^I) G_I - (\bar Y^I+2 i \phi^I) 
\bar G_I \right] \\
&=& \frac{i\pi}{2}\left[ F(Y)-\bar F(\bar Y) \right]
+ \frac{\pi}{2} \phi^I 
\left[ G_I+\bar G_I \right]
\label{osvman}
\eea
where, in going from the second to the third line, we used the homogeneity
of the prepotential, $Y^I G_I=2F(Y)$. 
On the other hand, the second stabilization equation
yields
\be
q_I = \frac12 \left( G_I + \bar G_I\right) 
= \frac{1}{2i} 
\left( \frac{\pa F}{\pa \phi^I}-\frac{\pa \bar F}{\pa \phi^I} \right)
\ee
Thus, defining
\be
\label{freef}
{\cal F}(p^I,\phi^I) = -\pi~ \Im\left[ F(p^I+i\phi^I) \right]
\ee
the last equation in \eqref{osvman} becomes
\be
\label{sbhosv}
S_{BH}(p^I,q_I) 
= \langle {\cal F}(p^I,\phi^I) + \pi~\phi^I q_I \rangle_{\phi^I}
\ee
where the r.h.s. is evaluated at its extremal value with respect to $\phi^I$.
In usual thermodynamical terms, this implies that ${\cal F}(p^I,\phi^I)$
should be viewed as the free energy of an ensemble of black holes in which
the magnetic charge $p^I$ is fixed, but the electric charge $q_I$ is free
to fluctuate at an electric potential $\pi \phi^I$. The implications
of this simple observation will be profound in Section \ref{osvconj}, 
when we discuss
the higher-derivative corrections to the Bekenstein-Hawking entropy.

\begin{exo} 
Apply this formalism to show that the entropy of a D0-D4 bound state
in type IIA string theory compactified on a Calabi-Yau three-fold, 
in the large charge regime, is given by
\be
\label{sd0d4}
S_{BH} = 2\pi \sqrt{-C_{ABC} p^A p^B p^C q_0}
\ee
and compare to \eqref{msw2}.
\end{exo}

\begin{exo} Show that the
Bekenstein-Hawking entropy \eqref{stabeq2} can be obtained by
extremizing
\be
\Sigma_{p,q}(Y,\bar Y) = -\frac{\pi}{4} \left[ K(Y,\bar Y) + 2 i 
[W(Y) - \bar W(\bar Y)] \right] 
\ee
with respect to $Y,\bar Y$, where $K(Y,\bar Y)$ and $W(Y)$
are defined in \eqref{kywy} \cite{Ooguri:2005vr,LopesCardoso:2006bg}. 
Observe that \eqref{sbhosv}
is recovered by extremizing over $\Re(Y)$.\label{altvar}
\end{exo}

\begin{exo}
Define the Hesse potential $\Sigma(\phi^I,\chi_I)$ 
as the Legendre transform of the 
topological free energy with respect to the magnetic charges $p^I$,
\be
\Sigma(\phi^I,\chi_I) = \langle\
{\cal F}(p^I,\phi_I) + \pi\, \chi_I p^I \rangle_{p^I}
\ee
Show that the dependence of $\Sigma$ on the electric and
magnetic potentials $(\phi^I,\chi_I)$ is identical (up to a sign)
to that of the black hole entropy $S_{BH}$ on the charges 
$(p^I,q_I)$. Compare to $\Sigma_{p,q}$ in the previous Exercise.
\label{exohesse}
\end{exo}

\subsection{Very Special Supergravities and Jordan Algebras \label{veryspe}}
In the remainder of this section, we illustrate the previous results
on a special class of $\CN=2$ supergravities, whose vector-multiplet
moduli spaces are given by symmetric spaces. These are interesting
toy models, which arise in various truncations of string compactifications.
Moreover, they are related to by analytic continuation to $\CN>2$ theories,
which will be further discussed in Section \ref{qatt}.

The simplest way to construct these models is to start from 5 
dimensions \cite{Gunaydin:1983bi}:
the vector multiplets consist of one real scalar for each vector, and their
couplings are given by
\be
S=\int d^5x~\sqrt{-g}~\left( R - 
G_{ij} \pa_\mu \phi^i \pa_\mu \phi^j\right) 
- \overset{\circ}{a}_{AB}~ F^A \wedge \star \ F^B + 
\frac{1}{24}\int C_{ABC}~ A^A \wedge 
F^B \wedge F^C
\ee
where the Chern-Simons-type couplings $C_{ABC}$ are constant, for gauge
invariance. $\CN=2$ supersymmetry requires the real scalar fields $\phi^i$
to take value in the cubic hypersurface ${\cal M}_5=\{
\xi , N(\xi)=1\}$ in an ambient space $\xi\in \IR^{n_V+1}$, where
\be
\label{nxi}
N(\xi) = \frac16 C_{ABC}~\xi^A~\xi^B~\xi^C
\ee
The metric
$G_{ij}$ is then the pull-back of the ambient space 
metric $a_{AB} d\xi^A d\xi^B$ to ${\cal M}_5$, where 
\be
\label{addn}
a_{AB}=-\frac12 \pa_{\xi^A}\pa_{\xi^B}N(\xi)
\ee
The gauge couplings $\overset{\circ}{a}_{AB}$ are instead given by 
the restriction
of $a_{AB}$ to the hypersurface ${\cal M}_5$.
Upon reduction from 5 dimensions to 4 dimensions, using the
standard Kaluza-Klein ansatz
\be
ds^2_5 = e^{2\sigma} ( dy + B_\mu dx^\mu)^2 + e^{-\sigma} g_{\mu\nu}
dx^\mu dx^\nu
\ee
the Kaluza-Klein gauge field $B_\mu$ provides the graviphoton,
while the constraint $N(\xi)=1$ is relaxed to $N(\xi)=e^{3\sigma}$.
Moreover, $\xi^A$ combine with the fifth components $a^A$ of the gauge fields
$A^A$ into complex scalars $t^A=a^A + i \xi^A=X^A/X^0$, which are the
special coordinates of a special K\"ahler manifold ${\cal M}_4$
with prepotential
\be
\label{5d4df}
F = N(X^A)/ X^0
\ee

In general, neither ${\cal M}_5$ nor ${\cal M}_4$ are 
symmetric spaces. The conditions for ${\cal M}_5$ to be a symmetric space
were analyzed in \cite{Gunaydin:1983bi}, and found to have a remarkably simple
interpretation in terms of Jordan algebras: these are commutative,
non-associative algebras $J$ satisfying the ``Jordan identity''
\be
x \circ ( y \circ x^2) = (x \circ y) \cdot x^2 
\ee
where $x^2= x \circ x$ (see e.g. \cite{MR0466235} for a nice review).

\begin{exo} Show that the algebra of $n\times n$ hermitean matrices with
product $A\circ B=\frac12(AB+BA)$ is a Jordan algebra.
\end{exo}

Jordan algebras were introduced and completely classified
in \cite{Jordan:1933vh} 
in an attempt to generalize quantum mechanics beyond the field 
of complex numbers. The ones relevant here are those which admit a norm
$N$ of degree 3 -- rather than giving the axioms of the norm, we shall 
merely list the allowed possibilities:
\begin{itemize}
\item[i)] One trivial case: $J=\IR$, $N(\xi)= \xi^3$
\item[ii)] One infinite series: $J=\IR \oplus \Gamma$ 
where $\Gamma$ is the Clifford algebra of $O(1,n-1)$, 
$N(\xi \oplus \gamma)= \xi \gamma^a \gamma^b \eta_{ab}$ 
\item[iii)] Four exceptional cases: 
$J={\rm Herm}_3(\mathbb{D})$, the algebra of $3\times 3$ 
hermitean matrices $\xi=\begin{pmatrix}
\alpha_1 & x_3 & \bar x_2 \\
\bar x_3 & \alpha_2 & x_1 \\
x_2 & \bar x_1 & \alpha_3\end{pmatrix}$
where $\alpha_i$ are real and $x_i$ are in one 
of the four ``division algebras'' $\mathbb{D}=\IR,\IC$, the 
quaternions $\IH$ or octonions $\IO$.
In each of these cases, the cubic norm is the ``determinant'' of $\xi$
\be
\label{ndet}
N(\xi) = 
\alpha_1\alpha_2\alpha_3
-\alpha_1 x_1\bar x_1
-\alpha_2 x_2\bar x_2
-\alpha_3 x_3\bar x_3+2\Re(x_1 x_2 x_3)
\ee
For $J_3^\IC$, this is equivalent to the determinant of an
unconstrained $3\times 3$ real matrix, and for $J_3^\IH$ to
the Pfaffian of a $6\times 6$ antisymmetric matrix.
\end{itemize}
To each of these Jordan algebras, one may attach several invariance groups,
summarized in Table \ref{tabgro}: 
\begin{itemize}
\item[a)] $\Aut(J)$,
the group of automorphisms of $J$, which leaves invariant the structure
constants of the Jordan product; 
\item[b)] $\Str(J)$, the ``structure'' group, which leaves invariant the
norm $N(\xi)$ up to a rescaling; and 
the ``reduced structure group'' $\Str_0(J)$, where
the center has been divided out;
\item[c)] $\Conf(J)$, the ``conformal'' group, such that the norm
of the difference of two elements $N(\xi-\xi')$ is multiplied
by a product $f(\xi)f(\xi')$; as a result, the ``cubic light-cone''
$N(\xi-\xi')=0$ is invariant;
\item[d)] $\QConf(J)$, the ``quasi-conformal group'', which we will
describe in Section \ref{quasiref}.
\end{itemize}

\begin{table}
\begin{equation*}
\begin{array}{|c||c|c|c|c|}
\hline
J & \Aut(J) & \Str_0(J) & \Conf(J) & \QConf(J) \\
\hline\hline
\IR & 1 & 1 & Sl(2,\IR) & G_{2(2)} \\
\hline
\IR\oplus \Gamma_{n-1,1} &  SO(n-1)& SO(n-1,1) 
  & Sl(2) \times SO(n,2) & SO(n+2,4) \\
\hline
J_3^{\IR} & SO(3) & Sl(3,\IR) & Sp(6)  & F_{4(4)}\\
\hline
J_3^{\IC} & SU(3) & Sl(3,\IC) & SU(3,3) & E_{6(+2)}\\
\hline
J_3^{\IH} & USp(6)& SU^*(6)& SO^*(12) & E_{7(-5)}\\
\hline
J_3^{\IO} & F_4   & E_{6(-26)} & E_{7(-25)} & E_{8(-24)}\\
\hline
\end{array}
\end{equation*}
\caption{Invariance groups associated to degree 3 Jordan algebras.
The lower $4\times 4$ part is known as the ``Magic Square'', due
to its symmetry along the diagonal \cite{Gunaydin:1983rk}.
\label{tabgro}}
\end{table}

In the case ii) above, $\Aut(J),\Str(J)$ and $\Conf(J)$ are just the 
orthogonal group $SO(n-1)$, Lorentz group $SO(n-1,1)$ and conformal
group $SO(n,2)$ times an extra $Sl(2)$ factor.

The relevance of these groups for physics is as follows: choosing
$N(\xi)$ in  \eqref{nxi} to be equal to the norm form
of a Jordan algebra $J$, the vector-multiplet 
moduli spaces for the resulting $\CN=2$ supergravity in $D=5$ and $D=4$ 
are symmetric spaces
\be
\label{verymod}
{\cal M}_5 = \frac{\Str_0(J)}{\Aut(J)}\ ,\quad
{\cal M}_4 = \frac{\Conf(J)}{\widetilde{\Str}_0(J) \times U(1)}\ ,\quad
\ee
where $\widetilde{\Str}_0(J)$ denotes the compact real form of $\Str_0(J)$.
In either case, the group 
in the denominator is the
maximal subgroup of the one in the numerator, which guarantees
that the quotient has positive definite signature.
The resulting spaces are shown in Table \ref{tabmod}, together with the ones
which appear upon reduction to $D=3$ on a space-like and
time-like direction respectively, to be discussed in
Section \ref{quasiref} below. The first column
indicates the number of supercharges in the corresponding supergravity:
the above discussion applies strictly speaking to cases with
8 supercharges (i.e. $\CN=2$ supersymmetry in 4 dimensions),
but other cases can also be reached with similar techniques,
using different real forms of the Jordan algebras above\footnote{For 
example, the cubic invariant of $E_{6(6)}$ appearing
in $\CN=8$ supergravity can be obtained from \eqref{ndet}
by replacing the usual octonions $\IO$ by the split
octions $\IO_s$, whose norm $x\bar x$ has split 
signature (4,4), see \cite{Ferrara:2006yb} for a recent discussion.}.

\begin{table}
\begin{equation*}
\begin{array}{|c|c|c|c|c|c|}
\hline
Q & J & D=5 & D=4 & D=3 & D=3^* \\ \hline\hline
8   & & & \frac{SU(n,1)}{SU(n) \times U(1)}
      & \frac{SU(n+1,2)}{SU(n+1) \times SU(2)\times U(1)}
      & \frac{SU(n+1,2)}{SU(n,1) \times Sl(2)\times U(1)} \\
\hline
8 & \Gamma_{n-1,1} & \IR \times \frac{SO(n-1,1)}{SO(n-1)} 
  & \frac{SO(n,2)}{SO(n)\times SO(2)} \times \frac{Sl(2)}{U(1)}
  & \frac{SO(n+2,4)}{SO(n+2)\times SO(4)}
  & \frac{SO(n+2,4)}{SO(n,2)\times SO(2,2)}\\ \hline
8  & &  & \frac{Sl(2)}{U(1)}              &\frac{SU(2,1)}{SU(2)\times U(1)} 
             & \frac{SU(2,1)}{Sl(2)\times U(1)} \\ \hline
8 & \IR & \varnothing & \frac{Sl(2)}{U(1)} & \frac{G_{2(2)}}{SO(4)} 
                                 & \frac{G_{2(2)}}{SO(2,2)} \\ \hline
8 & J_3^{\IR} & \frac{Sl(3)}{SO(3)} & \frac{Sp(6)}{SU(3)\times U(1)} 
                                 & \frac{F_{4(4)}}{USp(6)\times SU(2)}
                                 & \frac{F_{4(4)}}{Sp(6)\times Sl(2)}\\
\hline
8 & J_3^{\IC} 
& \frac{Sl(3,C)}{SU(3)} & \frac{SU(3,3)}{SU(3)\times SU(3)\times U(1)}  
                          & \frac{E_{6(+2)}}{SU(6)\times SU(2)}
                          & \frac{E_{6(+2)}}{SU(3,3)\times Sl(2)}\\
\hline
24 & J_3^{\IH} 
& \frac{SU^*(6)}{USp(6)} & \frac{SO^*(12)}{SU(6)\times U(1)}      
                           & \frac{E_{7(-5)}}{SO(12)\times SU(2)}
                           & \frac{E_{7(-5)}}{SO^*(12)\times Sl(2)}\\
\hline
8 & J_3^{\IO}
& \frac{E_{6(-26)}}{F_4}  & \frac{E_{7(-25)}}{E_6\times U(1)}
                           & \frac{E_{8(-24)}}{E_7\times SU(2)}
                           & \frac{E_{8(-24)}}{E_{7(-25)}\times Sl(2)}\\
\hline
10 & & & &  \frac{Sp(2n,4)}{Sp(2n)\times Sp(4)} & \\
\hline
12 & & & &  \frac{SU(n,4)}{SU(n)\times SU(4)} & \\
\hline
16 & \Gamma_{n-5,5}
  & \IR  \times \frac{SO(n-5,5)}{SO(n-5)\times SO(5)}  & \frac{Sl(2)}{U(1)} \times \frac{SO(n-4,6)}{SO(n-4)\times SO(6)} 
  & \frac{SO(n-2,8)}{SO(n-2)\times SO(8)}
  & \frac{SO(n-2,8)}{SO(n-4,2)\times SO(2,2)}\\
\hline
18 & & & & \frac{F_{4(-20)}}{SO(9)} & \\
\hline
20  & M_{1,2}(\IO) & & \frac{SU(5,1)}{SU(5)\times U(1)}
        & \frac{E_{6(-14)}}{SO(10)\times SO(2)} &
\frac{E_{6(-14)}}{SO^*(10)\times SO(2)}  \\
\hline
32 & J_3^{\IO_s}
  & \frac{E_{6(6)}}{USp(8)} & \frac{E_{7(7)}}{SU(8)}
                            & \frac{E_{8(8)}}{SO(16)}
                            & \frac{E_{8(8)}}{SO^*(16)}\\
\hline
\end{array}
\end{equation*}
\caption{Moduli spaces for supergravities with symmetric moduli spaces.
The last column refers to the reduction from 4 dimensions to 3 along
a time-like direction, which will become relevant in Section \ref{qatt}. 
\label{tabmod}}
\end{table}

The $\Str_0(J)$  invariance of the metric on ${\cal M}_5$ 
is indeed
obvious from \eqref{addn} above. The $\Conf(J)$ invariance of
the metric on the special K\"ahler space ${\cal M}_4$ is manifest too,
since the K\"ahler potential following from \eqref{5d4df} is
the proportional to the log of the ``cubic light-cone'',
\be
\cK(z,\bar z) = - \log N( z_i - \bar z_i)\ ,
\ee
invariant under $\Conf(J)$ up to K\"ahler transformations. 
Such special K\"ahler spaces are known as hermitean symmetric tube domains,
and are higher-dimensional analogues of Poincar\'e's upper half plane.

It should be pointed out that 
there also exist $D=4$ SUGRAs with symmetric moduli space which
do not descend from 5 dimensions: they may be described by 
a generalization of Jordan algebras known as ``Freudenthal
triple systems'', but we will not discuss them in any detail here.
Similarly, there exist $D=3$ supergravity theories with symmetric moduli 
spaces which cannot be lifted to 4 dimensions.

In general, it is not known whether these very special supergravities
arise as the low-energy limit of string theory. All except the 
exceptional $J_3^{\IO}$ case can be obtained formally by truncation of 
$\cN=8$ supergravity, but it is in general unclear how to 
consistently enforce this truncation. A notable exception is the 
case based on $J=\Gamma_{9,1}$, which is realized in type IIA string
theory compactified on a freely acting orbifold 
of $K3\times T^2$, or a CHL orbifold of the heterotic string
on $T^6$ \cite{Ferrara:1995yx}. The model with $J=J_3^{\IC}$ 
arises in the untwisted sector of type IIA compactified on 
the ``Z-manifold'' $T^6/\IZ_3$ \cite{Ferrara:1988fr}, but there are 
also massless fields from the twisted sector. We shall mostly
use these theories at toy models in the sequel, and assume 
that discrete subgroups $\Str_0(J,\IZ)$ and $\Conf(J,\IZ)$
remain as quantum symmetries of the full quantum theory, if it 
exists.

\subsection{Bekenstein-Hawking Entropy in Very Special Supergravities}

As an illustration of the simplicity of these models, we shall now proceed 
and compute the Bekenstein-Hawking entropy for BPS black holes with arbitrary
charges, following \cite{Pioline:2005vi}. 
A key property which renders the computation tractable is the fact
that the prepotential \eqref{5d4df} obtained from any Jordan algebra is 
invariant (up to a sign) under Legendre transform in all variables,
namely
\be
\label{nleg}
\langle N(X^A)/X^0 + X^A Y_A + X^0 Y_0 \rangle_{X^I} = - N(Y)/Y^0
\ee

\begin{exo} Show 
that \eqref{nleg} is equivalent to the ``adjoint identity'' for
Jordan algebras, $X^{\sharp\sharp}=N(X) X$ where $X^\sharp_A = 
\frac12 C_{ABC}
X^B X^C$ is the ``quadratic map'' from $J$ to its dual.
\end{exo}

In fact, just imposing \eqref{nleg} leads to the same classification
i),ii),iii) as above. This was shown independently in \cite{etingof},
as a first step in finding cubic analogues of the Gaussian, invariant
under Fourier transform (see \cite{Pioline:2003uk} for a short account).

\begin{exo}
Check by explicit computation that
for the ``STU'' model, $(1/X^0)e^{N(X^A)/X^0}$ is invariant
under Fourier transform, namely
\begin{equation}
\label{eleg}
\int \frac{dX^0dX^1dX^2dX^3}{X^0} \exp\left[
i\frac{X^1X^2X^3}{\hbar X^0} + i X^I Y_I \right]
=
 \frac{\hbar}{Y^0}
 \exp\left[ {i\hbar}  \frac{Y_1 Y_2 Y_3}{Y_0} \right]
\end{equation}
Conclude that the semi-classical approximation to this integral is exact.
{\it Hint: perform the integral over $X^1,X^2,X^0,X^3$ in this order.}
\end{exo}

In order to compute the Bekenstein-Hawking entropy, we start from the
``free energy'' \eqref{freef}
\be
\label{fvery}
{\cal F}(p,\phi) = \frac{\pi}{(p^0)^2+(\phi^0)^2}
\left\{
p^0 \left[ \phi^A p_A^\sharp  - N(\phi) \right]
+ \phi^0 \left[ p^A \phi_A^\sharp - N(p) \right]
\right\}
\ee
To eliminate the quadratic term  in $\phi^A$,
let us change variables to
\be
\label{xax0}
x^A= \phi^A - \frac{\phi^0}{p^0} p^A\ ,\quad
x^0= [(p^0)^2+(\phi^0)^2] / p^0
\ee
Moreover, we introduce an auxiliary variable $t$, such that,
upon eliminating $t$, we recover \eqref{fvery}:
\be
S_{BH} = \pi \langle  - \frac{N(x^A)}{x^0} 
+ \frac{p_A^\sharp  + p^0 q_A}{p^0} x^A
-\frac{t}{4} 
\left(\frac{x^0}{p^0}-1\right) -  
\frac{(2 N(p) + p^0 p^I q_I)^2}{t\ (p^0)^2}  \rangle_{\{x^I,t\}} 
\ee
Extremizing over $x^I$ now amounts to Legendre transforming
$N(x)/x^0$, which according to \eqref{nleg} reproduces $-N(y)/y^0$
where $y^I$ are the coefficients of the linear terms in $x^I$, so
\be
\label{sinter}
S_{BH} = \pi \langle
4 \frac{N[ p_A^\sharp +p^0 q_A]}{(p^0)^2 t}
-  \frac{[2 N(p) + p^0 p^I q_I]^2}{t\ (p^0)^2} 
+ \frac{t}{4} \rangle_{t}
\ee
Finally, extremizing over $t$ leads to
\be
S_{BH}=\frac{\pi}{p^0} \sqrt{ 4 N[ p_A^\sharp + p^0 q_A] 
- [2 N(p) + p^0 p^I q_I]^2 } 
\ee
The pole at $p^0=0$ is fake: upon Taylor expanding 
$N[ p_A^\sharp +p^0 q_A]$ in the numerator and further using the
homogeneity of $N$, its coefficient cancels. The final result
gives the entropy as the square root of a quartic polynomial in the
charges,
\be
S_{BH}= \pi \sqrt{I_4(p^I,q_I)}
\ee
where
\be
\label{i4jor}
I_4(p^I,q_I) = 4 p^0 N(q_A) - 4 q_0 N(p^A) + 4  q^A_\sharp p_A^\sharp
- ( p^0 q_0 + p^A q_A )^2
\ee
The fact that this quartic polynomial is invariant under the linear action of 
the four-dimensional ``U-duality'' group $\Conf(J)$ on the symplectic
vector of charges $(p^I,q_I)$, follows from Freudenthal's ``triple
system construction''. Several examples are worth mentioning:
\begin{itemize}
\item For the ``STU'' model with $N(\xi)=\xi^1\xi^2\xi^3$, 
the electric-magnetic charges transform
as a $(2,2,2)$ of $\Conf(J)=Sl(2)^3$, so can be viewed as sitting at the
8 corners of a cube; the quartic invariant is known as
Cayley's ``hyperdeterminant''
\be
I_4 = -\frac12 \epsilon^{AB}\epsilon^{CD}
\epsilon^{ab}\epsilon^{cd}
\epsilon^{\alpha\gamma}\epsilon^{\beta\delta}
q_{Aa\alpha}
Q_{Bb\beta}
Q_{Cc\gamma}
Q_{Dd\delta}
\ee
This has recently been related to the ``three-bit entanglement'' in 
quantum information theory
\footnote{According to Freudenthal's construction, the electric and magnetic
charges naturally arrange themselves into a square (rather than a cube)
$\begin{pmatrix} p^0 & p^I \\ q_I & q_0\end{pmatrix}$, where the diagonal elements are in $\IR$ while the off-diagonal
ones are in the Jordan algebra $J$. This suggests that the ``three-bit''
interpretation of the STU model may be difficult to generalize.}
 \cite{Duff:2006uz,Kallosh:2006zs,Levay:2006kf}.
\item More generally, for the infinite series, where the charges
transform as a $(2,n)$ of $Sl(2)\times SO(2,n)$, the quartic
invariant is
\be
I_4 = (\vec q_e \cdot \vec q_e)(\vec q_m \cdot \vec q_m) -
(\vec q_e \cdot \vec q_m)^2
\ee
Up to a change of signature of the orthogonal group, this is the
quartic invariant which appears in the entropy of 1/4-BPS black
holes in $\CN=4$ theories \eqref{sqrpp}.
\item In the exceptional $J_3^{\IO}$ case,
$I_4$ is the quartic invariant of the 56 representation of $E_{7(-25)}$.
Replacing $\IO$ by the split octonions $\IO_s$, one
obtains the quartic invariant of $E_{7(7)}$, which appears
in the entropy $S=\pi\sqrt{I_4}$ 
of 1/8-BPS states in $\CN=8$ supergravity \cite{Kallosh:1996uy},
\be
\label{sbhn8}
I_4(P,Q)= -\Tr(QPQP) 
+ \frac14 \left( \Tr QP \right)^2 
-4 \left[ \Pf(P) + \Pf(Q)  \right]
\ee
where the entries in the antisymmetric
$8\times 8$ matrices $Q$ and $P$ may be identified as \cite{Pioline:2005vi}:
\be
Q=\begin{pmatrix} 
[D2]^{ij} & [F1]^i & [kkm]^i \\ 
-[F1]^i & 0 & [D6]  \\ 
- [kkm]^i & - [D6] &  0
\end{pmatrix}\ ,\quad
P=\begin{pmatrix} 
[D4]_{ij} & [NS5]_i & [kk]_i \\ 
-[NS5]_i & 0 & [D0]  \\ 
- [kk]_i & - [D0] &  0
\end{pmatrix}\ ,\quad
\ee
Here, $[D2]^{ij}$ denotes a D2-brane wrapped along the directions
$ij$ on $T^6$, $[D4]_{ij}$ a D4-branes wrapped on all directions
{\it but} $ij$, $[kk]_i$ a momentum state along direction $i$,
$[kkm]^i$ a Kaluza-Klein 5-monopole localized along the direction
$i$ on $T^6$, $[F1]^i$ a fundamental string winding along direction $i$,
and $[NS5]_i$ a NS5-brane wrapped on all directions but $i$.
\end{itemize}

\begin{exo}
Show that in the $\CN=4$ truncation where
only the $[F1],[kk],[NS5],[kkm]$ charges are retained,
\eqref{sbhn8} reduces to the quartic invariant \eqref{sqrpp} 
under $Sl(2)\times SO(6,6)$. Similarly, in the $\CN=2$ truncation
where only $[D0],[D2],[D4],[D6]$ are kept, show that one obtains
the quartic invariant of a spinor of $SO^*(12)$, based on the
Jordan algebra $J_3^\IH$.
\end{exo}

The intermediate equation \eqref{sinter} also has an interesting 
interpretation: it is recognized as $1/p^0$ times the entropy 
$S_{5D}=\pi\sqrt{N(Q)-J^2}$ of a five-dimensional BPS black hole with
electric charge and angular momentum
\bea
Q_A &=& p^0 q_A + C_{ABC} p^B p^C \label{4d5dlift}\\
2 J_L &=& (p^0)^2 q_0 + p^0 p^A q_A +2 N(p)
\eea
The interpretation of these relations is as follows: 
when the D6-brane charge $p^0$ is non-zero,
the 4D black hole in Type IIA compactified on $\tilde Y$ may be lifted 
to a 5D black hole in M-theory on $\tilde Y \times
TN_{p^0}$, where $TN$ denotes the 4-dimensional Euclidean Taub-NUT space
with NUT charge $p^0$; at spatial infinity, this asymptotes to $\IR^3\times
S^1$, where the circle is taken to be the M-theory direction. Translations
along this direction at infinity, conjugate to the D0-brane charge $q_0$,
become $SU(2)$ rotations at the center of $TN$, where the black hole is
assumed to sit. The remaining factors of $p^0$ are accounted for by taking
into account the $\IR^4/\IZ_{p^0}$ singularity at the 
origin of $TN$ \cite{Gaiotto:2005gf}.
The formulae \eqref{4d5dlift} extend this lift to an arbitrary choice
of charges, in a manifestly duality invariant manner.

\begin{exo}
Using the fact that the degeneracies of five-dimensional black holes 
on $K3\times S^1$ are given by the Fourier coefficients
of the elliptic genus of ${\rm Hilb}(K3)$, equal to $1/\Phi_{10}$,
show that the DVV conjecture \eqref{dvv} holds for at least one
U-duality orbit of 4-dimensional dyons in type II/$K3\times T^2$ 
with one unit of D6-brane and some amout of D0,D2-brane charge.
You might want to seek help from \cite{Shih:2005uc}.
\end{exo}

\section{Topological String Primer\label{topprim}}
In the previous sections, we were concerned exclusively with low energy
supergravity theories, whose Lagrangian contains at most two-derivative 
terms. This is sufficient in the limit of infinitely large charges, but
not for more moderate values, where higher-derivative corrections
start playing a role. In this section, we give a self-contained introduction
to topological string theory, which offers a practical way of
to compute an infinite series of such corrections. Sections
\ref{topsigma} and \ref{topst} draw heavily from \cite{Witten:1991zz}.
Other valuable reviews of topological string theory
include \cite{Marino:2004uf,Marino:2004eq,
Neitzke:2004ni,Vonk:2005yv,Cordes:1994fc}.

\subsection{Topological Sigma Models\label{topsigma}}
Type II strings compactified on a K\"ahler manifold $X$ of
complex dimension $d$ are described by a
$N=(2,2)$ sigma model
\be
\label{22sigma}
S = 2t \int d^2 z \left( 
g_{i\bar j} \pa\phi^i \bar\pa\phi^{\bar j}+
g_{i\bar j} \bar \pa\phi^i \pa\phi^{\bar j}+
i \psi^{\bar i}_- D \psi_-^i g_{i\bar i} +
i \psi^{\bar i}_+ \bar D \psi_+^i g_{i\bar i} +
R_{i\bar i j\bar j} \psi_+^i\psi_+^{\bar i}
\psi_-^j \psi_-^{\bar j} \right)
\ee
where $\phi$ is a map from a two-dimensional genus $g$ 
Riemann surface $\Sigma$ to $X$,
$\psi_\pm^i$ is a section of $K^{1/2}_\pm \otimes \phi^*(T^{1,0}X)$, 
$\psi_\pm^{\bar i}$ is a section of $K^{1/2}_\pm \otimes \phi^*(T^{0,1}X)$,
and we denoted by $K_+$ the canonical bundle on $\Sigma$ (i.e. the 
bundle of (1,0) forms) and $K_-$ the anti-canonical bundle (of (0,1) forms).
The factor of $t$ (the string tension) is to keep track on the dependence 
on the overall volume of $X$.

This model is invariant under $N=(2,2)$ superconformal transformations
generated with sections $\alpha_\pm$ and $\tilde\alpha_\pm$ of 
$K_\pm^{1/2}$, acting e.g. as 
\be
\delta \phi^i = i ( \alpha_- \psi_+^i + \alpha_+ \psi_-^i )\ ,\quad
\delta \phi^{\bar i} = i ( \tilde\alpha_- \psi_+^{\bar i} 
+ \tilde \alpha_+ \psi_-^{\bar i} )
\ee 
This implies chirally conserved supercurrents $G^{\pm}$ of conformal dimension
3/2, which together with $T$ and the current $J$ generate the $\CN=2$ 
superconformal algebra,
\bea
G^+(z)~G^-(0)&=& \frac{2c}{3} \frac{1}{z^2} 
+ \left( \frac{2J}{z^2} + \frac{\pa J + 2 T }{z} \right) + {\rm reg}
\label{gpgm} \\
J(z) J(0) &=& \frac{c}{3} \frac{1}{z^2} + {\rm reg}
\eea
The current $J$ appearing in the OPE \eqref{gpgm} generates a $U(1)$ symmetry,
such that $G_\pm$ have charge $Q=\pm 1$ while $T$ and $J$ are neutral.
In the (doubly degenerate) Ramond sectors $R_\pm$, the zero-modes of the 
supercurrents generate a supersymmetry algebra
\be
(G_0^+)^2=(G_0^-)^2=0\ ,\quad
\left\{ G_0^+, G_0^- \right\} = 2 \left( L_0^{R_\pm} - \frac{c}{24} \right)
\ee
Unitarity forces the right-hand side to be positive on any state.
Moreover, the $\CN=2$ algebra admits an automorphism known as spectral flow,
which relates the NS and R sectors:
\be
\label{specflow}
J_0^{R_\pm} = J_0^{NS} \mp \frac{c}{6}\ ,\quad
L_0^{R_\pm} = L_0^{NS} \mp \frac12 J_0^{NS} + \frac{c}{24}
\ee
The unitary bound $\Delta \geq c/24$ in the R sector therefore 
implies a bound $\Delta \geq |Q|/2$ after spectral flow. States which
saturate this bound have no short distance
singularities when brought together, and thus form 
a ring under OPE, known as the {\it chiral ring} of the $\CN=2$ SCFT.
Applying the spectral flow twice maps the NS sector back to itself,
with $(\Delta,Q)\to(\Delta-Q+\frac{c}{6}, Q\mp \frac{c}{3})$. In particular,
the NS ground state is mapped to a state with $(\Delta,q)=(\frac{c}{6},
\mp \frac{c}{3})$ in the chiral ring. For a Calabi-Yau three-fold, starting
from the identity we thus obtain two R states 
with  $(\Delta,q)=(3/8,\pm 3/2)$, and one NS state with
 $(\Delta,q)=(3/2,\pm 3)$: these are identified geometrically as the 
covariantly constant spinor and the holomorphic $(3,0)$ form, respectively.

The spectral flow \eqref{specflow} above can be used to ``twist''
the $\CN=2$ sigma model into a topological sigma model: for this,  
bosonize the $U(1)$ current $J = i\sqrt{3}\partial H$, so that the 
spectral flow operator becomes
\be
\label{spesig}
\Sigma_{\pm}= \exp\left( \pm i\frac{\sqrt{3}}{2} H(z) \right)
\ee
with $(\Delta=3/8,Q=\pm 3/2)$.
The topological twist then amounts to adding a  background charge
$\pm \int \frac{\sqrt{3}}{2} H ~R^{(2)}$: its effect is to
change the two-dimensional spin $L_0$ into a linear combination
$L_0\mp\frac12 J_0$ of the spin and the $U(1)$ charge.
Under this operation,
choosing the $+$ sign, 
$\psi_+^i$ becomes a section of $\phi^*(T^{1,0}X)$, i.e. a 
worldsheet scalar, whereas $\psi_+^{\bar i}$ becomes a section of 
$K_+ \otimes \phi^*(T^{0,1}X)$, i.e. a worldsheet one-form; simultaneously,
the supersymmetry parameters $\alpha_-$ and $\tilde\alpha_-$ become
a scalar and a section of $K^{-1}$, respectively. Alternatively,
we may choose the $-$ sign in \eqref{specflow}, where instead
$\psi_+^i$ would become a section of $K_+ \otimes \phi^*(T^{1,0}X)$, 
while  $\psi_+^{\bar i}$ would turn into a worldsheet scalar. 
In either case, it is necessary that the canonical bundle $K$
be trivial, in order for that the correlation functions be unaffected
by the twist : this is achieved only when computing particular ``topological
amplitudes'' in string theory, which we will discuss in Section \ref{agnt}.

Since the sigma model \eqref{22sigma} has $(2,2)$ superconformal invariance,
it is possible to twist both left and right-movers by a spectral flow
of either sign. Only the relative choice of sign is important, leading
to two very distinct-looking theories, which we discuss in turn:

\subsubsection{Topological A-model} 
Here, both $\psi^i_+$ and $\psi_-^{\bar i}$
are worldsheet scalars, and can be combined in a scalar $\chi\in
\phi^*(TX)$. On the other hand, $\psi^i_{-}$ and $\psi^{\bar i}_+$ become
(0,1) and (1,0) forms $\psi^i_{\bar z}$ and $\psi^{\bar i}_z$ on the 
worldsheet. The action is rewritten as
\be
\label{amodac}
S = 2t \int d^2z \left( 
g_{i\bar j} \pa\phi^i \bar\pa\phi^{\bar j}+
g_{i\bar j} \bar \pa\phi^i \pa\phi^{\bar j}+
i \psi^{\bar i}_z \bar D \chi^i g_{i\bar i} +
i \psi^{\bar i}_{\bar z} D \chi^{\bar i} g_{i\bar i} -
R_{i\bar i j\bar j} \psi^i_{\bar z}\psi^{\bar i}_z
\chi^j \chi^{\bar j} \right)
\ee
It allows for a conserved ``ghost'' charge where $[\phi]=0,[\chi]=1,[\psi]=-1$,
and is invariant under the scalar nilpotent operator $Q=G_+$,
\be
\{ Q, \phi^I \} = \chi^I\ ,\quad \{Q,\chi^I\} =0\ ,
\{ Q, \psi^i_{\bar z} \} = i \bar\pa \phi^i 
- \chi^j \Gamma^i_{jk} \psi_{\bar z}^k
\ee
The action \eqref{amodac} is in fact $Q$-exact, up to a total derivative
term proportional to the pull-back of the K\"ahler form $\omega_K=
i g_{i\bar j} d\phi^i\wedge d\phi^{\bar j}$, complexified into 
$J=B+i \omega_K$ by including the coupling to the NS two-form:
\be
S = -i \{Q, V\} - t \int_\Sigma \phi^*(J)
\ee
where $V$ is the ``gauge fermion''
\be
V = t \int d^2z ~ g_{i\bar j} \left( \psi^i_{\bar z} \pa \phi^{\bar j}
+ \psi^{\bar j}_{z} \bar \pa \phi^{i} \right)
\ee
This makes it clear that the theory is independent of the worldsheet metric,
since the energy momentum tensor is $Q$-exact:
\be
T_{\alpha\beta} = \{ Q , b_{\alpha\beta} \}\ ,\quad
b_{\alpha\beta} = \frac{\pa V}{\pa g^{\alpha\beta}}
\ee
Moreover, the string tension $t$ appears only in the total derivative 
term so, in a sector with fixed homology class $\int_\Sigma \phi^*(J)$, 
the semi-classical limit $t\to 0$ is exact. The path integral thus
localizes\footnote{Localization is a general feature of integrals with
a fermionic symmetry $Q$: decompose the space of fields into orbits of
$Q$, parameterized by a Grassman variable $\theta$,  times its orthogonal
complement; since the integrand is independent of $\theta$ by assumption,
the integral $\int d\theta$ vanishes by the usual rules of Grassmannian 
integration. This reasoning breaks down at the fixed points of $Q$, which
is the locus to which the integral localizes.}
to the moduli space of $Q$-exact configurations,
\be
\pa_{\bar z}\phi^i=0, \quad \pa_{z}\phi^{\bar i}=0
\ee
i.e.  {\it holomorphic maps} 
from $\Sigma$ to $X$. Moreover, the local observables 
of the A-model $O_{\cal W}=
W_{I_1\dots I_n} \chi^{I_1}\dots\chi^{I_n}$, where $W_{I_1\dots I_n}
d\phi^{I_1}\dots d\phi^{I_n}$ is a differential form on $X$ of degree $n$, 
are in one-to-one correspondence with the de Rham cohomology of $X$,
since $\{Q, {\cal O}_W\}=-{\cal O}_{dW}$. Due to an anomaly in the
conservation of the ghost charge, correlators of $l$ observables 
vanish unless
\be
\sum_{k=1}^{l} {\rm deg}(W_k) = 2 d (1- g) + 2 \int_\Sigma 
\phi^*(c_1(X))
\ee
The last term vanishes when the Calabi-Yau condition $c_1(X)$ is obeyed.
For Calabi-Yau threefolds, at genus 0 the only correlator involves three
degree 2 forms,
\be
\langle {\cal O}_{W_1} {\cal O}_{W_2} {\cal O}_{W_3 } \rangle
= \int W_1 \wedge W_2 \wedge W_3
+ \sum_{\beta\in H^{2+}(X)} ~  e^{-t \int_\beta J} ~ \int_\beta W_1 ~
\int_\beta W_2 ~ \int_\beta W_3
\ee
At genus 1, only the vacuum amplitude, known as the
elliptic genus of $X$ is non-zero. In Section \ref{topst},
we will explain the prescription to construct non-zero amplitudes
at any genus, by coupling to topological gravity.

\subsubsection{Topological B-model}
The other inequivalent choice consists in twisting $\psi^{\bar i}_\pm$
into worldsheet scalars valued in $TX^{0,1}$, while $\psi_+^i$ and
$+\psi_-^i$ are $(0,1)$ and $(1,0)$ forms valued in $TX^{1,0}$. Defining
$\eta^{\bar i}=\psi_+^{\bar i}+\psi_-^{\bar i}$, $\theta_i=g_{i\bar i}
(\psi_+^{\bar i}-\psi_-^{\bar i}$, and taking 
$\psi^i_\pm$ as the two components of a one-form $\rho^i$, 
the action may be rewritten as
\be
S = i~t \{Q, V\} + t~W
\ee
where 
\bea
V &=& \int_\Sigma d^2z~g_{i\bar j} \left( \rho^i_z \bar\pa \phi^{\bar j}
+\rho^i_z \bar\pa \phi^{\bar j} \right) \\
W &=& - \int_\Sigma d^2z~ \left( \theta_i D\rho^i + \frac{i}{2}
R_{i\bar i j \bar j}~ \rho^i \wedge \rho^j~\eta^{\bar i} \theta_k g^{k\bar j}
\right)
\eea
and the nilpotent operator $Q=G_-$ acts as
\be
\{Q,\phi^i\}=0\ ,\quad \{Q,\phi^{\bar i}\}=-\eta^{\bar i}\ ,\quad
\{Q,\eta^{\bar i}\}=\{Q,\theta_i\}=0\ ,\quad \{Q,\rho^i\}=-id\phi^i
\ee
Again, the energy-momentum tensor is $Q$-exact, so that the model 
is topological. It is also independent of the K\"ahler structure of $X$,
and has a trivial dependence on $t$, since (apart from contributions
from the $Q$-exact term) $t$ may be reabsorbed by rescaling 
$\theta\to\theta/t$. The semi-classical limit $t\to \infty$ is therefore
again exact, and the path integral localizes on the fixed points of $Q$,
which are now {\it constant maps}, $d\phi^i=0$. 
After localization, the path integral then reduces to an integral over $X$.

The observables of the B-model are in one-to-one correspondence with
degree $(p,q)$ polyvector fields
\be
V= V_{\bar i_1 \dots \bar i_p}^{j_1\dots j_q} 
~d\bar z^{\bar i_1}\dots d\bar z^{\bar i_p} 
~\pa_{z^{j_1}} \dots \pa_{z^{j_q}}
\in H^p\left(X,\Lambda^q T^{1,0} X\right)
\ee 
via $d\bar z^{\bar i}\sim\eta^{\bar i}, \pa_{z^j}\sim\theta_j$,
since $\{Q,{\cal O}_V\}=-{\cal O}_{\bar\pa V}$. There are now two conserved
ghost charges, and the anomaly in the ghost number conservation requires that
\be
\sum_{k=1}^l p_k = \sum_{k=1}^l q_k = d(1-g)
\ee
For example, at genus 0, the only vanishing correlator on 
a Calabi-Yau three-fold involves three (1,1)
polyvector fields $V^i_{\bar j}$. Using the holomorphic $(3,0)$ form,
these are related to $(2,1)$ forms $\Omega_{ijl} V^l_{\bar k}$ parameterizing
the complex structure of $X$. The three-point function is
\be
\langle {\cal O}_{V_1} {\cal O}_{V_2} {\cal O}_{V_3} \rangle
= \int_X ~V^{i_1}_{\bar j_1}~V^{i_2}_{\bar j_2}~V^{i_3}_{\bar j_3}
\Omega_{i_1i_2i_3} ~
d\bar z^{\bar j_1}\wedge d\bar z^{\bar j_2}\wedge d\bar z^{\bar j_3}
\wedge \Omega
\ee
giving access to the third derivative of the prepotential.

\subsection{Topological Strings\label{topst}}
Due to the  conservation of the ghost number, we have seen that, from 
the sigma model alone, the only non-vanishing topological correlators are the 
three-point function on the sphere,
and the vacuum amplitude on the torus. It turns out that the coupling
to topological gravity allows to lift this constraint, and define
arbitrary $n$-point amplitudes at any genus.

Recall that in bosonic string theory, genus $g$ amplitudes are obtained
by introducing $6g-6$ insertions of the dimension 2 ghost (or, rather, 
``antighost'') $b$ of 
diffeomorphism invariance, folded with Beltrami differentials
$\mu_k \in H^1(\Sigma,T^{1,0}\Sigma)$:
\be
F_g = \int_{{\cal M}_g}~\langle 
\prod_{k=1}^{6g-6} (b,\mu_k) \rangle
\ee
where 
\be
(b,\mu) = \int_\Sigma d^2 z 
\left[ b_{zz} \mu_{\bar z}^z + b_{\bar z\bar z} ~ \bar\mu_z^{\bar z} \right]
\ee
This effectively produces the Weil-Peterson volume element on the moduli space 
${\cal M}_g$ of complex structures on the genus $g$ Riemann surface $\Sigma$
(compactified \`a la Deligne-Mumford). Since $b$ has ghost number $-1$,
this exactly compensates the anomalous background charge.

After the topological twist, which identifies the BRST charge $Q$
with (say) $G_+$, it is natural to identify $b$ with $G_-$, in such a way 
that the energy-momentum tensor is given by $T=\{Q,b\}=\{G_+,G_-\}$. Hence, the
genus $g$ vacuum topological amplitude may be written as
\be
\label{topvac}
F_g = \int_{{\cal M}_g}~\langle 
\prod_{k=1}^{3g-3} (G_-,\mu_k)~(G_\pm,\bar\mu_k) \rangle
\ee
where the upper (resp., lower) sign corresponds to 
the A-model (resp., B-model). Scattering amplitudes may be obtained by
inserting vertex operators with zero ghost number; these
may be obtained by ``descent'' from a ghost number 2 operator 
${\cal O}^{(0)}$,
\be
d{\cal O}^{(0)} = \{ Q, {\cal O}^{(1)} \}\ ,\quad
d{\cal O}^{(1)} = \{ Q, {\cal O}^{(2)} \}\ 
\ee
Prominent examples of ${\cal O}^{(0)}$ are of course 
$W_{i\bar i}\chi^i\chi^{\bar i}$ in the A-model, and
$V_j^{\bar i} \eta^{\bar i}\theta_j$ in the B-model. These describes
the deformations of the K\"ahler and complex structures, respectively.
Arbitrary numbers of integrated vertex operators $\int d^2z~  {\cal O}^{(2)}$ 
can then be inserted in \eqref{topvac} without spoiling the conservation
of ghost charge number. 

Weighting the contributions of different genera by powers of the ``topological
string coupling'' $\lambda$, namely
\be
\label{topamp}
F_{\rm top}= \sum_{g=0}^{\infty} \lambda^{2g-2} F_g
\ee
we obtain obtain a perturbative definition of the A and B-model
topological strings. Since the worldsheet is topological, the target space
theory has only a finite number of fields, so is really more a field theory
than a string theory. In fact, the tree-level
scattering amplitudes can be reproduced
by a simple action $X$, known as ``holomorphic Chern-Simons'' in the A-model,
and ``Kodaira-Spencer'' in the B-model; these describe the fluctuations
of K\"ahler and complex structures, respectively. We refer the reader
to \cite{Bershadsky:1993cx} for an extensive discussion of these theories.

\subsection{Gromov-Witten, Gopakumar-Vafa and Donaldson-Thomas Invariants
\label{gwgkdt}}
We now concentrate on the topological vacuum amplitude \eqref{topamp}
of the A-model on a Calabi-Yau threefold $X$. Up to holomorphic anomalies
that we discuss in the next section, $F_{\rm top}$ can be viewed as a function
of the complexified K\"ahler moduli $t^A=\int_{\gamma^A} J$. 
In the large volume limit (or more
generally, near a point of maximal unipotent monodromy), it has an
asymptotic expansion
\be
\label{ftopa}
F_{\rm top} = -i \frac{(2\pi)^3}{6\lambda^2} C_{ABC} t^A t^B t^C
- \frac{i\pi}{12} c_{2A} t^A + F_{GW}
\ee
where $C_{ABC}$ are the triple intersection numbers of the
4-cycles $\gamma_A$ dual to $\gamma^A$, and $c_{2A}=\int_{\gamma_A}
c_2(T^{(1,0)}X)$ are the second Chern classes of these 4-cycles.
The first two terms in \eqref{ftopa} are perturbative in $\alpha'$,
while $F_{GW}$ contains the effect of worldsheet instantons at
arbitrary genus,
\be
\label{fgw}
F_{GW} = \sum_{g\geq 0} \sum_{\beta\in H_2^+(X)}
N_{g,\beta}~e^{2\pi i \beta_A t^A} \lambda^{2g-2}
\ee 
where the sum runs over effective curves $\beta=\beta_A\gamma^A$
with $\beta_A \geq 0$,
and $N_g^\beta$ are (conjecturally) rational numbers known as 
the Gromov-Witten (GW) invariants of $X$. It is possible to re-organize the
sum in \eqref{fgw} into
\be
\label{fgwgv}
F_{GW} = \sum_{g\geq 0} \sum_{\beta\in H_2^+(X)} \sum_{d\geq 1}
n_{g,\beta}~\frac{1}{d} \left[ 2 \sin\left( \frac{d\lambda}{2} \right)
\right]^{2g-2} e^{2\pi i d \beta_A t^A} 
\ee
The coefficients $n_{g,\beta}$ are known as the Gopakumar-Vafa (GV) invariants,
and are conjectured to always be integer: indeed, one may show that 
the contribution of a fixed $\beta_A$ in \eqref{fgwgv}
arises from the one-loop contribution of a M2-brane
wrapping the isolated 
holomorphic curve $\beta^A \gamma_A$ in $X$ \cite{Gopakumar:1998ii,
Gopakumar:1998jq}. The GV invariants
can be related to the GW invariants by expanding
\eqref{fgwgv} at small $\lambda$ and matching on to \eqref{fgw}, e.g.
at leading order $\lambda^{-2}$,
\be
N_{0,\beta} = \sum_{d|\beta_A} d^3 ~ n_{0,\beta^A/d}
\ee
which incorporates the effect of multiple coverings for an isolated
genus 0 curve. 

It should be noted that the sum in \eqref{fgw} or \eqref{fgwgv} 
includes the term $\beta=0$, which corresponds to degenerate 
worldsheet instantons. It turns out that the only non-vanishing
GV invariant at genus 0 is $n_{0,0}=-\frac12 \chi(X)$, hence
\be
F_{\rm deg} = -\frac12 \chi(X)~  
\sum_{d\geq 1}
\frac{1}{d} \left[ 2 \sin\left( \frac{d\lambda}{2} \right)
\right]^{2}
\equiv -\frac12 \chi(X)~  f(\lambda)
\ee
The function $f(\lambda)$, known as the Mac-Mahon function, 
may be formally manipulated into
\be
\label{mmahondef}
f(\lambda) = - \sum_{d\geq 1}
\frac{e^{id\lambda}}{d(1-e^{id\lambda})^2} = -
\sum_{d=1}^{\infty} \sum_{n=1}^{\infty}  \frac{n~q^{n d}}{d}
= \sum_{n=1}^{\infty} n~\log(1-q^n)
\ee
where $q=e^{i\lambda}$.
The last expression converges in the upper half plane $\Im(\lambda)>0$,
and may be taken as the definition of the Mac-Mahon function, suitable
in the large coupling limit $\lambda\to i\infty$.

\begin{exo} Check that the coefficient of $q^N$ in the Taylor expansion
of $\exp(-f)$ counts the number of three-dimensional 
Young tableaux with $N$ boxes.
\end{exo}

In order to analyze its contributions at weak coupling $t=-i\lambda\to 0$, 
let us compute its Mellin transform\footnote{The following argument, due
to S. Miller (private communication), considerably streamlines the 
computation in \cite{Dabholkar:2005dt}.} 
\be
M(s) = \int_{0}^{\infty} \frac{dt}{t^{1-s}} f(t) = 
- \int_{0}^{\infty} \frac{dt}{t^{1-s}} \sum_{d=1}^{\infty} \sum_{n=1}^{\infty}
\frac{n}{d} e^{-n d t} 
\ee
Exchanging the integral and sums, 
the result is simply expressed in terms of Euler $\Gamma$ and
Riemann $\zeta$ functions, 
\be
- \sum_{d=1}^{\infty} \sum_{n=1}^{\infty}  \frac{n}{d} (nd)^{-s} \Gamma(s)
= - \zeta(s-1)\zeta(s+1)\Gamma(s)
\ee
The function $f(t)$ itself may be obtained conversely by
\be
f(t) = \frac{1}{2\pi i} \int_{\Re(s)=s_0} M(s)\  t^{-s}
\ee
where the contour is chosen to lie to the right of any pole of $M(s)$.
Moving the contour to the left and crossing the poles  generate the
Laurent series expansion of $f(t)$. 

To perform this computation, recall that $\Gamma(s)$ has simple
poles at $s=-n,
n=0,1,\dots$ with residue $(-1)^n/n!$. Moreover, $\zeta(s)$ has a simple 
pole
at $s=1$, and ``trivial'' zeros at $s=-2,-4,-6,\dots$. The trivial
zeros of $\zeta(s-1)$ and $\zeta(s+1)$ 
cancel the poles of $\Gamma(s)$ at odd negative integer, leaving
only the simple poles at even strictly negative integer, a double
pole at $s=0$ and a single pole at $s=2$. Altogether, returning
to the variable $\lambda=i t$,we obtain the Laurent series expansion
\be
\label{maclog}
f(\lambda) = \frac{\zeta(3)}{\lambda^2} + \frac{1}{12} \log(i\lambda)
-\zeta'(1) + \sum_{g=2}^{\infty} \frac{B_{2g} B_{2g-2} \lambda^{2g-2}}
{(2g-2)!(2g-2)(2g)}
\ee 
where we further used the relation $\zeta(3-2g)=-B_{2g-2}/(2g-2)$ $(g\geq 2)$
between the values of $\zeta$ and Bernoulli numbers. 

The leading term, proportional to $\zeta(3)$, leads to a constant shift
$-1/2 \chi(X) \zeta(3)$ in the tree-level prepotential, 
and can be traced back to the
tree-level $R^4$ term in the 10-dimensional effective action, reduced
along $X$ \cite{Grisaru:1986kw,Candelas:1990rm,Antoniadis:1997eg}. The terms
with $g\geq 2$ were first computed using heterotic/type II 
duality \cite{Marino:1998pg},
and impressively agree with an independent computation of the
integral over the moduli space ${\cal M}_g$ \cite{faber},
\be
\int_{{\cal M}_g} c_{g-1}^3 = - \frac{B_{2g} B_{2g-2} \lambda^{2g-2}}
{(2g-2)!(2g-2)(2g)}
\ee
The logarithmic correction in \eqref{maclog} originates from the double
pole of $M(s)$ at $s=0$, and has no simple interpretation yet. It is
nevertheless forced if one accepts that the correct non-perturbative
completion of the degenerate instanton series is the Mac-Mahon 
function \cite{Dabholkar:2005by,Dabholkar:2005dt}.

For completeness, let us finally mention the relation to a third type
of topological invariants, known as Donaldson-Thomas invariants
$n_{DT}(q_A,m)$ \cite{thomas-thesis}: these count ``ideal sheaves'' on $X$,
which can be understood physically
as bound states of $m$ D0-branes, $q_A$ D2-branes wrapped on 
$q^A \gamma_A \in H_2(\IZ)$ and a single D6-brane. S-duality implies
\cite{Nekrasov:2004js,Kapustin:2004jm} that the partition function 
of Donaldson-Thomas invariants is related
to the partition function of Gromov-Witten invariants by
\cite{gw-dt,gw-dt2}
\be
\sum_{q^A\in H_2(\IZ),m\in\IZ} ~n_{DT}(q_A,m)\ e^{i t_A q^A} q^m
= \exp\left[ F_{GW} (t,\lambda) -\frac{\chi}{2} f(\lambda) \right]
\ee
where $q=-e^{i\lambda}$. Such a relation may be understood from
the fact that a curve may be represented either by a set of
a equations (the Donaldson-Thomas side), or by an explicit parameterization
(the Gromov-Witten side).
This conjecture has been recently proven for any toric 
three-fold $X$ \cite{okounkov}.

\subsection{Holomorphic Anomalies and the Wave Function Property\label{holanomsec}}
In the previous subsection, we assumed that the topological amplitude
was a function of the holomorphic moduli $t^i$ only. This is naively
warranted by the fact that the variation of the anti-holomorphic
moduli $\bar t^{\bar i}$ results in the insertion of an
(integrated) $Q$-exact operator,
$\phi_{\bar i}=\{ G^+, [\bar G^+,\bar \phi_{\bar i}]\}$.
By the
same naive reasoning, one would expect that the $n$-point functions
$C_{i_1 \dots i_n}^{(g)}$ be independent of $\bar t$, and 
equal to the $n$-th order derivative of the vacuum amplitude $F_g$
with respect to $t^{i_1},\dots t^{i_n}$. Both of these expectations
turn out to be wrong, due to boundary contributions in the integral
over the moduli space of genus $g$ Riemann surfaces. By analyzing
these contributions carefully, Berschadsky, Cecotti, Ooguri and 
Vafa \cite{Bershadsky:1993cx} (BCOV) have shown that the $\bar t^i$
derivative of $F_g$ is related to $F_{h<g}$ at lower genera via\footnote{
When $g=1$, the holomorphic equation becomes second order, and can
be read off from \eqref{BCOV1} below.}
\be
\label{hae0}
\bar\pa_{\bar i} F_g = 
\frac12 e^{2\cK} \bar C_{\bar i\bar j\bar k}
g^{j\bar j} g^{k\bar k} \left( D_j D_k F_{g-1}
+ \sum_{h=1}^{g-1}(D_j F_h) (D_k F_{g-h}) \right)
\ee
where $D_i F_g=(\pa_i - (2-2g) \pa_i \cK) F_g$, as appropriate for
a section of ${\cal L}^{2-2g}$, where ${\cal L}$ is the Hodge bundle
defined below \eqref{kahl}. In \eqref{hae0}, the first term
on the r.h.s. originates from the boundary of ${\cal M}_g$ where
one non-contractible handle of $\Sigma$ is pinched, whereas the second term 
corresponds to the limit where a homologically trivial cycle vanishes,
disconnecting $\Sigma$ into two Riemann surfaces with genus $h$ and $g-h$.
A similar identity can be derived for $n$-point functions.
Moreover, the latter are indeed obtained from
the vacuum amplitude by derivation with respect to $t^i$, provided
one uses a covariant derivative taking into account the Levi-Civita
and K\"ahler connections:
\be
\label{dcf}
C^{(g)}_{i_1\dots i_n}= \left\{ \begin{array}{ccl}
D_{i_1}\dots D_{i_n} F_g & \mbox{for} & g\geq 1, n\geq 1 \\
D_{i_1}\dots D_{i_{n-3}} C_{i_{n-1} i_{n-1} i_n}  & \mbox{for} & g=0, n\geq 3 
\\
0  & \mbox{for} & 2g-2+n \leq 0
\end{array} \right.
\ee
where $C_{ijk}$ is the tree-level three-point function.
The resulting identities may be summarized by defining
the ``topological wave-function''
\be
\label{BCOVW}
 \Psi_{\rm BCOV} = \lambda^{\frac{\chi}{24}-1} \exp\left[ \sum_{g=0}^{\infty}
\sum_{n=0}^{\infty} \frac{1}{n!} \lambda^{2g-2} C^{(g)}_{i_1\dots i_n} x^{i_1}
\dots x^{i_n} \right] 
\ee
Note that $\Psi_{\rm BCOV}$ does 
{\it not} incorporate the genus 1 vacuum amplitude.
In terms of this object, the identities \eqref{hae0} 
(or rather their generalization
to $n$-point functions) and \eqref{dcf} are summarized by the two 
equations
\bea
\pa_{\bar t^i} &=& 
\frac{\lambda^2}{2} e^{2\cK} \bar C_{\bar i\bar j\bar k} g^{j\bar j} g^{k\bar k}
\frac{\pa^2}{\pa x^j\pa x^k}
- g_{\bar i j} x^j \left( 
\lambda \frac{\pa}{\pa \lambda} + x^k \frac{\pa}{\pa x^k} \right) 
\label{BCOV1}\\
\pa_{t^i} &=& \Gamma_{ij}^k x^j \frac{\pa}{\pa x^k}
- \pa_i \cK \left( \frac{\chi}{24}-1 - \lambda \frac{\pa}{\pa\lambda} \right)
+ \frac{\pa}{\pa x^i} - \pa_i F_1 - \frac{1}{2\lambda^2}
C_{ijk} x^j x^k\label{BCOV2}
\eea
By rescaling $x^i\to \lambda x^i, \Psi\to e^{f_1(t)} \Psi_V$ where $f_1(t)$
is the holomorphic function in the general solution
\be
F_1 = -\frac12 \log|g| + \left( \frac{n_V+1}{2} 
- \frac{\chi}{24} + 1 \right) \cK + f_1(t) + \bar f_1(\bar t)
\ee
of the holomorphic anomaly equation for $F_1$, E. Verlinde 
\cite{Verlinde:2004ck} was able to
recast \eqref{BCOV1},\eqref{BCOV2} in a form involving only 
special geometry data,
\bea
\pa_{\bar t^i} &=& 
\frac{1}{2} e^{2\cK} \bar C_{\bar i\bar j\bar k} g^{j\bar j} g^{k\bar k}
\frac{\pa^2}{\pa x^j\pa x^k}
+ g_{\bar i j} x^j \frac{\pa}{\pa \lambda^{-1}} \label{Ver1b}\\
\nabla_i - \Gamma_{ij}^k x^j \frac{\pa}{\pa x^k}&=& 
\frac12 \partial_{t^i} \log|g| +
\frac{1}{\lambda}\frac{\pa}{\pa x^i} - \frac{1}{2} e^{-2\cK} C_{ijk} x^j x^k
\label{Ver2b}
\eea
where 
\be
\nabla_i = \partial_{i} + \pa_i \cK
\left( x^k \frac{\pa}{\pa x^k} - \lambda \frac{\pa}{\pa\lambda} 
+ \frac{n_V+1}{2} \right)
\ee
Here, $|g|=\det(g_{i\bar j})$.
The implications of these equations were understood in \cite{Witten:1993ed} 
and further
clarified in \cite{Gerasimov:2004yx,Verlinde:2004ck,Gunaydin:2006bz}: 
$\Psi(t,\bar t;x,\lambda)$ 
should be thought of
as a single state $|\Psi\rangle$ in a Hilbert space, 
expressed on a $(t,\bar t)$-dependent
basis of coherent states,
\be
\Psi_V(t,\bar t;x,\lambda) = _{(t,\bar t)}\! \langle x^i,\lambda | \Psi \rangle
\ee
This is most easily explained in the B-model, where $(x,\lambda^{-1})$
and their complex conjugate can be viewed as the coordinates of a 3-form
$\gamma\in H^3(X,\IR)$ on the Hodge decomposition
\be
\label{hdec}
\gamma = \lambda^{-1} \Omega + x^i D_i \Omega +  x^{\bar i} D_{\bar i} 
\bar \Omega +  \bar\lambda^{-1} \bar\Omega 
\ee
The space $H^3(X,\IR)$ admits a symplectic structure 
\be
\omega= i\ e^{-\cK} ~ \left( g_{i\bar j} dx^i \wedge
d\bar x^{\bar j} - d\lambda^{-1} \wedge d\bar\lambda^{-1} \right)
\ee
inherited from
the anti-symmetric pairing $(\alpha,\beta)=\int_X \alpha\wedge \beta$,
which leads to the Poisson brackets between the coordinates
\be
\label{poissbs}
\left\{ \lambda^{-1},\bar\lambda^{-1} \right\} = i~ e^{\cK}\ ,\quad
\left\{ x^i, \bar x^{\bar j} \right\} = -i g^{i\bar j}
\ee
The phase $H^3(X,\IR)$ may be quantized by considering functions 
(or rather half-densities, to account for the zero-point energy)
of $(\lambda^{-1},x^i)$ 
and representing $\bar\lambda^{-1}$ and $\bar x^{\bar i}$ 
as derivative operators,
\be
\bar\lambda^{-1} = - e^\cK \frac{\pa}{\pa\lambda^{-1}}\ ,\quad
\bar x^{\bar i} = e^\cK g^{\bar i j} \frac{\pa}{\pa x^j} 
\ee
The resulting wave 
function $\Psi(t,\bar t;\lambda,x)$ carries
a dependence on the ``background'' variables $(t,\bar t)$ since the 
decomposition \eqref{hdec} does depend on these variables via $\Omega$.
A variation of $t$ and $\bar t$ generically mixes $(\lambda^{-1},x)$ with
their canonical conjugate, and so may be compensated by an infinitesimal
Bogolioubov transformation, reflected in \eqref{Ver1b},\eqref{Ver2b}.
In fact, we can check that these two equations are hermitean conjugate
under the inner product
\be
\label{vinner}
\begin{split}
\langle \Psi' | \Psi\rangle =&
\int dx^i d\bar x^{\bar i}  d\lambda^{-1}d\bar \lambda^{-1}
|g|\  e^{-\frac{n_V+1}{2}\cK} \\
&\exp\left( -e^{-\cK} x^i g_{i\bar j} \bar x^j + e^{-\cK}
\lambda^{-1}\bar\lambda^{-1} \right)
\Psi^{'*}(t,\bar t;\bar x,\bar\lambda) \Psi(t,\bar t;x,\lambda) 
\end{split}
\ee
which is the natural inner product arising in K\"ahler 
quantization. In contrast to $\Psi$ and $\Psi'$ separately,
the inner product is background independent (and, in fact,
a pure number), by virtue of the anomaly equations.

\begin{exo} Show that in the harmonic oscillator Hilbert space,
the wave functions in the real and oscillator polarizations are
related by (abusing notation)
\be
f(q) = \int da^\dagger\ e^{i a^\dagger q \sqrt{2} + q^2/2 - 
(a^\dagger)^2/2} f(a^\dagger)
= \int da\ e^{- i a q \sqrt{2} - q^2/2 + 
a^2/2} f(a)
\ee 
Conclude that the inner product in oscillator basis
is given by 
\be
\int dq~f^*(q) g(q) = \int da da^{\dagger}\  
e^{-a a^{\dagger}} f^*(a) g(a^\dagger)
\ee
\end{exo}

This observation suggests that there exists a different, background 
independent polarization obtained by  choosing a real symplectic basis 
$\gamma^I,\gamma_I$ of three-cycles in $H_3(X,\IZ)$, and expanding
\be
\gamma = p^I \gamma_I + q_I \gamma^I
\ee
The symplectic form is now just $\omega=dq_I\wedge dp^I$, so $H_3(X,\IR)$ can
be quantized by considering functions of $p^I$, and representing
$q_I$ as $i\pa/\pa p^I$; equivalently, one may introduce a set 
of coherent states $|p^I\rangle$, and define  the wave function
in the ``real'' polarization,
\be
\Psi_{\IR}(p^I) = \langle p^I | \Psi\rangle\ .
\ee
This is related to 
the wave function in the K\"ahler polarization by a finite Bogolioubov 
transformation\footnote{A precursor of this formula was already 
found in \cite{Bershadsky:1993cx}, although not recognized as such.}
\be
\label{psixpsip}
\Psi_{\IR}(p^I)
= \int dx^i\ d\lambda\ 
\langle p^I | x^i,\lambda\rangle\ 
\Psi_{V}(t,\bar t; \lambda, x) 
\ee
The overlap of coherent states
$\langle p^I | x^i, \lambda\rangle$
is a solution of the equations hermitian-conjugate to 
\eqref{Ver1b}, \eqref{Ver2b} 
\cite{Verlinde:2004ck,Gunaydin:2006bz},
\be
\label{twinerdvv}
\begin{split}
\langle p^I | x^i, \lambda\rangle
&= e^{-(n_V+1)\cK/2} \sqrt{\det{g_{i\bar j}}}
\exp\left[ -\frac12 p^I \bar\tau_{IJ} p^J 
+2 i p^I [\Im\tau]_{IJ}
( \lambda^{-1} X^I + e^{-\cK/2} x^i f_i^I ) \right.\\
& \left. 
+i  \left( \lambda^{-2} X^I [\Im\tau]_{IJ} X^J
+2 \lambda^{-1} e^{-\cK/2} x^i f_i^I [\Im\tau]_{IJ} X^J
+ e^{-\cK} x^i f_i^I [\Im\tau]_{IJ} f_j^J x^j \right) \right]
\end{split}
\ee
While the topological wave function in the real polarization has the
great merit of being background independent, it is nevertheless not
canonical, since it depends on a choice of symplectic basis. As usual
in quantum mechanics, changes of  symplectic basis are
implemented by the metaplectic representation of $Sp(2n_V+2)$ (or rather,
of its metaplectic cover). In particular, upon exchanging 
$A$ and $B$ cycles, $\Psi_{\IR}(p^I)$ is turned into its Fourier
transform, which is the quantum analogue of the classical property
discussed in Exercise \ref{exoleg} on page \pageref{exoleg}.

For completeness, let us mention that there exists a different
``holomorphic'' polarization, intermediate between the K\"ahler and real
polarizations, where the topological amplitude is a purely holomorphic
function of the background moduli $t^i$, satisfying a heat-type 
equation analogous to the Jacobi theta series \cite{Gunaydin:2006bz}.
Moreover, for ``very special supergravities'', the holomorphic anomaly
equations can be traced to operator identities in the ``minimal''
representation of the three-dimensional duality group $\QConf(J)$;
this is analogous to the case of the Jacobi theta series, where
the Siegel modular group $Sp(4,\IZ)$ plays the role of $\QConf(J)$.
This hints at the existence of a one-parameter generalization of the
topological string amplitude, which we return to
in Section \ref{minrep}.

\section{Higher Derivative Corrections and Topological Strings \label{highder}}
In this section, we return to the realm of physical string theory, and
explain how a special class of higher-derivative terms in the low-energy
effective action can be reduced to a topological string computation.
We then discuss how these terms affect the Bekenstein-Hawking entropy
of black holes, and formulate the Ooguri-Strominger-Vafa conjecture,
which purportedly relates the topological amplitude to the microscopic
degeneracies.

\subsection{Gravitational F-terms and Topological Strings \label{agnt}}
In general, higher-derivative and higher-genus corrections in string
theory are very hard to compute: the integration measure on supermoduli
space is ill-understood beyond genus 2 (see \cite{D'Hoker:2005ia}
for the state of the art at genus 2), and the current computation
schemes (with the exception of the pure spinor superstring, see e.g.
\cite{Berkovits:2002zk}) are non-manifestly supersymmetric, 
requiring to evaluate
many different scattering amplitudes at a given order in momenta.

Fortunately, $\CN=2$ supergravity coupled to vector multiplets has 
an off-shell superspace description, which greatly reduces
the number of diagrams to be computed, and also provides a
special family of ``F-term'' interactions, which can be efficiently
computed. The most convenient formulation
starts from $\CN=2$ conformal supergravity and fixes the conformal gauge
so as to reduce to Poincar\'e supergravity (see \cite{Mohaupt:2000mj}
for an extensive review of this approach). The basic objects are 
the Weyl and matter chiral superfields,
\bea
W_{\mu\nu}(x,\theta) = T_{\mu\nu}- \frac12 R_{\mu\nu\rho\sigma} 
\epsilon_{\alpha\beta} \theta^\alpha \sigma_{\lambda\rho} \theta^\beta 
+ \dots\\
\Phi^I(x,\theta) = X^I + \frac12 {\cal F}_{\mu\nu}^{I} \epsilon_{\alpha\beta}
\theta^\alpha \sigma^{\mu\nu} \theta^\beta + \dots
\eea
where $\alpha,\beta=1,2$. $T_{\mu\nu}$ is an auxiliary anti-selfdual tensor,
identified by the (tree-level) equations of motion
as the graviphoton \eqref{tgrav}.  From $W$ one may construct the 
scalar chiral superfield
\be
W^2(x,\theta) = T_{\mu\nu} T^{\mu\nu}- 2 \epsilon_{ij}
\theta^i \sigma^{\mu\nu} \theta^j  R_{\mu\nu\lambda\rho} T^{\lambda\rho}
- (\theta^i)^2 (\theta^j)^2  R_{\mu\nu\lambda\rho}  R^{\mu\nu\lambda\rho}
+ \dots
\ee
where the anti-self dual parts of $R$ and $T$ are understood.
Starting with any holomorphic, homogeneous of degree 
two function $F(\Phi^I, W^2)$,
regular at $W^2=0$,
\be
F(\Phi^I, W^2)
\equiv  \sum_{g=0}^{\infty} F_g( \Phi^I) W^{2g}
\ee
(where $F_g$ is homogeneous of degree $2-2g$)
one may construct the chiral integral
\be
\label{wilsonc}
\int d^4\theta d^4x ~ F(\Phi, W^2) 
=S_{\rm tree} + \int \sum_{g=1}^{\infty}
F_g(X^I) \left( g\ R^2 T^{2g-2} + 2 g(g-1) (RT)^2 T^{2g-4} \right) 
+ \dots
\ee
which reproduces the tree-level $\CN=2$ supergravity action based
on the prepotential $F_0$, plus
an infinite sum of higher-derivative ``F-term'' gravitational
interactions (plus non-displayed terms).  $F(\Phi^I,W^2)$
is known as the generalized prepotential.

In order to compute the coefficients $F_g(X^I)$, one should compute the
scattering amplitude of 2 gravitons and $2g-2$ graviphotons in 
type II (A or B) string theory at leading order in momenta
compactified on a Calabi-Yau threefold $X$. This problem was studied
in \cite{Antoniadis:1993ze}, where it was shown (as anticipated in 
\cite{Bershadsky:1993cx}) that it reduces to a computation
in topological string theory. We now briefly review the argument.

The graviphoton originates from the Ramond-Ramond sector; taking into 
account the peculiar couplings of RR states to the dilaton, $F_g$ is 
identified as a genus $g$ amplitude\footnote{When $X$ is K3-fibered,
and in the limit of a large base, one can obtain the
generalized prepotential from a one-loop heterotic computation \cite{Antoniadis:1995zn,Marino:1998pg}.}. 
Perturbative contributions from
a different loop order or non-perturbative ones are forbidden, since
the type II dilaton is an hypermultiplet. The graviton vertex operator
(in the 0 superghost picture) is
\be
\label{vg}
V_g^{(0)}= h_{\mu\nu}  (\partial X^\mu + i p\cdot \psi \ \psi^\mu )
(\bar \partial X^\mu + i p\cdot \tilde \psi \ \tilde \psi^\mu ) e^{ip X}
\ee
The vertex operator of the
graviphoton (in the -1/2 superghost picture) is
\be
V_T^{(-1/2)}= \epsilon_\mu p_\nu e^{-(\phi+\tilde\phi)/2} 
\left( S \sigma_{\mu\nu} \tilde S\ \Sigma_+ \tilde \Sigma_\mp + cc \right) e^{ipX}
\ee
where $S,\tilde S$ are spin fields in the 4 non-compact dimensions, 
and $\Sigma_\pm$ is the spectral flow operator \eqref{spesig} in the 
$N=(2,2)$ SCFT. 
The insertion of $2g-2$ graviphotons induces a background charge
$\int \frac{\sqrt{3}}{2} H ~R^{(2)}$, which induces the topological
twist $L_0\to L_0-\frac12 J$. The same process takes place in the
SCFT describing the 4 non-compact directions. As a result, the
bosonic and fermionic fluctuation determinants cancel. Moreover,
choosing the polarizations of the graviton and graviphotons to be
anti-self-dual, only the $\psi\psi\tilde\psi\tilde\psi$ terms
in \eqref{vg} contribute after summing over spin structures, 
and cancel against the contractions of the spin fields $S \tilde S$.

Now we turn to the cancellation of the superghost charge: the integration
over supermoduli brings down $2g-2$ powers of the picture-changing operator
$e^\phi T_F \times cc$, where $T_F=G_++G_-$ is the supercurrent. 
In order to cancel the superghost background
charge $2g-2$, it is therefore necessary to transform $g-1$ of
the $2g-2$ graviphoton vertex operators in the $+1/2$ picture. In total,
we thus have $3g-3$ insertions of $T_F$. By conservation of the $U(1)$
charge, it turns out that only the $G_-$ and $\tilde G_\pm$ parts
of $T_F$ and $\tilde T_F$ contribute. Finally, we reach
\be
A_g = (g!)^2 \int_{{\cal M}_g} \langle
\prod_{a=1}^{3g-3} ( \mu_a G_-)  ( \tilde\mu_a \tilde G_{\pm}) \rangle
= (g!)^2 F_g 
\ee
where the upper (lower) sign corresponds to type IIB (resp. IIA).
We conclude that the generalized prepotential $F_g(X)$ in type IIA (B)
string theory compactified on $X$ is equal to the all genus
vacuum amplitude \eqref{topamp} of the A (resp. B)-model topological string.
The precise identification of the variables is
\be
\label{precid}
F_{\rm top}= \frac{i\pi}{2} F_{SUGRA}\ , \quad
t^A = \frac{X^A}{X^0}\ ,\quad
\lambda = \frac{\pi}{4} \frac{W}{X^0}
\ee
To be more precise, the vacuum topological amplitude $F_g(t,\bar t)$,
computes the physical $R^2 T^{2g-2}$ coupling; it differs from the
the holomorphic ``Wilsonian'' coupling $F_g(X)$ appearing in 
\eqref{wilsonc}
due to the contributions of massless particles. It is often assumed that
these contributions are removed by taking $\bar t\to \infty$ keeping
$t$ fixed; it would be interesting to 
determine whether this is indeed equivalent
to going to using the real polarized topological wave 
function \eqref{psixpsip}.

For completeness and later reference, let us mention that, by a similar
reasoning, the topological B-model (resp. A) in type IIA (resp. B) 
computes higher-derivative interactions between the hypermultiplets,
of the form \cite{Antoniadis:1993ze}
\be
\label{ftilde}
\tilde S = \int d^4x \sum_{g=1}^{\infty} \tilde
 F_g(X) \left[ g (\partial\partial S)^2 (\partial Z)^{2g-2}
+  2g(g-1) (\partial\partial S \partial Z)^2 (\partial Z)^{2g-4} \right]
\ee
where $(S,Z)$ describes the universal hypermultiplet. It is also an
interesting open problem to construct an off-shell superfield formalism
which would describe all these interactions at once as F-terms. 

\subsection{Bekenstein-Hawking-Wald Entropy}
In general, higher-derivative corrections affect the macroscopic entropy
of black holes in two ways: 
\begin{itemize}
\item[i)] they affect the actual solution, and in particular the
relation between the horizon geometry and the data measured at infinity;
\item[ii)] by modifying the stress-energy tensor, they change the relation 
between geometry and entropy. 
\end{itemize}
Moreover, since subleading contributions to the statistical entropy 
are non-universal, comparison with the microscopic result requires
\begin{itemize}
\item[iii)] specifying the statistical ensemble implicit in
the low-energy field theory.
\end{itemize}
As far as i) is concerned, and provided that we restrict to BPS black holes, 
the fact that the generalized $\CN=2$ 
supergravity has an off-shell description simplifies the 
computation drastically: the supersymmetry transformation rules are 
the same as at tree-level;  Cardoso, de Wit and Mohaupt 
\cite{LopesCardoso:1998wt,LopesCardoso:1999cv,LopesCardoso:1999ur,
LopesCardoso:1999xn} (CdWM) have shown 
that the horizon geometry is still $AdS_2 \times S^2$,  
while the value of the moduli is 
governed by the a generalization of the stabilisation 
equations \eqref{stabeq},
\be
\label{defstab}
\Re(Y^I)=p^I\ ,\quad \Re(G_I)=q_I\ ,\quad W^2=2^8
\ee
where $G_I$ is now the derivative of the generalized prepotential,
$G_I=\pa F(Y,W^2)/\pa Y^I$. 

As far as ii) is concerned, Wald \cite{Wald:1993nt}
has given a general prescription
for obtaining an entropy functional that satisfies the first law
\footnote{The validity of the zero-th and second law 
was discussed in \cite{Jacobson:1993vj,Jacobson:1995uq}.} of
thermodynamics, in the context of a 
Lagrangian ${\cal L}(R)$ with a general dependence on the Riemann tensor:
\be
\label{sbhw}
S_{BHW} = 2\pi \int_{\Sigma} \frac{\partial {\cal L}}{\partial 
R_{\mu\nu\rho\sigma}} \epsilon^{\mu\nu}
\epsilon^{\rho\sigma} ~\sqrt{h} ~d\Omega 
\ee
where $h$ is the induced metric on the horizon $\Sigma$, and 
$\epsilon^{\mu\nu}$ is the binormal. 

\begin{exo} Show that for ${\cal L}=-\frac{1}{16\pi G} R$,
\eqref{sbhw} reduces to the usual Bekenstein-Hawking area law.
\end{exo} 

While the $\CN=2$ corrected Lagrangian does not have such a simple form, 
CdWM adapted Wald's construction and found a simple result
generalizing \eqref{stabeq2}
\be
\label{defsb}
S_{BHW}=\frac{i\pi}{4} \left( \bar Y^I G_I - Y^I \bar G_I \right)
-\frac{\pi}{2} \Im\left[ W \pa_W F \right]
\ee
where the r.h.s. should be evaluated at the attractor point \eqref{defstab}.

It should be emphasized that this result takes into account the contributions
of the F-terms only; at a given order in momenta, there surely are other
``D-terms'' interactions which would contribute to the thermodynamical
entropy. The results below suggest that such contributions should cancel
for BPS black holes: a beautiful
proof has been given in \cite{Kraus:2005vz}, but assumes that the black hole
can be lifted to 5 dimensions.

\subsection{The Ooguri-Strominger-Vafa Conjecture \label{osvconj}}
As noticed in \cite{Ooguri:2004zv}, 
using the homogeneity relation $Y^I G_I + W \pa_W F = 2F$,
it is possible to perform the same manipulation as in \eqref{osvman},
and rewrite
the entropy \eqref{defsb} as a Legendre transform
\bea
S_{BHW} &=& \frac{i\pi}{4} \left[ (Y^I-2i\phi^I) G_I 
- (\bar Y^I+2i\phi^I) \bar G_I \right]
+\frac{i\pi}{4} \left[ W \pa_W F - \bar W \pa_{\bar W} \bar F \right]
\\
&=& \frac{i\pi}{2} (F - \bar F ) + \frac{\pi}{2}
\phi^I(G_I+\bar G_I)\\
&=& {\cal F}(p^I,\phi^I) + \pi \phi^I q_I \label{osvcor}
\eea
of the ``topological free energy'' ${\cal F}(p^I,\phi^I)$, which 
now incorporates the infinite series of higher derivative F-term corrections,
\be
\label{freeftop}
{\cal F}(p^I,\phi^I) = 
-\pi \ \mbox{Im} 
\left[ F( Y^I=p^I+i\phi^I; W^2=2^8) \right]
\ee
In fact, there are now general arguments \cite{Sen:2005wa,Kraus:2005vz}
to the effect that the Bekenstein-Hawking-Wald entropy 
is equal the Legendre transform
of the Lagrangian evaluated on the near-horizon geometry; 
in the case of $\CN=2$ supergravity, the 
equality of this Lagrangian with the topological
free energy ${\cal F}(p,\phi)$ was
checked recently in \cite{Sahoo:2006rp}.

As argued by OSV, the simplicity of \eqref{osvcor} strongly suggests
that the thermodynamical ensemble implicit in the
BHW entropy is a ``mixed'' ensemble, where 
magnetic charges are treated micro-canonically 
but electric charges are treated canonically; the thermodynamical
relation \eqref{osvcor} should then perhaps be viewed as an approximation
of an exact relation between two different statistical ensembles
\be
\sum_{q_I \in \Lambda_{el}} \Omega(p^I, q_I) e^{- \pi \phi^I q_I}
\stackrel{?}{=}
 e^{{\cal F}(p^I,\phi^I)} 
\ee
where $\Omega(p^I, q_I)$ are the ``microcanonical'' degeneracies of
states with fixed charges $(p^I,q_I)$, and
the sum runs over the lattice $\Lambda_{el}$ of electric charges.
Making use of \eqref{freeftop}, the right-hand side may be 
rewritten as
\be
\label{osvdir}
\sum_{q_I \in \Lambda_{el}} \Omega(p^I, q_I) e^{- \pi \phi^I q_I}
\stackrel{?}{=}
\left| \Psi_{\rm top}(p^I + i \phi^I, 2^8) \right|^2
\ee
or, conversely,
\be
\label{osvinv}
\Omega(p^I, q_I) \stackrel{?}{=} \int d\phi^I 
\ |\Psi_{\rm top}(p^I + i \phi^I, 2^8)|^2
e^{\pi \phi^I q_I}
\ee  
It should be stressed that going from the ``OSV fact'' \eqref{sbhosv}
to the OSV conjecture \eqref{osvdir} involves a considerable leap of
faith which should not be taken lightly.

In its strongest form, the conjecture  provides a way to compute 
the exact microscopic degeneracies $\Omega(p^I, q_I)$  from the 
topological string amplitude $F(X,W^2)$. However, this would most likely
require extending the definition of $F(X,W^2)$ to include non-perturbative
contributions in $W$. Conversely, one may hope to understand the 
non-perturbative completion of the topological string
from a detailed knowledge of black hole degeneracies.
The weaker, more concrete form of the OSV conjecture states that 
the relation \eqref{osvinv} should hold 
asymptotically to all orders in inverse charges. 
 
The conjecture calls for some immediate remarks: 
\begin{itemize}
\item While the formula \eqref{osvinv} at first sight seems to treat electric
and magnetic charges differently, it is nevertheless invariant under 
electric-magnetic duality, provided that the topological amplitude 
$\Psi_{\rm top}$ transforms in the metaplectic representation of
the symplectic group (see Exercise \ref{wigfour} on page 
\pageref{wigfour} below). 
Thus, $\Psi_{\rm top}$ should be understood as
the topological wave function $\Psi_{\IR}(p^I)$ in the real polarization
\cite{Verlinde:2004ck}, which may be different from 
the $\bar t\to\infty$ limit, as
stressed below \eqref{precid}.
\item Upon analytically continuing $\phi^I=i\chi^I$, 
the r.h.s. of  \eqref{osvinv} defines the Wigner 
function associated to the quantum state $\Psi_{\rm top}$ (we shall
return to this observation in Section \ref{qatt}). As it is well known
in quantum mechanics, it is not definite positive, so if the strong
conjecture is to hold, $\Omega(p,q)$ should probably refer to an index rather
than to an absolute degeneracy of states. This fits well with the fact that
$\Psi_{\rm top}$ contains only information about F-term interactions, which
is probably insufficient to encode the absolute degeneracies.
\item Due to charge quantization, the l.h.s. of \eqref{osvdir}
is formally periodic under imaginary shifts $\phi^I\to \phi^I + 2 i k^I$,
$k^I\in \IZ$, which is not the case of the r.h.s. $|\Psi_{\rm top}|^2$.
This can be repaired by replacing \eqref{osvdir} by
\be
\sum_{q_I \in \Lambda_{el}} \Omega(p^I, q_I) e^{- \pi \phi^I q_I}
\stackrel{?}{=} 
\sum_{k^I \in \Lambda_{el}^*} 
\Psi^*\left(p^I - 2 k ^I - i \phi^I\right)\ 
\Psi\left(p^I + 2 k ^I + i \phi^I\right)
\ee
without affecting the converse statement \eqref{osvinv}. 
This r.h.s. of this equation is reminiscent of a theta series.
Similar averaging have indeed been found to occur in some non-compact
models \cite{Vafa:2004qa,Aganagic:2004js}. Note however that 
this averaging renders the prospect of recovering the 
non-perturbative generalization
of $\Psi_{\rm top}$ from $\Omega(p^I,q_I)$ more uncertain.
\item The sum on the l.h.s. of \eqref{osvdir}
does not appear to converge, which reflects the
thermodynamical instability of the mixed ensemble.
Moreover, specifying the integration contour in \eqref{osvinv} would require
understanding the singularities of the topological amplitude.
These subtleties do not affect the weak form of the conjecture, since
the saddle point approximation to \eqref{osvinv} is independent
of the details of the contour.
\item A variant of the OSV conjecture \eqref{osvinv} has been proposed
in \cite{LopesCardoso:2006bg}, which involves an integral
over both $X^I$ and $\bar X^I$, or equivalently a thermodynamical
ensemble with fixed electric and magnetic potentials (see 
Exercise \ref{altvar} on page \pageref{altvar}). It would be
interesting to demonstrate the equivalence of this approach 
with the one based on the holomorphic polarization of the 
topological amplitude \cite{Verlinde:2004ck}.
\end{itemize}

The OSV conjecture has been successfully tested in the case of 
non-compact Calabi-Yau manifolds of the form
$O(-m) \oplus O(2g-2+m)\to \Sigma_g$, where 
$\Sigma$ is a genus $g$ Riemann surface  \cite{Vafa:2004qa,Aganagic:2004js}:
BPS states are counted by topologically
twisted SYM on $N$ D4-brane wrapped on a 4-cycle $O(-m)\to \Sigma$, 
which is equivalent to 2D Yang Mills (or a $q$-deformation
thereof, when $g\neq 1$). At large $N$, the partition
function of 2D Yang-Mills indeed factorizes into two 
chiral halves \cite{Gross:1993hu},
which indeed agree with the topological amplitude computed independently.
Exponentially suppressed corrections to the large $N$ limit
of 2D Yang-Mills have been studied in \cite{Dijkgraaf:2005bp}, 
and seem to call for a 
``second quantization'' of the r.h.s. of \eqref{osvdir}. 
For $\CN=4$ and $\CN=8$ compactifications on $K3\times T^2$ and $T^6$,
the formula \eqref{osvinv} has been compared to the prediction for
dyons degeneracies based on U-dualities, and agreement has been found
in the semi-classical approximation \cite{Shih:2005he}. 
More recently, several ``derivations''
of the weak form of the OSV conjecture have been given, using a $M2-{\bar M}2$ 
or $D6-{\bar D}6$ representation of the black hole, and some modular
properties of the partition function \cite{Gaiotto:2006ns,Mooretoappear,
Beasley:2006us,deBoer:2006vg}. These approaches make it clear that
the strong form of the conjecture cannot hold, and suggest possible
sources of deviations from the ``modulus square'' form.

In the next Section, we shall present a precision test of the OSV
conjecture in the context of small black holes in $\CN=4$
and $\CN=2$ theories, whose microscopic counting can be made exactly.

\section{Precision Counting of Small Black Holes\label{smallbh}}
In order to test the OSV conjecture, one should be able to compute
subleading corrections to the microscopic degeneracies $\Omega(p,q)$. 
Due to subtleties in the ``black string'' CFT description of 4-dimensional 
black holes, it has not been hitherto possible to reliably compute subleading 
corrections to \eqref{cardy256} for generic BPS black holes.

On the other hand, the heterotic string has a variety of BPS excitations
which can be counted exactly using standard wordsheet techniques. Since these
states are only charged electrically (in the natural heterotic polarization),
their Bekenstein-Hawking entropy evaluated using tree-level supergravity
vanishes. This means that higher-derivative corrections cannot
be neglected, and indeed, upon including $R^2$ corrections to the 
effective action, a smooth horizon with finite area is obtained.
We refer to these states as ``small black holes'', to be contrasted
with ``large black holes'' which have non-vanishing entropy already at
tree level. This section is based on \cite{Dabholkar:2004yr,Dabholkar:2005by,
Dabholkar:2005dt}.

\subsection{Degeneracies of DH states and the Rademacher formula}

The simplest example to study this phenomenon is  the heterotic string
compactified on $T^6$. A class of perturbative BPS states, known as 
``Dabholkar-Harvey'' (DH) states, can be constructed by tensoring the 
ground state of the right-moving superconformal theory with a level $N$ 
excitation of the 24 left-moving bosons, and adding momentum $n$
and winding $w$ along one circle in $T^6$ such 
that the level matching condition $N-1= n w$ is 
satisfied \cite{Dabholkar:1989jt,Dabholkar:1990yf}.
The number of distinct DH states with fixed charges $(n,w)$ is 
$\Omega(n,w)= p_{24}(N)$, where $p_{24}(N)$ is the number of partitions
on $N$ into the sum of 24 integers (up to an overall factor of 16
corresponding
to the size of short $\CN=4$ multiplets, which we will always drop).
Accordingly, the generating function of the degeneracies of DH states is
\begin{equation}
\label{partition}
\sum_{N=0}^{\infty} p_{24}(N)\ q^{N-1} = \frac{1}{\Delta(q)},
\end{equation}
where  $\Delta(q)$ is Jacobi's discriminant function
\be
\Delta(q)
=\eta^{24}(q)
=q\prod_{n=1}^{\infty}(1-q^n)^{24}
\ee
In order to determine the asymptotic density of states at large $N-1=nw$,
it is convenient to extract $d(N)$ from the partition function
\eqref{partition} by an inverse Laplace transform,
\begin{equation}\label{density}
    p_{24}(N) = \frac{1}{2\pi i} \int_{\eps-i\pi}^{\eps+i\pi}
    d\beta\ e^{\beta (N-1)}
    \frac{16}{\Delta(e^{-\beta})}.
\end{equation}
where the contour $C$ runs from $\eps-i\pi$ to $\eps+i\pi$,
parallel to the imaginary axis. One may now take the high temperature
limit $\eps\rightarrow 0$, and use the modular property of the discriminant
function 
\begin{equation}\label{modular}
    \Delta(e^{-\beta}) = \left(\frac{\beta}{2\pi}\right)^{-12}
    \Delta(e^{-4 \pi^2 /\beta}).
\end{equation}
As $e^{-4 \pi^2 /\beta} \rightarrow 0$, we can approximate
$\Delta(q) \sim q$ and write the integral as
\begin{equation}\label{asymptotic}
   p_{24}(N) = \frac{16}{2\pi i} \int_C d\beta\ \left(\frac{\beta}
   { 2\pi} \right)^{12} e^{\beta (N-1) + 4 \frac{\pi^2}{\beta}}
\end{equation}
This integral may be evaluated by steepest descent:
the saddle point occurs at $\beta = 2\pi / \sqrt{N-1}$, leading to
the characteristic Hagedorn growth 
\be 
p_{24}(N)  \sim \exp{( 4 \pi \sqrt{nw})}
\ee
for the spectrum of DH states.

To calculate the sub-leading terms systematically in an asymptotic
expansion at large $N$,
one may recognize that \eqref{asymptotic} is proportional
to the integral representation of a modified Bessel function,
\begin{equation}\label{Besselint}
    \hat I_\nu(z) = 
-i (2\pi)^\nu  \int_{\epsilon
    -i\infty}^{\epsilon +i\infty} \frac{dt}{t^{\nu +1}} e^{(t
    +z^2/4t)} 
\hat I_\nu(z)
\end{equation}
We thus obtain
\begin{equation}\label{asymptotic2}
    \Omega(n,w) = p_{24}(N) \sim 2^4\ \BesselI{13}{4}{nw} \ .
\end{equation}
Using the standard asymptotic expansion of $\hat I_\nu(z)$ at large $z$
\be\label{inuas}
\hat I_\nu(z) \sim 2^\nu \left(\frac{z}{2\pi}\right)^{-\nu-\frac12}
\left[ 1- \frac{(\mu
    -1)}{8z} + \frac{(\mu
    -1)(\mu -3^2)}{2!(8z)^2} - \frac{(\mu
    -1)(\mu -3^2)(\mu -5^2)}{3!(8z)^3}+ \ldots \right],
\ee
where $\mu = 4 \nu^2$, we can compute the subleading corrections
to the microscopic entropy of DH states to arbitrary high order,
\begin{equation}\label{boltzentropy}
\log \Omega(n,w) \sim
 4\pi \sqrt{|nw|} -\frac{27}{4} \log|nw|
    +\frac{15}{2}\log 2 -\frac{675}{32\pi\sqrt{|nw|}} -\frac{675}{2^8 \pi^2
 |nw|}-\ldots
\end{equation}

This is still {\it not} the complete asymptotic expansion of $\Omega(n,w)$
at large charge. Exponentially suppressed corrections to \eqref{asymptotic2} 
can be computed by using the Rademacher formula 
(see \cite{Dijkgraaf:2000fq} for a physicist account)
\be
\begin{split}
\label{radi} F_\nu(n) = & \sum_{c=1}^\infty\sum_{\mu=1}^{r}
  c^{w-2} {\rm Kl} (n,\nu,m,\mu;c)
\sum_{m+ \Delta_\mu < 0} F_\mu(m) \\
&\vert m+\Delta_\mu
\vert^{1-w}  \hat I_{1-w}
 \biggl[ \frac{4\pi}{c} \sqrt{\vert m+\Delta_\mu\vert(n + \Delta_\nu)}
\biggr] .
\end{split}
\ee
In this somewhat formidable expression, $F_\mu(m)$ denote the Fourier
coefficients of a vector of modular forms 
\be
\label{collec}
f_\mu(\tau) = q^{\Delta_\mu} \sum_{m \geq 0} F_\mu(m) q^m \qquad
\mu = 1, \dots, r
\ee
which transforms as a finite-dimensional unitary 
representation of the modular group of weight $w<0$, with
\bea\label{rep}
f_\mu(\tau+1) & = &e^{2\pi i \Delta_\mu} f_\mu(\tau)\\
f_\mu(-1/\tau) & = &(-i \tau)^w S_{\mu\nu} f_\nu(\tau)
\eea
The coefficients ${\rm Kl}(n,\nu,m,\mu;c)$ are
generalized Kloosterman sums, defined as
\be
\label{kloos}
{\rm Kl}(n,\nu;m,\mu;c)\equiv 
\sum_{0<d<c; d\wedge c=1}
e^{2\pi i \frac{d}{c} (n+\Delta_\nu)}\  M(\gamma_{c,d})^{-1}_{\nu\mu}\ 
e^{2\pi i \frac{a}{c} (m+\Delta_\mu)}
\ee
where
\be
\gamma_{c,d} = \begin{pmatrix} a & (ad-1)/c \\ c & d \end{pmatrix}
\ee
is an element of $Sl(2,\IZ)$ and $M(\gamma)$ its matrix representation.
For $c=1$ in particular, we have:
\be\label{radiii}
{\rm Kl}(n,\nu,m,\mu;c=1)   = S^{-1}_{\nu\mu}
\ee
Going back to \eqref{radi}, we see that the growth of the 
Fourier coefficients is determined only by the Fourier coefficients
of the ``polar'' part $F_{\mu}(m)$ where $m+\Delta_\mu<0$, as well as
some modular data. The Ramanujan-Hardy formula 
\be
\label{cardyform}
F_{\mu}(n) \sim \exp\left[ 2\pi \sqrt{\frac{c_{\rm eff}}{6} n} \right]
\ee
is reproduced by keeping the
leading term $c=1$ only, using  $\Delta=c_{\rm eff}/24, w=-c_{\rm eff}/2$ 
and the
asymptotic behavior  \eqref{inuas}.
The terms with $c>1$ also grow exponentially,
but at a slower rate than the term with $c=1$. They therefore contribute
exponentially suppressed contributions to $\log F_\nu(n)$. 

Applying \eqref{radi} to the case at hand, we have the convergent
series expansion
\be
\label{rade24}
\Omega(n,w) = 2^4 \sum_{c=1}^{\infty} c^{-14}\ \mbox{Kl}(nw+1,0;c)\
\BesselI{13}{\frac{4}{c}}{|nw|}
\ee

\subsection{Macroscopic entropy and the topological amplitude\label{bhw}}
We now turn to the macroscopic side, and determine the 
Bekenstein-Hawking-Wald entropy for a BPS black hole with the above
charges. Since the attractor formalism is tailored for
$\CN=2$ supergravity, one should first decompose the spectrum
in $\CN=2$ multiplets: the $\CN=4$ spectrum decomposes into 
one $\CN=2$ gravity multiplet, 2 gravitino multiplets and
$n_V=23$ vector multiplets (not counting the graviphoton). 
Provided the charges under the 
4 vectors in the gravitino multiplets vanish, the $\CN=2$
attractor mechanism applies. 

The topological amplitude $F_1$ has been computed in \cite{Harvey:1996ir},
and can be obtained as the holomorphic part of the
$R^2$ amplitude at one-loop,
\be \label{fr2} f_{R^2} = 24 \log ( T_2 |\eta(T)|^4 )
\ee
where $T,U$ denote the K\"ahler and complex
structure moduli of the torus $T^2$. 
All higher topological
amplitudes $F_g$ for $g>1$  vanish for models with $\CN=4$ supersymmetry.
We therefore obtain the generalized prepotential
\be
F(X^I,W^2)=  - \frac12 \sum_{a,b=2}^{23} C_{ab} \frac{X^a X^b X^1}{X^0}
-\frac{W^2}{128\pi i} \log\Delta(q) \ee
where $C_{ab}$ is the intersection matrix on $H^2(K3)$,
$T=X^1/X^0$ and $q=e^{2\pi i T}$. The appearance of the
same discriminant function $\Delta(q)$ as in the 
microscopic heterotic counting
\eqref{partition} is at this stage coincidental.

Identifying $p^1=w$, $q_0=n$ and allowing for arbitrary electric charges
$q_0,q_{i=2..23}$, the black hole free energy \eqref{freeftop} reduces to
\be
\label{free622}
{\cal F}(\phi^I,p^I) = - \frac{\pi}{2} C_{ab} \frac{\phi^a \phi^b
p^1}{\phi^0} -\log|\Delta(q)|^2
\ee
where
\be
\label{qq}
q=\exp\left[ \frac{2 \pi}{\phi^0} \left( p^1 + i \phi^1 \right) \right].
\ee
The Bekenstein-Hawking-Wald entropy
is then obtained by performing a Legendre transform over
all electric potentials  $\phi^I, I=0,\dots23$. The Legendre
transform over $\phi^{a=2..23}$ sets $\phi^a=(\phi^0/p^1) C^{ab} q_b$,
where $C^{ab}$ is the inverse of the matrix $C_{ab}$. We will
check {\it a posteriori} that in the large charge
limit, it is consistent to approximate $\Delta(q)\sim q$,
whereby all dependence on $\phi^1$ disappears.
We thus obtain
\be
\label{sext}
S_{BHW} \sim  \langle
\left[ - \frac{\pi}{2}  \frac{C^{ab} q_a q_b }{p^1} \phi^0
+ 4\pi \frac{p^1}{\phi^0} + \pi \phi^0 q_0  \right] \rangle_{\phi^0}
\ee
The extremum of the bracket lies at
\be
\label{phi0st}
\phi^0_*= \frac12\sqrt{-p^1/\hat q_0}\ ,\qquad
\hat q_0 \equiv  q_0 + \frac{1}{2p^1} C^{ab} q_a q_b
\ee
so that at the horizon the
K\"ahler class $\Im T \sim \sqrt{ - p^1 \hat q_0}$ is very large,
justifying our assumption. Evaluating \eqref{sext} at the extremum,
we find
\be
\label{sbhl}
S_{BH} \sim 4\pi \sqrt{Q^2/2} \ ,\quad
Q^2 = 2p^1 q_0 + C^{ab} q_a q_b
\ee
in agreement with the leading exponential behavior in
\eqref{boltzentropy}, including the precise numerical factor.
Note that the argument up to this stage is independent of  the OSV conjecture,
and relies only on the classical attractor mechanism in the presence of
higher-derivative corrections. The fact that
the Bekenstein-Hawking entropy of small black holes
comes out proportional to $\sqrt{Q^2/2}$ was
argued in \cite{Sen:1995in,Sen:1997is,Sen:2004dp}, based ongeneral
scaling arguments. The precise numerical agreement was demonstrated
in \cite{Dabholkar:2004yr}, although with hindsight it could also have
been observed by the authors of \cite{Maldacena:1997de}. 
This agreement indicates
that the tree-level $R^2$ coupling in the effective action of the
heterotic string on $T^6$ (or, equivalently,
large volume limit of the 1-loop $R^2$ coupling in type IIA/$K3\times
T^2$) is sufficient to cloak the singularity of the small black hole
behind a smooth horizon. This is in fact confirmed by a study of the
corrected geometry \cite{Sen:2004dp,Dabholkar:2004dq,Hubeny:2004ji}.

\subsection{Testing the OSV Formula}

We are now ready to test the proposal \eqref{osvinv} and evaluate
the inverse Laplace transform of $\exp({\cal F})$ with respect
to the electric potentials,
\be
\label{omosv1}
\Omega_{OSV}(p)
= \int d\phi^0~d\phi^1~d^{22}\phi^a
\frac{1}{\vert\Delta(q)\vert^2}
\exp\left[ - \frac{\pi}{2} C_{ab} \frac{\phi^a \phi^b p^1}{\phi^0}
+ \pi \phi^0 q_0+ \pi \phi^a q_a \right]
\ee
Due to the non-definite signature of $C_{ab}$,
the integral over $\phi^a$ diverges for real values.
This may be avoided by rotating the integration contour to $\eps+i \IR$ for all
$\phi$'s. The integral over $\phi^a$ is now a Gaussian, leading to
\be
\label{osvde}
\Omega_{OSV}(Q)
= \int d\phi^0~d\phi^1 \left( \frac{\phi_0}{p^1} \right)^{11}
\frac1{\vert \Delta(q) \vert^2}
\exp\left( - \frac12  \frac{C^{ab} q_a q_b}{p^1} \phi^0 +  q_0 \phi^0 \right)
\ee
where we dropped numerical factors and used the fact that $\det C=1$.
The asymptotics of $\Omega$ is independent of the details of the contour, 
as long as it selects
the correct classical saddle point \eqref{phi0st} at large charge.
Approximating again $\Delta(q)\sim q$, we find the quantum version
of \eqref{sext},
\be
\label{osvde2}
\Omega_{OSV}(Q)
\sim \int d\phi^0~d\phi^1 \left( \frac{\phi_0}{p^1} \right)^{11}
\exp\left( - \frac12  \frac{C^{ab} q_a q_b}{p^1} \phi^0
-4\pi \frac{p^1}{\phi^0} +  q_0 \phi^0 \right)
\ee
The integral over $\phi^1$ superficially leads to an infinite result.
However, since the free energy is invariant under $\phi^1\to\phi^1+\phi^0$,
it is natural to restrict the integration to a single period
$[0,\phi^0]$, leading to an extra factor of $\phi^0$ in \eqref{osvde2}.
The integral over $\phi^0$ is now of Bessel type, leading to
\be
\label{osvfin}
\Omega_{OSV}(Q) = (p^1)^2\ \BesselI{13}{4}{Q^2/2}
\ee
in impressive agreement with the microscopic result \eqref{asymptotic2}
at all orders in $1/Q$.

Some remarks on this computation are in order:
\begin{itemize}
\item Note that the 
extra factor $(p^1)^2$ in Eq.~\eqref{osvfin}
is inconsistent with $SO(6,22,\IZ)$ duality, which requires
the exact degeneracies to be a function of $Q^2$ only. 
Moreover, the agreement depends
crucially on discarding the
non-holomorphic correction proportional to $\log T_2$ in $F_1$.
Both of these issues call for a better understanding of the
relation between the physical amplitude and the 
topological wave function in the real polarization. It should
be mentioned that an alternative approach has been developped
by Sen, keeping the non-holomorphic corrections but using
a different statistical ensemble \cite{Sen:2005ch,Sen:2005pu}.

\item The ``all order'' result \eqref{osvfin} depends only on
the number of $\CN=2$ vector multiplets, as well as on the
leading large volume behavior of $F_1 \sim q /(128 \pi i)$.
By heterotic/type II duality, this term is mapped to a
tree-level $R^2$ interaction on the heterotic
side, which is in fact universal. We thus conclude that in all
$\CN=2$ models which admit a dual heterotic description, provided
higher genus $F_{g>1}$ and genus 0,1
Gromov-Witten instantons can be neglected, the
degeneracies of small black holes predicted by \eqref{osvinv}
are given by
\be
\label{unideg}
\Omega_{OSV}(Q) \propto \BesselI{\frac{n_V+3}{2}}{4}{Q^2/2}\ ,
\ee
where $n_V$ is the number of Abelian gauge fields, including the
graviphoton. We return to the validity of the assumption in the next 
subsection.

\begin{exo}
By applying a similar argument to large black holes with $p^0=0$,
assuming that only the large-volume limit of $F_1$ contributes,
show that the OSV conjecture \eqref{osvinv}, in the saddle
point approximation, predicts \cite{Dabholkar:2005by,Dabholkar:2005dt}
\be
\label{ibessl}
\Omega(p^A, q_A) \sim
\pm \frac12 \vert \det C_{ab}(p)\vert^{-1/2}
\Bigl({\hat C(p) / 6} \Bigr)^{\frac{n_V+2}{2}}\ \times \
\hat I_{\frac{n_V+2}{2}} \Biggl(2\pi \sqrt{- \hat C(p) \hat q_0/ 6} \Biggr)
\ee
where
\be\label{dfsn}
C_{AB}(p)  = C_{ABC} p^C, \quad C(p)
= C_{ABC} p^A p^B p^C , \quad \hat C(p) = C(p) + c_{2A}p^A \ ,
\ee
and compare to the microscopic counting \eqref{msw}. \label{exomsw}
\end{exo}

\item In order to see if the strong version of the OSV conjecture has
a chance to hold, it is instructive to change variable to $\beta=\pi/t$
in \eqref{density} and rewrite the exact microscopic result as
\be
\Omega_{\rm exact}(Q) = \int dt\ t^{-14} \ \frac{\exp\left( \frac{\pi n w}{t}
  \right)}{\Delta\left( e^{-4\pi t} \right)}
\ee 
On the other hand, it is convenient to change variables in the
OSV integral \eqref{osvde} to $\tau_1=\phi^1/\phi^0,
\tau_2=-p^1/\phi^0$, with Jacobian $d\phi^0 d\phi^1 = 8 (p^1)^2
d\tau_1 d\tau_2/\tau_2^3$, leading to
\be
\label{osvtry}
\Omega_{OSV}(Q)
\sim \int d\tau_1 ~ d\tau_2 ~ \tau_2^{-14}
\frac{\exp\left(\frac{\pi n w}{\tau_2} \right)} {\vert
\Delta\left( e^{-2\pi \tau_2 + 2\pi i \tau_1} \right) \vert^2} 
\ee
Despite obvious similarities, it appears unlikely that the two
results are equal non-perturbatively.

\item Just as the perturbative result \eqref{asymptotic2}, the
result \eqref{osvfin} misses subleading terms in the Rademacher
expansion \eqref{rade24}. It does not seem possible to interpret
any of the terms with $c>1$ as the contribution of a subleading
saddle point in either \eqref{asymptotic} or \eqref{osvde}.
\end{itemize}

Despite these difficulties, it is remarkable that the
black hole partition function in the OSV ensemble, obtained from
purely macroscopic considerations, reproduces the entire
asymptotic series exactly to all orders in inverse charge.
Recent developments show that this agreement is largely
a consequence of supersymmetry and anomaly cancellation
for black holes which have an $AdS_3$ region \cite{Kraus:2005vz,Kraus:2005zm,
Kraus:2006nb} (see also the lectures by P. Kraus \cite{Kraus:2006wn} 
in this volume).

\subsection{$\CN=2$ Orbifolds}
We conclude this section with a few words on small black holes
in $\CN=2$ orbifolds, referring to \cite{Dabholkar:2005by,Dabholkar:2005dt} 
for detailed computations.
We find that the agreement found in $\CN=4$ models broadly continues
to hold, with the following caveats:
\begin{itemize}

\item In contrast to $\CN=4$ cases, the neglect of Gromov-Witten instantons 
is harder to justify rigorously: 
when $\chi(X)\neq 0$, the series of point-like
instantons contribution becomes strongly coupled in the regime
of validity of the Rademacher formula,
$\hat q_0 \gg \hat C(p)$. The strong coupling
behavior is controlled, up to a logarithmic term, by the Mac-Mahon
function \eqref{mmahondef}, which is exponentially suppressed in this regime. 
The logarithmic term in \eqref{maclog} may be reabsorbed
into a redefinition of the
topological string amplitude $\Psi_{\rm top}\to \lambda^{\chi/24}
\Psi_{\rm top}$. As for non-degenerate
instantons, they are exponentially suppressed provided all
magnetic charges are non zero. This is unfortunately not the case
for the small black holes dual to the heterotic DH states, whose
K\"ahler classes are attracted to the boundary of the K\"ahler cone at the
horizon. 

\item For BPS states in twisted sectors of $\CN=2$ orbifolds,
we find that  the instanton-deprived OSV proposal appears to successfully
reproduce the  {\it absolute degeneracies}, equal to the indexed
degeneracies, to all orders.
For untwisted DH states of the OSV prediction
appears to agree with the {\it absolute degeneracies} of untwisted DH
states to leading order ( which have the same exponential growth as
twisted DH states), but not at subleading order (as
the subleading corrections in the untwisted sector are moduli-dependent,
and uniformly smaller than in the twisted sectors).
The indexed degeneracies are exponentially
smaller than absolute degeneracies, due to cancellations of pairs
of DH states, so plainly disagree with the OSV prediction.
\end{itemize}

\section{Quantum Attractors and Automorphic Partition Functions \label{qatt}}
In this final chapter, we elaborate on an intriguing proposal by Ooguri,
Verlinde and Vafa \cite{Ooguri:2005vr}, to interpret the OSV conjecture 
as a holographic duality between the usual Hilbert space of black hole 
micro-states quantized with respect to global time, and the Hilbert space of 
stationary, spherically symmetric geometries quantized with respect 
to the radial direction. Although we shall find some difficulties in 
implementing this proposal literally, this line of thought will prove
fruitful in suggesting non-perturbative extensions of the OSV conjecture.
In particular, we shall find tantalizing hints of a one-parameter 
generalization of the topological string amplitude in $\CN=2$ theories, 
and obtain a natural framework for constructing automorphic black hole 
partition functions (in cases with suitably large U-duality groups)
which go beyond the Siegel modular forms discussed in Section \ref{dvvform}. 
This chapter is based on \cite{Pioline:2005vi,Gunaydin:2005mx,
Gunaydin:2006bz,Neitzke:2007ke} 
and on-going work \cite{gnpw-in-progress,gnopw-in-progress}.

\subsection{OSV Conjecture and Quantum Attractors \label{osvqatt}}
In order to motivate this approach, recall that, 
after analytically continuing $\phi^I=i\chi^I$ to the imaginary axis,
the r.h.s. of the OSV conjecture \eqref{osvinv} 
\be
\Omega(p^I, q_I) \sim \int d\chi^I 
\Psi_{\rm top}^*( p^I + \chi^I) 
\Psi_{\rm top}( p^I - \chi^I) 
e^{i \pi \chi^I q_I} 
\equiv  W_{\Psi_{\rm top}}(p^I,q_I)
\ee
could be interpreted as the Wigner distribution associated to the
wave function $\Psi_{\rm top}$. In usual quantum mechanics, the Wigner 
distribution  $W_\psi(p,q)$ is a function on phase space associated to
a wave function $\psi(q)$, such that quantum averages of 
Weyl-ordered operators on $\psi$ are equal to classical averages of 
their symbols with respect to $W_\psi$,
\be
\label{osvwig}
\langle \psi | {\cal O}(\hat p,\hat q) | \psi \rangle
= \int dp~dq ~ W_\psi(p,q) ~{\cal O}(p,q)
\ee
Moreover, when $\psi$ satisfies the Schr\"odinger equation, $W$
satisfies the classical Liouville equation to leading order 
in $\hbar$; the Wigner distribution is thus a useful tool 
to study the semi-classical limit. The above observation thus begs
the question: what is the physical quantum system of which $\Psi_{\rm top}$
is the wave function\footnote{Or, to paraphrase Ford Prefect, what is the
Question to the Answer $\Psi_{\rm top}$ ?}, and how come does it encode the
black hole degeneracies ?

\begin{exo}
Show that 
\be
W_{\tilde\psi}(p^I,q_I) = 
W_{\psi}\left(\frac{q_I}{2},2 p^I\right)
\ee
where $\tilde\psi(\phi)=\int d\chi\ e^{i\pi \chi\phi}\ \psi(\chi)$ 
is the Fourier transform of $\psi$.
\label{wigfour}
\end{exo}

\begin{figure}
\centerline{\hfill\includegraphics[height=3cm]{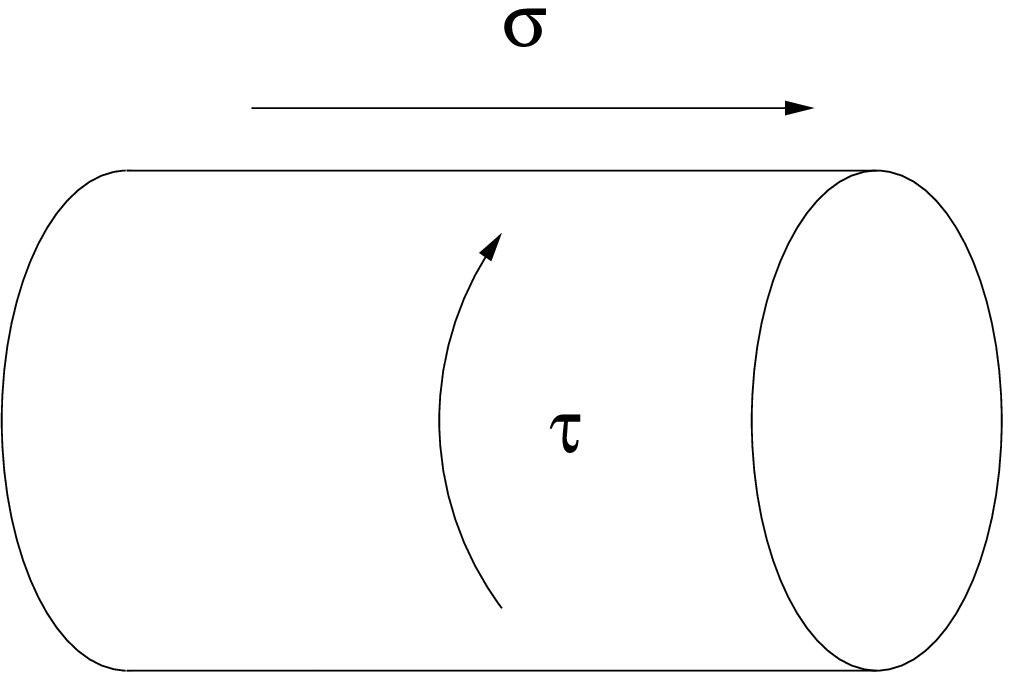}\hfill
\includegraphics[height=5cm]{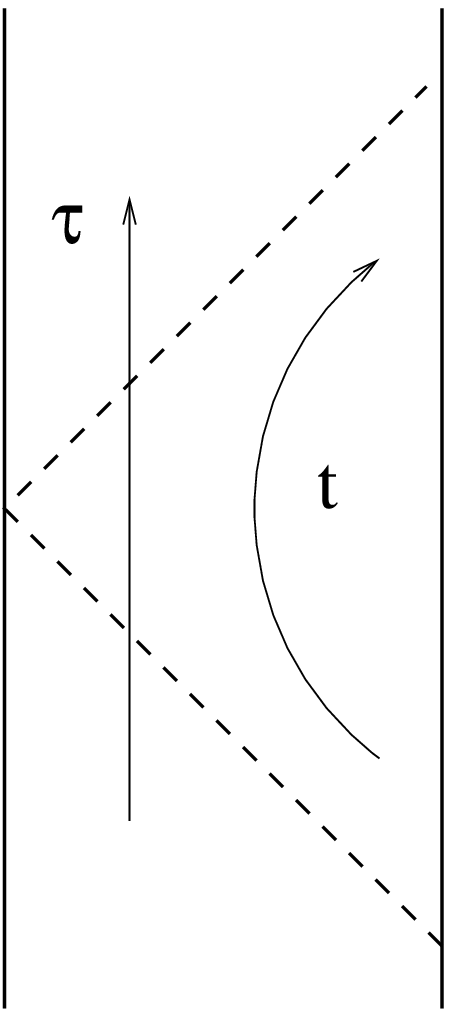}\hfill}
\caption{{\it Left:} the cylinder amplitude in string theory
can be viewed either as 
a trace over the open string Hilbert space (quantizing along $\tau$)
channel) or as an inner product between two wave functions in the
closed string Hilbert space (quantizing along $\sigma$). 
{\it Right:}
The global geometry of Lorentzian $AdS_2$ has the topology of a strip; 
its Euclidean continuation at finite temperature becomes a cylinder.
$\tau$ and $t$ are the global and Poincar\'e time, respectively.
\label{cylpic}}
\end{figure}

In order to try and answer this question, it is useful to reabsorb 
the dependence on the charges $(p^I,q_I)$ into the state itself, by defining
\be
\label{osvwig2}
\Psi_{p,q}^\pm(\chi) \equiv  
e^{\pm i \pi q \chi} \Psi^\pm_{\rm top}(\chi\mp p) 
\equiv  V^\pm_{p,q}\cdot \Psi_{\rm top}(\chi)
\ee
Equation \eqref{osvwig} is then rewritten more suggestively as 
an overlap of two wave functions,
\be
\label{osvlap}
\Omega(p,q) \sim \int d\chi\ [\Psi^-]^*_{p,q}(\chi) \ \Psi^+_{p,q}(\chi)
\ee
On the other hand, recall that the near horizon geometry $AdS_2\times S^2$,
written in global coordinates as
\be
\label{ads2glob}
ds^2 = |Z_*|^2 \left( \frac{-d\tau^2+d\sigma^2}{\cos^2\sigma}
+ d^2\Omega \right)
\ee
has two distinct conformal boundaries at $\sigma=0,\pi$, respectively; its
Euclidean sections at finite temperature have the topology of a cylinder (see
Figure \ref{cylpic}).

\begin{exo} Check that the metric \eqref{ads2glob} is equivalent to
\eqref{ads2s2} upon changing coordinates $
\tau=\arctan(z+t)-\arctan(z-t)\ , \quad
\sigma=\arctan(z+t)+\arctan(z-t)$\ .
Map out the portion of the global geometry covered by the Poincar\'e
coordinates $z,t$.
\end{exo}

With this in mind, it is tempting to view \eqref{osvlap} as an analogue of 
open/closed duality for conformal field theory on the cylinder, 
\be
\label{openclosed}
\Tr~ e^{-\pi t H_{\rm open}} = \langle B' | e^{-\frac{\pi}{t} H_{\rm closed}}
| B \rangle 
\ee
where $| B \rangle$ and $| B' \rangle$ are closed string boundary states.
The right-hand side of \eqref{osvlap}, analogue of the closed string channel,
is identified with the partition function
of quantum gravity on $AdS_2\times S^2$ in radial quantization along the
space-like coordinate $\sigma$, 
with boundary conditions at $\sigma=0,\pi$ specified by the 
``boundary states'' $\Psi^\pm_{p,q}$, while the left-hand side,
analogue of the open string channel, is recognized as a trace of
the identity operator 
in a sector of the Hilbert space for quantization along the global time 
coordinate $\tau$, with fixed charges $p^I,q_I$ (the absence of 
an analogue of the Hamiltonians $H_{\rm open}$ and $H_{\rm closed}$ 
can be traced to diffeomorphism invariance, which requires 
physical states to be solutions of the Wheeler-De Witt equation
$H|\psi\rangle=0$). It should be stressed that 
the Hilbert spaces for time-like and radial quantization are distinct, 
just like the open string and closed string Hilbert spaces are different. 

For this interpretation to make sense, it should of course be possible
to view $\Psi_{\rm top}$ as a state in the Hilbert space for radial 
quantization. This is, at least superficially, consistent with the
wave function interpretation
of $\Psi_{\rm top}$ discussed  in Section \ref{holanomsec}, 
and would in fact provide a nice physical interpretation of this otherwise 
mysterious quantum mechanical behavior. 
Moreover, the functional 
dimension, $n_V+1$, of the Hilbert space hosting $\Psi_{\rm top}$, 
is roughly in accordance with the number of complex scalars $z^i$
varying radially in the black hole geometry. This leads one to expect that
$\Psi_{\rm top}$ may provide a radial wave function for the 
vector-multiplet scalars, in a truncated Hilbert space 
where only static, spherically symmetric BPS configurations are kept.
Such a ``mini-superspace'' truncation is usually hard to justify, but
may hopefully be suitable for the purpose of 
computing indexed degeneracies of BPS black holes, in the same way
as the Ramond-Ramond ground states in the closed string 
channel control the growth of the index in the open string Ramond sector.

This brings us to the problems of (i) quantizing the attractor flow
\eqref{att1},\eqref{att2}, (ii) showing that the resulting Hilbert space 
is the correct habitat for $\Psi_{\rm top}$, and (iii) finding 
a physical principle that selects $\Psi_{\rm top}$ among the continuum
of states in that BPS Hilbert space. Answering these questions will
be the subject of the rest of this chapter. Before doing so,
several general remarks are in order:
\begin{itemize}

\item The idea of radial quantization of static black holes
has a long history in the canonical gravity literature, e.g. 
\cite{Thiemann:1992jj,Kuchar:1994zk,
Cavaglia:1994yc,Hollmann:1996cb,Hollmann:1996ra,Breitenlohner:1998yt}. 
The main new ingredients here are supersymmetry, which may
provide a better justification for the mini-superspace
approximation,  and holography, which offers the possibility 
to reconstruct the spectrum of the global time Hamiltonian
from the overlap of two radial wave functions. The quantization 
of BPS configurations has been considered recently in various 
set-ups and found to agree with gauge theory computations \cite{Mandal:2005wv,
Maoz:2005nk,Rychkov:2005ji,Grant:2005qc,Biswas:2006tj,Mandal:2006tk}.

\item The ``channel duality'' argument is in line
with the usual AdS/CFT philosophy that the black hole
micro-states should be described by ``gauge theoretical'' 
degrees of freedom
living on the boundary of $AdS_2$. Contrary to higher dimensional AdS
spaces, the conformal quantum mechanics describing $AdS_2$ 
is still largely mysterious, and the above approach is 
a possible indirect route towards determining its spectrum.

\item One usually assumes that black hole micro-states can be
described only in terms of the near horizon geometry. The above
proposal to quantize the whole attractor flow seems to be at odds with
this idea. A possible way out is that the topological wave function
be a fixed point of the quantum attractor flow. In the sequel, 
we will study the full quantum attractor flow, from asymptotic
infinity to the horizon, as a function of the Poincar\'e 
radial coordinate $r$ (rather than the ``global radial coordinate $\sigma$'',
which is well defined only near the horizon).

\item The analogy between global $AdS_2$ and open strings 
explained below Eq.~\eqref{openclosed} can be pushed quite a bit further: 
due to pair production of charged particles,
a black hole may fragment in different throats, analogous to the joining
and splitting interactions of open strings \cite{Maldacena:1998uz}
(see \cite{Pioline:2005pf} for a perturbative approach to this problem).
The study of exponentially suppressed corrections to the partition
function in 
certain non-compact Calabi-Yau threefolds suggests that the
attractor flow should be ``second quantized'' to allow for this
possibility \cite{Dijkgraaf:2005bp}. Note that the process whereby two
ends of an open string join to form a closed string does not
seem to have a black hole analogue.

\item  Finally, let us mention that further interest for the
quantization of attractor flows stems from the 
relation between black hole attractor equations and the
equations that determine supersymmetric vacua in flux 
compactifications (see e.g. \cite{Kallosh:2005ax} for a recent
discussion). 
Upon double analytic continuation, one may hope to relate 
the black hole wave function to the wave function of the Universe,
and address vacuum selection in the Landscape \cite{Ooguri:2005vr}. 
There are however
many difficulties with this idea that we shall not discuss here. 
At any rate, it will be clear that our discussion of radial quantization
bears many similarities with ``mini-superspace'' approaches to 
quantum cosmology.
\end{itemize}


\subsection{Attractor Flows and Geodesic Motion}
The most convenient route to quantize the attractor flow, or more generally
perform the radial quantization of stationary, spherically symmetric
black holes, is to use the equivalence between the equations governing
the radial evolution of the fields in four dimensions, and the geodesic
motion of a fiducial particle on an appropriate pseudo-Riemannian
manifold \cite{Breitenlohner:1987dg}. This equivalence holds irrespective 
of supersymmetry, so we
consider the general two-derivative action for four-dimensional gravity 
coupled to scalar fields $z^i$ and gauge fields $A_4^I$,
\be
\label{4dact}
S_4 = \frac12 \int \left[ \sqrt{-\gamma} \ 
R[\gamma] \  d^4 x +  g_{ij}\ dz^i \wedge {}\star 
dz^{j} 
-F^I \wedge\Big( t_{IJ} {}\star F^J +\theta_{IJ}\wedge F^J\Big) 
\right] .
\ee
Here,~$\gamma$ denotes the four-dimensional metric,~$g_{ij}$ the
metric on the moduli space~${\cal M}_4$ where the (real) scalars~$z^i$ take
their values,~$F^I=dA_4^I$ 
and the (positive definite) gauge couplings~$t_{IJ}$ and angles 
$\theta_{IJ}$ are in general functions of~$z^i$. In \eqref{4dact}, we
have dropped the contribution of the fermionic fields, but we shall reinstate
them in Section \ref{sec-susygeo} below 
when we return to a supersymmetric setting.
Moreover, since the pseudo-Riemannian manifold already arises under
the sole assumption of stationarity, we begin
by relaxing the assumption of spherical symmetry.

\subsubsection{Stationary solutions and $KK^*$ reduction}
A general ansatz for stationary metrics and gauge fields is
\be
\label{stationan}
\gamma_{\mu\nu} dx^\mu dx^\nu = 
-e^{2U} (dt + \omega)^2  + e^{-2U} \gamma_{\ri\rj} dx^\ri dx^\rj 
\ ,\quad
A_4^I = \zeta^I dt + A^I_3\ .
\ee
where the three-dimensional metric $\gamma_{\ri\rj}$, one-forms 
$A^I_3, \omega$ and scalar $U, \zeta^I,z^i$, are general functions of
the coordinates $x^\ri$ on 
the three-dimensional spatial slice. Since all these fields are independent of
time, one may reduce the four-dimensional action \eqref{4dact}
along the time direction and obtain a field theory in three Euclidean 
dimensions. This process is analogous to the usual Kaluza-Klein reduction, 
except for the time-like signature of the Killing vector~$\pa_t$, which leads
to unusual sign changes in the three-dimensional action. 

Just as in usual Kaluza-Klein reduction, the one-forms 
$A^I_3$ and~$\omega$ can be dualized into axionic scalars~$\tzeta_I, \sigma$, 
using Hodge duality between one-forms and  pseudo-scalars in three 
dimensions. Thus, the four-dimensional theory reduces 
to a gravity-coupled non-linear sigma model
\be
\label{3dsig}
S_3 = \frac12 \int\ \left( \sqrt{g_3}\ R[g_3]\ d^3 x
+ g_{ab} \ d\phi^a \wedge \star d\phi^b \right) 
\ee
whose target manifold~${\cal M}_3^*$ includes the
four-dimensional scalar fields~$z^i$ together with~$U$,  
$\zeta^I$, $\tzeta_I,\sigma$. The metric $g_{ab}$ on ${\cal M}_3^*$ has 
indefinite signature, and can be obtained by analytic 
continuation~$(\zeta^I,\tzeta_I) \to  i(\zeta^I,\tzeta_I)$ 
\cite{Breitenlohner:1987dg,Hull:1998br} from the 
(Riemannian) three-dimensional moduli space~${\cal M}_3$ arising 
in standard, spacelike, 
Kaluza-Klein reduction, (see {\it e.g.}~\cite{Ferrara:1989ik})   
\be
\label{noncmap}
\begin{split}
ds^2_{{\cal M}_3^*} &= 2\,dU^2+ g_{ij}\,dz^i dz^j
+\frac12 e^{-4U} 
\left(d\sigma+\zeta^I d\tilde \zeta_I - \tilde \zeta_I d\zeta^I\right)^2 \\
& - e^{-2U} \left[ t_{IJ}\, d\zeta^I d\zeta^J
+ t^{IJ} \left(d\tilde\zeta_I+ \theta_{IK} d\zeta^K\right)
\left(d\tilde\zeta_J+ \theta_{JL} d\zeta^L\right) \right]
\end{split}
\ee
where $t^{IJ}\equiv [t^{-1}]^{IJ}$.
Importantly,~${\cal M}_3^*$ 
always possesses (at least)~$2n+2$ isometries corresponding to the
gauge symmetries of~$A^I, \tilde A_I,\omega$, as well as rescalings of 
time~$t$. The Killing vector fields generating these isometries read
\bse
\be
\label{killingh}
p^I = \pa_{\tzeta_I} - \zeta^I \pa_\sigma\ ,\quad
q_I = -\pa_{\zeta^I} - \tzeta_I \pa_\sigma\ ,\quad
k = \pa_\sigma\ ,
\ee
\be
M=-\left(\pa_U + \zeta^I\pa_{\zeta^I} + \tzeta^I\pa_{\tzeta^I} + 2 \sigma \pa_\sigma\right)
\ee
\ese
and satisfy the Lie-bracket algebra
\bse
\be
\label{killal}
[ p^I, q_J ] = -2\delta^I_J\ k
\ee
\be
 [M, p^I] = p^I\ ,\quad \ [M,q_I] = q_I\ ,\quad [M,k]=2k
\ee
\ese
In general, stationary solutions in four dimensions are therefore
given by harmonic maps from the three-dimensional slice, with metric 
$\gamma_{\ri\rj}$, to the  three-dimensional moduli space ${\cal M}_3^*$,
such that Einstein's equation in three-dimension is fulfilled,
\be
R_{\ri\rj}[\gamma] = g_{ab}\left( \pa_\ri \phi^a \pa_\rj\,  \phi^b 
-\frac12 \pa_\rk \phi^a \,\pa_\rl \phi^b\,\gamma^{\rk\rl}\, 
\gamma_{\ri\rj} \right) 
\ee
Moreover, the Killing vectors $p^I,q_I,k,M$ give rise to conserved currents,
whose conserved charges will be identified with the overall
electric and magnetic charges, NUT charge and ADM mass of
the configuration.

\subsubsection{Spherical symmetry and geodesic motion}

Now, let us restrict to spherically symmetric, stationary solutions:
the spatial slices can be parameterized as
\be
\label{3dsli}
\gamma_{\ri\rj} dx^\ri dx^\rj = N^2 (\rho)\, d\rho^2 + r^2(\rho)\, 
(d\theta^2 + \sin^2\theta\, d\phi^2)
\ee
while all scalars on ${\cal M}_3^*$ become functions of $\rho$ only.
After dropping a total derivative term,
the three-dimensional sigma-model action reduces to classical
mechanics,
\be
\label{clasme}
S_1 = \int d\rho \left[ \frac{N}{2} + \frac{1}{2N} \left( 
r^{'2} - r^2 \, g_{ab}\,  \phi^{'a}\,  \phi^{'b} \right)  \right]
\ee
where the prime denotes a derivative with respect to $\rho$. This Lagrangian
describes the free motion of a fiducial particle on a cone\footnote{A
similar mechanical  arises in mini-superspace cosmology \cite{Damour:2002et,
Pioline:2002qz}.}
${\cal C}=\IR^+ \times {\cal M}_3^*$ over the three-dimensional moduli
space ${\cal M}_3$.
The lapse $N$ is an auxiliary field; its equation of motion 
enforces the mass shell condition
\be
r^{'2} - r^2 \ g_{ab}\ \phi^{'a} \ \phi^{'b} = N^2
\ee
or equivalently, the Wheeler-De Witt (or Hamiltonian) constraint
\be
\label{hwdw}
H_{\rm WDW} = (p_r)^2 - \frac{1}{r^2} g^{ab} p_a p_b - 1 \equiv 0
\ee
where $p_r,p_a$ are the canonical momenta conjugate to $r,\phi^a$.

Solutions are thus massive
geodesics on the cone, with fixed mass equal to 1. 
In particular, the phase space describing the set of stationary,
spherically symmetric solutions of \eqref{4dact} is the cotangent 
bundle $T^*{\cal C}$ of the cone ${\cal C}$. 

As is most easily seen in the gauge~$N=r^2$, 
the motion separates into geodesic motion on the base of the cone
${\cal M}_3^*$, with 
affine parameter~$\tau$ such that~$d\tau=d\rho/r^2(\rho)$,
and motion along the radial direction~$r$, 
\be
\label{rrhoc}
(p_r)^2 - \frac{C^2}{r^2} - 1 \equiv 0\ ,\quad  
g^{ab} p_a p_b \equiv C^2
\ee
where~$p_r=r'=\dot{r}/r^2$ and~$p_i=r^2 \phi^{'i}= \dot\phi^{i}$;
here the dot denotes a derivative with respect to~$\tau$. 
It is interesting to note that the radial motion is governed by 
the same Hamiltonian as in~\cite{deAlfaro:1976je}, and therefore
exhibits one-dimensional conformal invariance. This is a
consequence of the existence of the homothetic Killing vector $r\pa_r$
on the cone ${\cal C}$.

\subsubsection{Extremality and light-like geodesics}
The motion along~$r$ is easily integrated to 
\be
\label{eqrrho}
r=\frac{C}{\sinh(C\tau)}\ ,\quad \rho=\frac{C}{\tanh C\tau}
\ee
Assuming that the sphere $S^2$ reaches a finite area $A$ at the
horizon $\tau=\infty$, so that $e^{-2U} r^2 \to A/(4\pi)$, one may
rewrite the metric \eqref{stationan} as \cite{Kallosh:2006bt}
\be
ds^2 \sim \frac{C^2}{\sinh^2(C\tau)}
\left( - \frac{4\pi}{A} (dt+\omega)^2 + \frac{A}{4\pi} d\tau^2 \right)
+ \frac{A}{4\pi} d^2 \Omega 
\ee
The horizon at $\tau=\infty$ is degenerate for $C^2=0$, and non-degenerate
for $C^2>0$, corresponding to extremal and non-extremal black holes,
respectively. We conclude that extremal black holes correspond
to {\it light-like} geodesics on ${\cal M}_3^*$ (it is indeed fortunate 
that ${\cal M}_3^*$ is a pseudo-Riemannian manifold, so that
light-like geodesics do exist). 

\begin{exo}
Show that the extremality parameter $C$ is related to the 
Bekenstein--Haw-king entropy and Hawking temperature 
by $C=2 S_{BH}T_{H}$.
\end{exo}

Setting $C=0$ in \eqref{rrhoc}, we moreover see that $r=\rho=1/\tau$,
and therefore that the spatial slices in the ansatz \eqref{stationan} 
are flat. We could therefore have set $N=1, r=1/\tau$ from the start,
and obtained the action for geodesic motion on ${\cal M}_3$ in
affine parameterization,
\be
S_1' = \int d\tau ~\frac12~g_{ab}~\dot{\phi}^a~\dot{\phi^b}
\ee 
While one may dispose of the radial variable~$r$ altogether, it is
however advantageous to retain it for the purpose of defining observables 
such as the horizon area,~$A_H = 4\pi e^{-2U} r^2\vert_{U\to -\infty}$
and the ADM mass~$M= r (e^{2U}-1)\vert_{U\to 0}$.

\subsubsection{Conserved charges and black hole potential \label{bhpot}}

As anticipated by the notation in~\eqref{killal}, 
the isometries of~${\cal M}_3$ imply conserved 
Noether charges,
\bea
\label{consca}
q_I \ d\tau&=&-2 e^{-2U} \left[t_{IJ} d\zeta^J +\theta_{IJ} t^{ JL}
\left( d\tilde{\zeta}_L +\theta_{LM} d\zeta^M \right) \right] 
+ 2\,k \tilde \zeta_I \nn\\
p^I \ d\tau&=&-2 e^{-2U} t^{ IL}
\left(d\tilde{\zeta}_L +\theta_{LM} d\zeta^M \right)
-2\,k \zeta^I \label{conscha}\\
k \ d\tau&=& e^{-4U} \left(d\sigma+\zeta^I d\tilde \zeta_I - \tilde \zeta^I
    d\zeta_I\right)\nn
\eea
(as well as $M$, whose precise form we will not need)
identified as the electric, magnetic and NUT charges
$p^I,q_I,k$. Their algebra under Poisson bracket is
the same as algebra of the Killing vectors under Lie bracket, 
\be
\label{heis}
\{ p^I, q_J \}_{\rm PB} =- 2 \delta^I_J k 
\ ,\quad
\{M, p^I\}_{\rm PB} = p^I\ ,\quad \ \{M,q_I\}_{\rm PB} = q_I\ ,\quad 
\{M,k\}_{\rm PB}=2k
\ee
In particular, the electric and magnetic charges satisfy 
an Heisenberg algebra, the center of which is the NUT charge~$k$.
The latter is related to the 
off-diagonal term in the metric~\eqref{stationan} via 
$\omega=k\cos\theta\ d\phi$. When~$k\neq 0$, the metric 
\be
ds_4^2 = -e^{2U} (dt + k \cos\theta\, d\phi)^2  + e^{-2U}
\left( d\rho^2 + r^2(\rho) [d\theta^2 + \sin^2\theta\, d\phi^2] \right)
\ee
has closed timelike curves along the compact $\phi$ coordinates 
near~$\theta=0$, all the way from 
infinity to the horizon. Bona fide 4D black holes have~$k=0$, which
corresponds to a ``classical'' limit of the Heisenberg algebra~\eqref{heis}.

\begin{figure}
\centerline{\hfill\includegraphics[height=9cm]{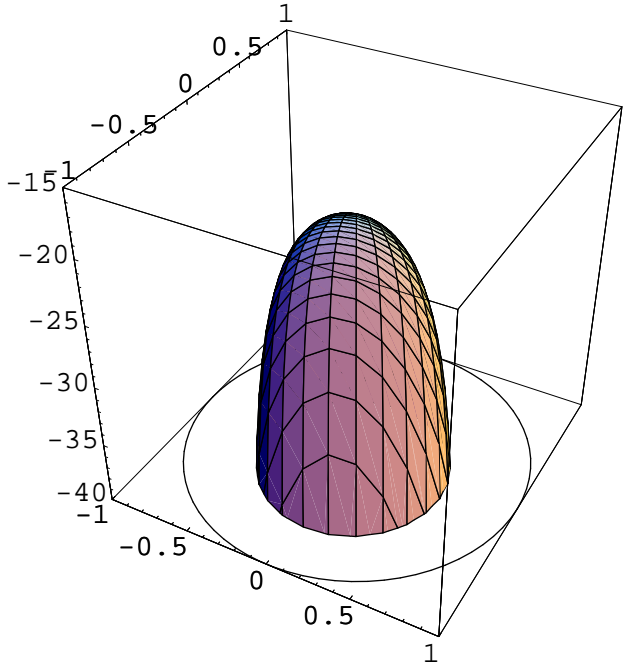}\hfill}
\caption{Potential governing the radial evolution of the 
complex scalar in the same model as in Figure \ref{su22fig0}
and same charges, at $U=0$. The potential has a global maximum at
$z_*=X^1/X^0=(1-3i)/10$.
\label{su22fig}}
\end{figure}

Using the conserved charges~\eqref{consca},
one may express the Hamiltonian 
for affinely parameterized geodesic motion on~${\cal M}_3^*$ as
\be
\label{hampqk}
H \equiv p^a g_{ab} p^b = \frac12 \left[ p_U^2 + \frac14 p_{z^i} g^{ij} 
p_{z^j} - e^{2U} V_{BH} + k^2 e^{4U} \right]
\ee
where~$p_U,p_{z^i}$ are the momenta canonically conjugate to~$U,z^i$,
\be
V_{BH}(p,q,z) = -\frac12 (\hat q_I - \theta_{IJ} \hat p^J)
t^{IK} (\hat q_K - \theta_{KL} \hat p^L)
-\frac12 \hat p^I t_{IJ} \hat p^J
\ee
and
\be
\label{hatpq}
\hat p^I = p^I + 2k \zeta^I\ ,\quad \hat q_I = q_I - 2k \zeta^I\ ,\quad 
\ee
For~$k=0$, the motion along 
$\zeta^I,\tzeta_I,\sigma$ separates from that along~$U,z^i$, effectively
producing a potential for these variables.
Following~\cite{Ferrara:1997tw}, we refer to~$V_{BH}$ 
as the ``black hole potential'', but it should be kept in mind that
it contributes negatively to the actual 
potential $V=-e^{2U}V_{BH}+k^2 e^{4U}$ governing the Hamiltonian motion.
In Figure \ref{su22fig}, we plot the potential $V$ for $\cN=2$
supergravity with one minimally coupled vector multiplet.


\subsubsection{The Universal Sector\label{sec-uni}}
\begin{figure}
\centerline{
\hfill\includegraphics[height=5cm]{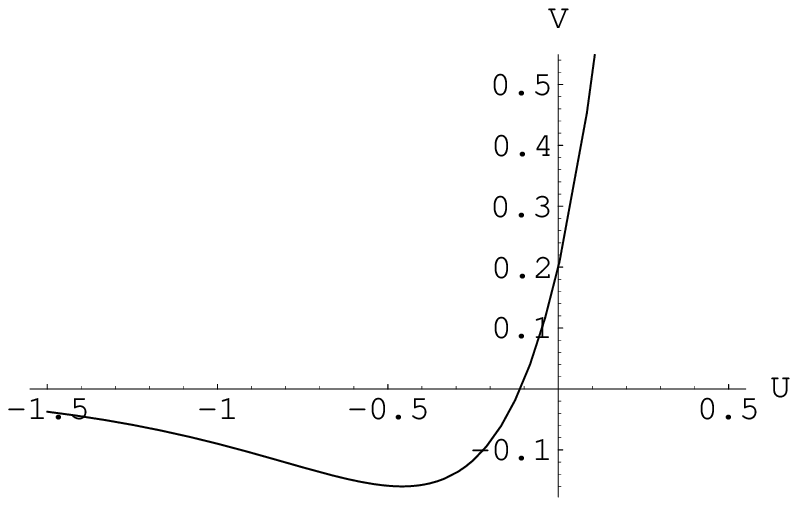}
\hfill\includegraphics[height=5cm]{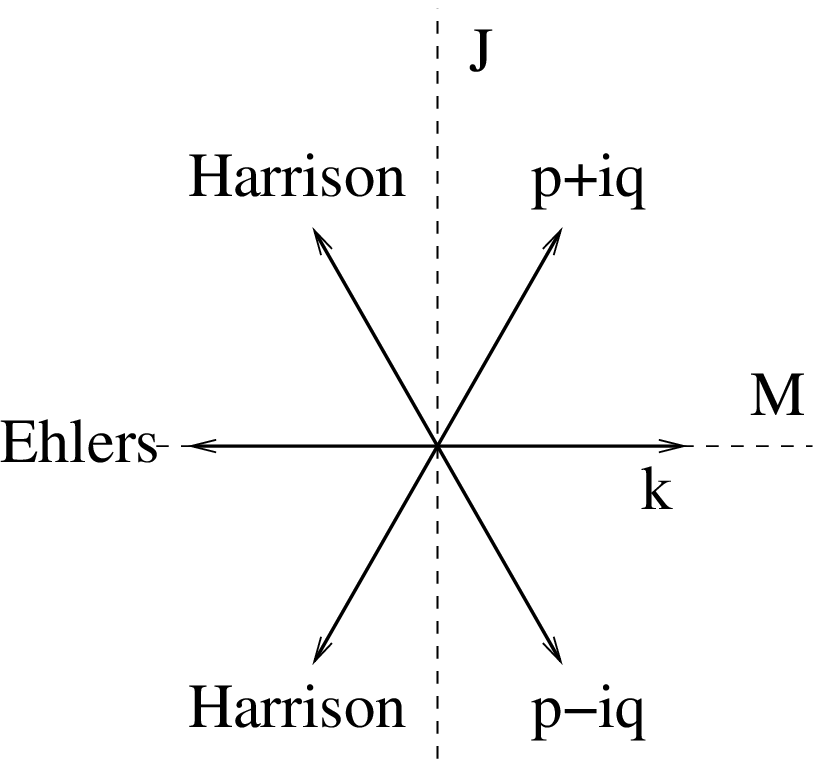}
\hfill}
\caption{{\it Left:} Potential governing the motion along
the $U$ variable in the universal sector. The horizon
is reached at  $U\to-\infty$. {\it Right}: Root
diagram of the $SU(2,1)$ symmetries in the universal sector.\label{su21pic}}
\end{figure}

As an illustration, and a useful warm-up for the symmetric case discussed in
Section \ref{quasiref} below, it is instructive to work out the dynamics 
in the ``universal sector'', which
encodes the scale $U$, the graviphoton electric and magnetic charges, and
the NUT charge $k$. The resulting pseudo-quaternionic-K\"ahler manifold is 
the symmetric space ${\cal M}_3^*=SU(2,1)/Sl(2)\times U(1)$, 
an analytic continuation of the quaternionic-K\"ahler space
${\cal M}_3=SU(2,1)/SU(2)\times U(1)$,
which describes the tree-level couplings of the universal 
hypermultiplet in 4 dimensions. It is obtained via c-map 
from a trivial moduli space $\cM_4$ corresponding to the prepotential $F=
-i (X^0)^2/2$. The Hamiltonian \eqref{hampqk} becomes 
\be
\label{hamuni}
H =  \frac18 (p_U)^2 - \frac14 e^{2U} 
\left[ ( p_{\tilde\zeta} - k \zeta )^2 + (p_\zeta + k \tilde\zeta)^2 \right]
+ \frac12 e^{4U} k^2
\ee
The motion separates 
between the $(\tzeta,\zeta)$ plane and the $U$ direction, while the
NUT potential $\sigma$ can be eliminated in favor of its conjugate momentum
$k=e^{-4U}(\dot \sigma + \zeta \dot\tzeta - \tzeta \dot\zeta)$.
The motion in the $(\tzeta,\zeta)$ plane is that of a 
charged particle in a constant magnetic field. The electric, magnetic charges
and the angular momentum $J$ in the plane (not to be confused with
that of the black hole, which vanishes by spherical symmetry)
\be
p = p_{\tilde \zeta} + \zeta k \ ,\quad q = p_{\zeta} - \tilde\zeta k
\ ,\quad J = \zeta p_{\tzeta} - \tzeta p_\zeta
\ee
satisfy the usual algebra of the Landau problem,
\be
\{p,q\}_{\rm PB} = 2k\ \quad, 
\{[J,p\}_{\rm PB} =q\ ,\quad \{J,q\}_{\rm PB} =-p
\ee
where $p$ and $q$ are the ``magnetic translations''.
The motion in the $U$ direction is governed effectively by
\be
H =  \frac18 (p_U)^2 + \frac12 e^{4U} k^2
- \frac14 e^{2U} \left[ p^2 + q^2 -4 k J\right] = C^2
\ee
The potential is depicted on Figure \ref{su21pic} (left).
At spatial infinity ($\tau=0$), one may impose the initial conditions
$U=\zeta=\tzeta=a=0$. The momentum $p_U$ at infinity is proportional
to the ADM mass, and $J$ vanishes, so the mass shell condition 
\eqref{hamuni} becomes
\be
\label{bpsuh}
M^2 + 2k^2 - (p^2 + q^2) = C^2
\ee
Extremal black holes correspond to $C^2=0$; in this low
dimensional example are automatically BPS, as we shall see
in the next Section.
Equation \eqref{bpsuh} is then the  BPS mass condition, generalized
to non-zero NUT charge. Note that for a given value of $p,q$,
there is a maximal value of $k$ such that $M^2$ remains positive.

At the horizon $U\to -\infty$, $\tau\to\infty$, the last term 
in \eqref{hamuni} is irrelevant, and one may integrate the
equation of motion of $U$, and verify that the metric \eqref{stationan}
becomes $AdS_2\times S^2$ with area
\be
A =  2\pi (p^2 + q^2) = 2\pi \sqrt{ (p^2 + q^2)^2}
\ee
in agreement with the Bekenstein-Hawking entropy of Reissner-Nordstr\"om
black holes \eqref{bhbps}.

Since the universal sector is a symmetric space, there must exist
3 additional conserved charges, so that the total set of conserved
charges can be arranged in an element $Q$ in the Lie algebra 
${\mathfrak g}_3=su(2,1)$ (or rather, in its dual ${\mathfrak g}_3^*$),
\be
Q = \begin{pmatrix}
M + i J/3 & {E_p} - i {E_q}, & i {E_k} \\ 
{E_{p'}} + i {E_{q'}} & -2i J/3 & -({E_p} + i {E_q})\\
-i {E_{k'}} & -({E_{p'}} - i {E_{q'}}) & -M + i J/3
\end{pmatrix}
\ee
where $M,E_p\equiv p^0,E_q\equiv q_0,E_k\equiv k$ 
have been given in \eqref{killingh} 
and $J= \zeta\pa_{\tzeta} - \tzeta\pa_{\zeta}$.
The remaining Killing vectors can be easily found \cite{gnpw-in-progress},
\begin{eqnarray*}
E_{p'} &=& -\tzeta \pa_U -(\sigma+2\zeta\tzeta)\pa_{\zeta}
+ \left[ e^{2U} + \frac12 ( 3 \zeta^2 -\tzeta^2) \right]\pa_{\tzeta}
+ \left[ \zeta \left( e^{2U}+ \frac12(\zeta^2 + \tzeta^2)\right) 
- \sigma\tzeta\right]
\pa_{\sigma} \\
E_{q'} &=& \zeta \pa_U
- \left[ e^{2U} + \frac12 ( 3 \tzeta^2 -\zeta^2) \right]\pa_{\zeta}
-(\sigma-2\zeta\tzeta)\pa_{\tzeta}
+ \left[ \tzeta \left( e^{2U}+ \frac12(\zeta^2 + \tzeta^2)\right) 
+ \sigma\zeta\right]
\pa_{\sigma} \\
E_{k'} &=&  - \sigma \pa_U 
+ \left[ \left( e^{2U} + \frac12 (\zeta^2+\tzeta^2) \right)^2 
- \sigma^2 \right]\pa_{\sigma} \\
&& - \left[ \tzeta \left(e^{2U} 
+ \frac12 (\zeta^2+\tzeta^2)\right)+ \sigma\zeta \right] \pa_{\zeta}
+ \left[   \zeta  \left( e^{2U}+ \frac12(\zeta^2 + \tzeta^2)
\right)- \sigma \tzeta \right]
\pa_{\tzeta}
\end{eqnarray*}
The physical origin of these extra symmetries are the Ehlers and Harrison 
transformations, well known to general relativists \cite{kinnersley}. 
It is easy to check that these Killing vectors satisfy
the Lie algebra of $SU(2,1)$, whose root diagram is depicted on 
Figure \ref{su21pic}. The Casimir invariants of $Q$ can be easily
computed:
\be
\Tr (Q^2)= H \ ,\quad \det(Q)=0
\ee
The last condition ensures that the conserved quantities do not 
overdetermine the motion. 
The co-adjoint action $Q\to h Q h^{-1}$ of $G_3$ on ${\mathfrak g}_3^*$
relates different trajectories with the same value of $H$.
The phase space, at fixed value of $H$, is therefore a 
generic co-adjoint orbit of $G_3$, of dimension 6 (the
symplectic quotient of the full 8-dimensional phase space
by the Hamiltonian $H$).
By the Kirillov-Kostant construction, it carries a
canonical symplectic form such that the Noether charges
represent the Lie algebra ${\mathfrak g}_3$. 

As we have just seen, extremal solutions have $H=0$. The standard property
of $3\times 3$ matrices
\be
Q^3 - \Tr(Q) Q^2 + \frac12[\Tr(Q^2)-(\Tr Q)^2] Q - \det(Q) = 0
\ee
then implies that $Q^3=0$, as a matrix equation in the fundamental 
representation; more intrinsically, in terms of the adjoint representation,
this is equivalent to
\be
\label{adq5}
[Ad(Q)]^5=0
\ee
Thus, $Q$ is a nilpotent element of order 5 in ${\mathfrak g}_3^*$.
This condition is invariant under the co-adjoint action of $G_3$.
We conclude that the classical phase space of
extremal configurations is a nilpotent coadjoint orbit\footnote{It is a 
peculiarity of this model that the dimension of this
nilpotent orbit is the same -- 6 -- as that of the generic 
semi-simple orbits. In general, nilpotent orbits can be much
smaller than the generic ones.} of $G_3$. By the general
``orbit philosophy'' \cite{MR1701415}, the quantum Hilbert space 
then furnishes a ``unipotent'' representation 
of $G_3$, obtained by quantizing this
nilpotent co-adjoint orbit. As we shall see in Section \ref{quasiref}, 
this fact extends to the BPS Hilbert space in very special supergravities,
where ${\cal M}_3^*$ is a symmetric space.

\subsection{BPS black holes and BPS geodesics\label{sec-susygeo}}
Up till now, our discussion did not assume any supersymmetry. In general
however, the KK$^*$ reduction of the fermions gives extra
fermionic contributions in \eqref{3dsig}, such that the resulting 
non-linear sigma model has the same amount of supersymmetry as its
four-dimensional parent. Moreover, the spherical reduction 
of the fermions preserves half of 
the supersymmetries. This leads to the action for 
a supersymmetric spinning particle
moving on $\cC$, schematically
\be
\label{s1ferm}
S_1 = \int d\tau ~ \left[ g_{ab} {\dot \phi}^a {\dot \phi}^b
+ g_{ab}\ \psi^a D_\tau \psi^b  
+ R_{abcd} \psi^a \psi^b \psi^c \psi^d \right]
\ee
This Lagrangian is supersymmetric for any  target space, but has 
$N$-fold extended supersymmetry when $\cC$ admits
$N-1$ complex structures $J^{(i)}$ ($i=1,\dots, N-1$).
The supersymmetry variations of the fermions are then of the form
\be
\label{dpsi}
\delta_{\eps} \psi^a = \sum_{i=0}^{N-1} \eps^{(i)} 
J^{(i)a}_{b}  \dot \phi^b + O(\psi^2)
\ee
with $J^{(0)a}_b=\delta^a_b$ the identity operator. Moreover, the
existence of a homothetic Killing vector $r\pa_r$ implies that the
action $S_1$ should be superconformally invariant.

BPS solutions in four dimensions correspond to special trajectories on 
${\cal M}_3^*$, for which there exist a non-zero $\epsilon^{(i)}$ such
that the right-hand side of \eqref{dpsi} vanishes. This puts a strong
constraint on the momentum $p_a=g_{ab}\dot\phi^b$ of the fiducial particle,
which defines a ``BPS'' subspace of the phase space $T^*(\cC)$. The symplectic
structure on this BPS phase space can then be obtained using Dirac's
theory of Hamiltonian constraints. Due to the existential quantifier
$\exists \eps^{(i)}\neq 0$, it is sometimes convenient to extend the phase
space by including the Killing spinor $\eps^{(i)}$, we shall see an example
of this in Section \ref{twisec}. In theories with 
$N\geq 2$ supersymmetry in 4 dimensions, black holes may preserve different 
fractions of supersymmetry, associated to different orbits of the 
momentum $p$ under the holonomy group of $\cC$. Correspondingly there will 
be different BPS phase spaces, nested into each other. 

\subsubsection{Attractor Flow and BPS Geodesic Flow in~$\CN=2$ 
SUGRA\label{cmap}}
After this deliberately schematic discussion, we now 
specialize to $\CN=2$ 
supergravity, and show that the attractor flow \eqref{att1},\eqref{att2} 
is indeed
equivalent to BPS geodesic flow on the three-dimensional
moduli space ${\cal M}_3^*$.

As explained in Section \ref{spegeo}, $\CN=2$ supersymmetry determines the 
metric on $\cM_4$ (now denoted $g_{i\bar \jmath}$, to take into
account the complex
nature of the vector multiplet moduli) and gauge couplings 
$\theta_{IJ} - i t_{IJ} \equiv {\cal N}_{IJ}$ in terms
of a prepotential $F(X)$ via \eqref{kahl},\eqref{nijtau}.
The scalar manifold~${\cal M}_3$ obtained by Kaluza-Klein reduction
to three dimensions is now a quaternionic-K\"ahler space, usually
referred to as the ``c-map'' of the special K\"ahler manifold~${\cal M}_4$  
\cite{Cecotti:1988qn,Ferrara:1989ik}.
The analytically
continued~${\cal M}_3^*$, with metric \eqref{noncmap} is a 
pseudo-quaternionic-K\"ahler space, which we shall refer to 
as the  ``c$^*$-map'' of~${\cal M}_4$.
While~${\cal M}_3$ has a Riemannian metric with 
special holonomy~$USp(2)\times USp(2n_V+2)$, ${\cal M}_3^*$
has a split signature metric with special holonomy~$Sp(2)\times Sp(2n_V+2)$.
For convenience, we will work with  the Riemannian space~${\cal M}_3$
and perform the analytic continuation at the end.

In order to determine the couplings of the corresponding fermions, 
one should in principle reduce the four-dimensional fermions along
the time direction, then further on the spherically symmetric 
ansatz~\eqref{3dsli}. For our present purposes however, it is sufficient to
recall that the quaternionic-K\"ahler space ${\cal M}_3$ equivalently
arises as the target space of a $N=2$ supersymmetric sigma model in 
3+1 dimensions, coupled to gravity~\cite{Bagger:1983tt}. 
Upon reducing the action and supersymmetry transformations 
of~\cite{Bagger:1983tt} along three flat spatial 
directions, one obtains a $N=4$ supersymmetric sigma model in 
0+1 dimensions, which must be identical to the result of the
spherical reduction. The supersymmetry variations are then simply 
\be
\label{dpsiq}
\delta_\eps \phi^a = O(\psi)\ ,\quad
\delta_\eps \psi^{AA'} = V^{AB'}_a \dot \phi^a \epsilon^{A'}_{B'} 
+ O(\psi^2)
\ee
Here,~$V^{AA'}$ ($A=1,..,2n_V+2$ and $A'=1,2$) is the
``quaternionic viel-bein'' afforded by the decomposition
\be
T_\IC {\cal M}_3 = E \otimes H
\ee
of the complexified tangent bundle of ~${\cal M}_3$, where $E$ 
and $H$ are complex vector bundles of respective dimensions $2n_V+2$ and $2$.
Similarly, the Levi-Civita connection decomposes into
its $USp(2)$ and~$USp(2n_V+2)$ parts $p$ and~$q$,
\be
\label{omab}
\Omega_{AA'}^{BB'} =  p^{A'}_{B'} \eps_{A}^{B} + q^A_B \eps_{A'}^{B'}
\ee
where~$\epsilon_{A'B'}$ and~$\eps_{AB}$ are
the antisymmetric tensors invariant under~$USp(2)$ and~$USp(2n)$.
The viel-bein $V$ controls both the
metric and the three almost complex structures on the quaternionic-K\"ahler
space,
\be
ds^2 =\epsilon_{A'B'}\ \eps_{AB}\ 
 V^{AA'}   \otimes V^{BB'} \ ,\quad 
\Omega^i = \epsilon_{A'B'}\ (\sigma^i)^{B'}_{C'} 
\ \eps_{AB}\ 
 V^{AA'}   \wedge V^{BC'}\label{metric}
\ee
(where $\sigma^i$, $i=1,2,3$ are the Pauli matrices)
and is covariantly constant with respect to the connection \eqref{omab}.

From~\eqref{dpsiq}, it is apparent that supersymmetric solutions
are obtained when~$V^{AA'}$ has a zero right-eigenvector,
\bse
\label{susycond}
\bea
\label{susycond1}
\mbox{SUSY}\quad &\Leftrightarrow& \quad
\exists \epsilon_{A'}\neq 0\ \slash \ 
V^{AA'} \eps_{A'} = 0\\
&\Leftrightarrow& \quad 
\forall\, A,B\ ,\quad 
\eps_{A'B'}\,V^{AA'}\, V^{BB'} = 0
\label{susycond2}
\eea
\ese
For fixed~$\epsilon^{A'}$, these are~$2n_V+2$ conditions on 
the velocity vector~$\dot\phi^a$ at any point along the geodesic,
removing half of the degrees of freedom from the 
generic trajectories. In particular, the conditions \eqref{susycond2}
imply that  
\be
\eps_{AB}\eps_{A'B'}V^{AA'}V^{BB'}=0 = H\ ,
\ee
and therefore that a BPS solution is automatically extremal.
For  the universal sector discussed in Section \ref{sec-uni}, where $n_V=0$, 
this is actually a necessary and sufficient condition for supersymmetry.

For the case of the~$c$-map~${\cal M}_3$, the quaternionic viel-bein 
was computed explicitly in~\cite{Ferrara:1989ik}. After analytic
continuation, one obtains
\be
V^{AA'} = \begin{pmatrix} i u & v \\ e^a & i E^a  \\
-i \bar E^{\bar a} & \bar e^{\bar a} \\ -\bar v & i \bar u\end{pmatrix} 
\ee
where~$e^a=e^a_i dz^i$ is a viel-bein of the special K\"ahler manifold, 
$e^a_i \bar e_{\bar a\overline \jmath} \delta_{a\bar a}=g_{i\overline \jmath}$, and
\bea
u&=& e^{\cK/2-U} X^I \left( d\tilde\zeta_I 
+{\cal N}_{IJ}d\zeta^J \right) \\
v&=&  -dU+ \frac{i}{2} e^{-2U}\left( 
  d\sigma+\zeta^Id\tilde\zeta_I-\tilde\zeta^Id \zeta_I\right) \\
E^a&=& e^{-U} e^{a}_{i} g^{i\overline \jmath} \bar f_{ \overline \jmath}^I 
\left( d\tilde\zeta_I + {\cal N}_{IJ}d\zeta^J \right)\label{vb}
\eea
Expressing~$d\zeta^I, d\tilde\zeta_I, d\sigma$ in terms of the conserved
charges~\eqref{conscha}, the entries in the quaternionic viel-bein may 
be rewritten as
\bea
u&=& -\frac{i}{2} e^{\cK/2 + U} X^I \left[ q_I - 2 k \tilde\zeta_I 
-{\cal N}_{IJ} (p^J + 2 k \zeta^J)  \right] d\tau \ ,\quad\\
v&=&  -dU+ \frac{i}{2} e^{2U} k \ d\tau\\
e^a &=&  e^a_i\ d z^i\ ,\quad\\
E^a&=& -\frac{i}{2} e^{U} e^{a i} g^{i\overline \jmath} \bar f_{ \overline \jmath}^I 
\left[ q_I - 2 k \tilde\zeta_I -  {\cal N}_{IJ} (p^J + 2 k \zeta^J) \right] 
d\tau
\eea
Now, return to the supersymmetry variation of the fermions~\eqref{dpsiq}:
the existence of~$\eps_{A'}^{B'}$ such that 
$\delta\psi^{AA'}$ vanishes implies that the first column of~$V$ has to
be proportional to the second, hence
\bea
-\frac{dU}{d\tau}+ \frac{i}{2} e^{2U} k &=& 
-\frac{i}{2}\ e^{i\theta}\ e^{\cK/2+U}\ X^I \left( q_I - k \tilde\zeta_I 
-{\cal N}_{IJ} (p^J + k \zeta^J)  \right) \\
\frac{d z^i}{d\tau} &=& -\frac{i}{2}\ e^{i\theta} \ e^{U}\ g^{i\overline \jmath} \bar f_{ \overline \jmath}^I 
\left( q_I - k \tilde\zeta_I -  {\cal N}_{IJ} (p^J + k \zeta^J) \right)
\eea
where the phase~$\theta$ is determined by requiring that~$U$
stays real. These equations may be rewritten as 
\be
-\frac{dU}{d\tau} + \frac{i}{2} e^{2U} k
 = -\frac{i}{2} e^{i\theta} e^{U} Z \ ,\quad
\frac{d z^i}{d\tau} = -i  e^{i\theta} \frac{|Z|}{Z} e^{U} g^{i\overline \jmath} 
\partial_{\overline \jmath} |Z| 
\ee
where~$\hat Z$ is the ``generalized central charge''
\be
\hat Z(p,q,k)=e^{\cK/2} \left[ \hat q_I X^I - \hat p^I F_I \right] 
\ee
and $\hat p^I, \hat q_I$ have been defined in \eqref{hatpq}.
For vanishing NUT charge, we recognize the attractor flow
equations \eqref{att1},\eqref{att2}.
The equivalence between 
the attractor flow equations on~${\cal M}_4$
and supersymmetric geodesic motion on~${\cal M}_3$ 
was in fact observed long ago in~\cite{Gutperle:2000ve}, and is
a consequence of T-duality between black holes and 
instantons, after compactifying to three 
dimensions~\cite{Behrndt:1997ch,deVroome:2006xu}.

This concludes our proof that BPS geodesics, characterized by the BPS 
constraints \eqref{susycond}, indeed describe stationary, spherically 
symmetric BPS black holes. 

\subsubsection{Swann space and twistor space\label{twisec}}
While the analysis in the previous section identified the BPS subspace
of the phase space $T^*\cM_3^*$ (namely, the solution to the quadratic 
constraints \eqref{susycond2}), the non-linearity of the BPS constraints
makes it difficult to obtain its precise symplectic structure. 
We now show that, by lifting the geodesic motion on the \qk $\cM_3^*$ to 
a higher-dimensional space, namely the Swann space $\cS$, it is possible 
to linearize these constraints. 

The Swann space is a standard construction, which relates
\qk geometry in dimension $4n_V+4$ to \hk geometry in 
$4n_V+8$ dimensions \cite{MR1096180}. Namely, 
let $\pi^{A'}$ ($A'=1,2$) be complex
coordinates in the vector bundle $H$ over ${\cal M}_3$, and 
$\cS$ be the total space of this bundle. $\cS$ admits a \hk metric
\be
\label{dssw}
ds^2_{\cS} =|D\pi|^2 + R^2\, ds^2_{{\cal M}_3}\, .
\ee
where 
\be
\label{DPi}
D\pi^{A'} = d\pi^{A'} + p^{A'}_{B'} \pi^{B'}\ ,\quad
R^2 \equiv |\pi|^2 =  |\pi^1|^2 + |\pi^2|^2
\ee
In fact, $R^2$ is the \hk potential of \eqref{dssw}, i.e. a K\"ahler potential
for all complex structures.
Being hyperk\"ahler, $\cS$ has holonomy $USp(2n_V+4)$; the
corresponding covariantly constant vielbein ${\cal V}^{\aleph}$ 
(where $\aleph \in\{A,A'\}$ runs over two more indices than $A$)
can be simply obtained from the quaternionic vielbein $V^{AA'}$ on 
the base $\cM_3$ via
\be
\label{vsw}
{\cal V}^{A} = V^{AA'} \pi_{A'}\ ,\quad {\cal V}^{A'} = D\pi^{A'}
\ee
The viel-bein ${\cal V}^{\aleph}$ gives a set of $(1,0)$-forms on $\cS$
(for a particular complex structure), which together with 
$\bar{\cal V}$, span the cotangent space of $\cS$.
The fermionic variations in the corresponding sigma model split into
\be
\label{dpsisw}
\delta_\eps \psi^{\aleph} = \cV^\aleph \epsilon + \dots \ ,\quad 
\delta_{\bar \eps} \bar\psi^{\bar \aleph} = 
\bar\cV^\aleph \bar \epsilon + \dots
\ee
Moreover, the metric \eqref{dssw} has a manifest $SU(2)$ isometry, 
and homothetic Killing vector $R\pa_R=\pi^{A'}\pa_{\pi^{A'}}+\bar\pi^{A'}
\pa_{\bar \pi^{A'}}$. Geodesic motion on $\cS$ is therefore 
equivalent to geodesic motion on the base $\cM_3$, provided one
restricts to trajectories with zero angular momentum under the
$SU(2)$ action (and disregard the motion along the radial direction
$R^2=|\pi|^2)$. By suitable $SU(2)$ rotation, BPS geodesics
on $\cS$ can be chosen to be annihilated by $\delta_\eps$, and so
correspond to 
\be
\label{susycondsw}
\forall\, \aleph\ ,\qquad \cV^{\aleph} = 0
\ee
Using \eqref{vsw}, this entails
\be
V^{AA'} \pi_{A'} = 0 \ ,\quad D\pi^{A'} = 0
\ee
The first condition reproduces the BPS condition \eqref{susycond1} on $\cM_3$
upon identifying\footnote{In particular, the 
radius $R$ of the Swann space $\cS$ is equal to the norm of the Killing 
spinor, and must be carefully distinguished from the radius $r$ of the 
cone $\cS$.} $\pi^{A'}$ with the Killing spinor $\epsilon_{A'}$, 
while the second can
be shown to follow from the Killing spinor conditions in four dimensions,
consistently with this identification. The condition \eqref{susycondsw}
shows that BPS trajectories are such that the momentum vector is 
anti-holomorphic at every point. These BPS constraints are clearly
first class, and therefore the extended BPS phase space is the 
Swann space $\cS$ itself, equipped with its K\"ahler form.

While the Swann space has a clear physical motivation, the fiber being
identified with the Killing spinor, the fact that one must 
restrict to $\IR^\times \times SU(2)$ invariant trajectories 
means that it is 
somewhat too large. In fact, one may perform a symplectic
reduction -- more precisely, a K\"ahler quotient -- with respect 
to $U(1)\subset SU(2)$ while keeping most
of the pleasant properties of the Swann space.  The
result, known as the twistor space $\cZ$, 
retains one of the three complex structures of $\cS$, which is
sufficient for exposing half of the $\cN=4$ supersymmetries of \eqref{s1ferm}.
To exhibit the structure of $\cZ$, it is useful to choose the following
coordinates on the unit sphere in $\IR^4$, 
\be
e^{i\varphi} =\sqrt{\pi^2/\bar \pi^2}\ ,\quad
z={\pi^1}/{\pi^2}\, .
\ee
where $\varphi$ is the angular coordinate for the Hopf fibration
$U(1)\to S^3\to S^2$ and $z$ is a stereographic coordinate on
$S^2=\IC\IP^1$. In these coordinates, the metric \eqref{dssw} 
rewrites as 
\be
\label{dsswsph}
ds^2_{\cS} = d R^2 + R^2 \left[ 
D\phi^2 + \frac{Dz D\bar z}{(1+z\bar z)^2} + 
ds^2_{\cM_3} \right]
\ee
where
\bea
D z &\equiv& dz -\frac12(p_1+i p_2) -2p_3 z 
-\frac12 (p_1-i p_2) z^2\ ,\\
D \phi &\equiv& 
d\phi+ \frac{i}{2(1+z\bar z)} \left( z[d\bar z-(p_1+i p_2)]-\bar z 
[dz-(p_1-i p_2)]
-2i p_3 (1-\bar z z) \right) \nn
\eea
and $p_i = \sigma_{(i)}^{A'B'} p_{(A'B')}$. The connection term
in $Dz$ is sometimes known as the projectivized $USp(2)$ connection.
The twistor space $\cZ$ is the K\"ahler quotient
of $\cS$ by $U(1)$ rotations along $\phi$ \cite{MR664330}; its metric is
therefore given by the last two terms in \eqref{dsswsph}
\be
\label{dsz}
ds^2_{\cZ}=\frac{|Dz|^2}{(1+\bar z z)^2}
+ ds^2_{\cM_3}\, .
\ee
The space $\cZ$ is itself an $S^2$ bundle over ${\cal M}_3$
and carries a canonical complex structure, which is an integrable 
linear combination of the triplet of almost complex structures on $\cM_3$. 
It will also be important that $\cZ$ carries a holomorphic contact
structure $X$ (proportional to the one-form $Dz$), 
inherited from the holomorphic symplectic structure on the
hyperk\"ahler cone $\cS$. 

For later purposes, it will be useful to have an explicit set of
$2n_V+3$ complex coordinates $(\xi^I,\txi_I,\alpha)$ 
on the twistor space $\cZ$, adapted to the Heisenberg symmetries,
i.e. such that the Killing vectors $p^I, q_I, k$ in \eqref{killingh}
take the standard form
\be
\label{cxkillingh}
p^I= \pa_{\txi_I} - \xi^I \pa_\alpha + \cc \ ,\quad
q_I = -\pa_{\xi^I} - \txi_I \pa_\alpha + \cc \ ,\quad
k = \pa_\alpha + \cc
\ee
while the holomorphic contact structure takes the canonical, Darboux
form,
\be
\label{contactf}
X = d\alpha + \txi_I d\xi^I - \xi_I d\txi^I
\ee
Such a coordinate system has been 
constructed recently in \cite{Neitzke:2007ke},
from which we collect the relevant formulae. 
The  complex coordinates $(\xi^I,\txi_I,\alpha)$  are related 
to the coordinates  $U,z^i,\bar z^{\bar i},\zeta^I,\tzeta_I,
\sigma$ on the \qk base, as well as the fiber coordinate $z\in\CP_1$,
via the ``twistor map'' 
\bse
\label{gentwi}
\bea
\xi^I &=& \zeta^I + 2i\ e^{U+{\cal K}(X,\bar X)/2} 
\left( z \bar X^{I} + z^{-1} X^{I} \right) \label{gentwi1}\\
\txi_I &=& \tzeta_I + 2i\ e^{U+{\cal K}(X,\bar X)/2}  
\left( z\ \bar F_I +  z^{-1}\ F_I \right) \label{gentwi2}\\
\alpha &=& \sigma + \zeta^I \txi_I -\tzeta_I  \xi^I
\label{gentwi3}
\eea
\ese
These formulae were derived in \cite{Neitzke:2007ke} by using 
the projective superspace description of the $c$-map found 
in \cite{Rocek:2005ij}. 
A key feature of these formulae is that, for a fixed point on the base, 
the complex coordinates $\xi^I,\txi_I,\alpha$ depend rationally on
the fiber coordinate $\cZ$; said differently, the fiber over
any point on the base is rationally in $\cZ$.
This is a general property of twistor spaces, which allows
for the existence of the Penrose transform relating holomorphic functions
on $\cZ$ to harmonic-type functions on $\cM_3$, a topic which we shall
return to in Section \ref{sec-penrose}. 

The K\"ahler potential on $\cZ$ in these coordinates was also computed in 
\cite{Neitzke:2007ke}, and reads
\be
\label{kz-hesse}
K_{{\cal Z}}= \frac12\log\left\{
 \Sigma^2 \left[\frac{i}{2}(\xi^I-\bar\xi^I),
\frac{i}{2}(\txi_I-\bar\txi_I)\right]
+\frac{1}{16} \left[
\alpha -\bar\alpha+ \xi^I \bar\txi_I-\bar\xi^I \txi_I
\right]^2 \right\}+\log 2\ .
\ee
where $\Sigma_{BH}(\phi^I,\chi_I)$ is the Hesse potential 
defined in Exercise \ref{exohesse} on page \eqref{exohesse}.
In particular, $K_{\cZ}$ is a symplectic invariant,
but, as we shall see in Section \ref{quasiref}, it can be
invariant under an larger group which mixes $\xi^I,\txi_I$ with $\alpha$.

The Swann space can be recovered from the twistor space $\cZ$ 
by supplementing  the coordinates $\xi^I,\txi_I,\alpha$ with one complex 
coordinate $\lambda$ (a coordinate in the $O(-1)$ bundle over $\cZ$).
The hyperk\"ahler potential on $\cS$ and the coordinates $\pi^{A'}$
in the $\IR^4$ fiber are then obtained by
\begin{equation} 
\label{hkc-coords}
R^2 = |\lambda|^2 \ e^{\cK_{\cZ}}\ ,\qquad 
\begin{pmatrix} \pi^1 \\ \pi^2 \end{pmatrix} = 2\,\lambda \, e^U\, 
\begin{pmatrix} z^\half \\ z^{-\half} \end{pmatrix}.
\end{equation}
Using the twistor map (and its converse, which can be found 
in \cite{Neitzke:2007ke}), it was shown that the holomorphy  
condition \eqref{susycondsw} 
for supersymmetric geodesics on $\cS$ allows to fully integrate the motion, 
reproducing known spherically symmetric black hole solutions.

\subsection{Quantum Attractors}
We now discuss the radial quantization of stationary,
spherically symmetric geometries in four dimensions, using the 
equivalence between the radial evolution equations
and geodesic motion of a fiducial particle on the cone~${\cal C}=
\IR^+ \times {\cal M}_3^*$. For brevity, we drop the cone direction
and restrict to motion along ${\cal M}_3^*$. 
We start with some generalities
in the non-supersymmetric set-up, and then restrict to the BPS
sector of $\cN=2$ supergravity.

\subsubsection{Radial Quantization of Spherically Symmetric 
Black Holes\label{sphgeoq}}
Based on the afore-mentionned equivalence, a natural path towards quantization
is to replace functions on the classical phase space~$T^*(\cM_3^*)$
by square integrable functions $\Phi$ on~$\cM_3^*$, 
and impose the quantum version of the mass-shell condition \eqref{rrhoc},
\be
\label{Depsi3}
\left[ \Delta_3  + C^2 \right]
\ \Phi_{C}( U, z^i, \zeta^I, \tzeta_I, \sigma) = 0
\ee
Here~$\Delta_3$ is the Laplace-Beltrami operator on~${\cal M}_3^*$,
the quantum analogue of the Hamiltonian $-H$.
In writing this, we have ignored the fermionic degrees of freedom,
which we shall discuss in the next Section \ref{bpstwi}, 
and possible
quantum corrections to the energy $C^2$.
In practice, we are interested in wave functions which are 
eigenmodes of the electric and magnetic charge operators, given by
the differential operators in \eqref{killingh},
\be
\Phi_{C}(U, z^i, \zeta^I, \tzeta_I, \sigma )  = \Phi_{C,p,q}(U, z^i)\ 
e^{i (p^I \tzeta_I- q_I \zeta^I)} 
\ee
which is then automatically a zero eigenmode of the NUT charge~$k$.
Note however that, due to the Heisenberg algebra \eqref{killal}, 
it is impossible to simultaneously diagonalize the
ADM mass operator~$M$, unless either~$p^I$ or~$q_I$ vanish. Equation
\eqref{Depsi3} then implies that the wave function  
$\Phi_{C,p,q}(U, z^i)$ should satisfy a quantum version of
\eqref{hampqk},
\be
\label{qHampqk}
\left[ -\pa_U^2 - \Delta_4 - e^{2U} V_{BH}(p,q,z) - C^2 \right]
\ \Phi_{C,p,q}(U, z^i)=0
\ee
where~$\Delta_4$ is now the Laplace-Beltrami on the four-dimensional
moduli space~${\cal M}_4$. The wave function~$\Phi_{C,p,q}(U, z^i)$ 
describes the 
quantum fluctuations of the scalars~$z^i$ as a function of the size~$e^{U}$
of the thermal circle ( {\it i.e.} effectively as a function of the distance to
the horizon). Importantly, the wave function is not uniquely specified
by the charges and extremality parameter, as  the condition~\eqref {qHampqk} 
leaves an infinite dimensional Hilbert space; this ambiguity 
reflects the classical freedom in choosing the values of the 4D moduli 
at spatial infinity.

An important aspect of any quantization scheme is the definition
of the inner product: as in similar instances of mini-superspace 
quantization, the~$L_2$ norm on the space of functions on~${\cal C}$
is inadequate for defining expectation values, since it involves
an integration along the ``time'' direction~$U$ at which one is supposed
to perform measurements. The customary approach around this problem is
to recall the analogy of~\eqref{qHampqk} with the usual Klein-Gordon equation,
and to replace the~$L_2$ norm on~$\cM_3^*$ by the Klein-Gordon norm 
(or Wronskian) at a fixed time~$U$: 
\be
\label{kg2}
\langle\Phi | \Phi \rangle =\int dz^i \ d\zeta^I \ d\tzeta_I \ d\sigma
\  \Phi^* \stackrel{\leftrightarrow}{\pa_{U}} \Phi 
\ee
By construction, this is independent of the value of $U$ chosen to evaluate it.
A severe drawback of this inner product is that it is not positive definite. 
This also has
a standard remedy in the case of the Klein-Gordon equation, which is
to perform a ``second quantization'' and replace the wave function~$\Phi$
by an operator; a similar procedure can be followed here, in analogy
with ``third quantization'' in quantum cosmology~\cite{Giddings:1988wv}. 
This procedure should presumably be relevant for describing multi-centered 
solutions.
Fortunately, for BPS states this problem is void, since,
as we shall see in Section \ref{sec-penrose}, the Klein-Gordon 
product~\eqref{kg2} is positive definite when restricted to this sector.

\subsubsection{Supersymmetric quantum mechanics and 
BPS Hilbert space\label{bpstwi}}
In the presence of fermionic degrees of freedom, the general discussion 
in the previous subsection must be slightly amended. Upon quantization,
the fermions $\psi^a$ in \eqref{s1ferm} become Dirac matrices on the
target space $\cM_{3}^*$, and the wave function is now valued in 
$L^2(\cM_{3}^*) \otimes \mathrm{Cl}$, where $\mathrm{Cl}$ is the
Clifford algebra of $\cM_{3}^*$. Equivalently,  one may 
represent the fermion $\psi^a$ as a differential $d\phi^a$ 
in the exterior differential algebra on $\cM_3^*$, and view the wave function
as an element of the de Rham complex of  $\cM_3^*$,
i.e. as a set of differential forms of arbitrary degree \cite{Witten:1982df}. 
The Wheeler-De Witt equation \eqref{Depsi3} now selects eigenmodes of the 
de Rham Laplacian $d\star d$ with eigenvalue $-C^2$; in particular,
for extremal black holes, the wave function becomes an element of
the de Rham cohomology of  $\cM_3^*$. These subtleties does not affect
the functional dimension of the Hilbert space, and there still
exist a continuum of states with given electric and magnetic charges.

In the presence of extended supersymmetry, however, it becomes possible to 
look for quantum states which preserve part of the supersymmetries. The 
simplest example is supersymmetric quantum mechanics on a K\"ahler
manifold \cite{Alvarez-Gaume:1983at,Friedan:1983xr,Gauntlett:1992yj}: 
the de Rham complex is 
refined into the Dolbeault complex, and states annihilated by one-half of 
the supersymmetries are elements of the Dolbeaut cohomology $H^{p,0}(X)$, 
isomorphic to the sheaf cohomology
group $H^0(X,\Omega^p)$. In more mundane terms, this means that the
BPS wave functions are holomorphic differential
forms of arbitrary degree, in particular, the functional dimension of the
BPS Hilbert space is now $\dim(X)/2$, half the dimension of the Hilbert
space for generic ground states.

We now turn to the case of main interest for us, supersymmetric
quantum mechanics on a quaternionic-K\"ahler manifold\footnote{This
system first appeared in the context of monopole dynamics 
in $\cN=2$ gauge theories \cite{Gauntlett:1993sh}}. Classically, we
have seen in \eqref{susycond1} that supersymmetric solutions are those for 
which the quaternionic viel-bein $V^{AA'}$ has a zero eigenvector 
$\epsilon_{A'}$. If we disregard the Killing spinor, the BPS condition
is summarized by the quadratic equations in \eqref{susycond2}. Since
$V^{AA'}/d\tau$ is equal to the momentum of the
fiducial particle, this is naturally quantized into
\be
\label{wavebpsqk}
\forall\, A,B\ ,\qquad 
\left[ \eps^{A'B'} \nabla_{AA'}\, \nabla_{BB'} + \kappa\, \eps_{AB} 
\right] \Phi = 0
\ee 
where we allowed for a possible quantum ordering ambiguity $\kappa$. 
Here, $\nabla_{AA'}=V_{AA'}^{a} \nabla_a$ is the covariant derivative on 
$\cM_3^*$, rotated by the inverse quaternionic viel-bein. 

On the other hand, we have seen that it was possible to work in
an extended phase space which includes the Killing spinor $\epsilon_{A'}$, 
and describes geodesic motion on the Swann space $\cS$. The supersymmetry 
condition
\eqref{susycondsw} is now linear in the momentum $V_{\aleph}$, and is
naturally quantized into
\be
\label{wavebpssw}
\forall \,\aleph\ ,\qquad \bar\partial_{\aleph} \Phi' = 0
\ee
where $\bar\partial_{\aleph}$ are partial derivatives with respect
to a set of antiholomorphic coordinates $\bar z^{\bar\aleph}$ on $\cS$.
Thus, wave functions on the extended phase space are just holomorphic
functions on $\cS$ (or more accurately, elements of the sheaf
cohomology of $\cS$). 

Since the classical geodesic motions on $\cM_3^*$ 
and $\cS$ are equivalent only for trajectories with vanishing $SU(2)$ 
momentum, it should be possible to generate a solution 
of the second order differential equation \eqref{wavebpsqk} from a 
holomorphic function on $\cS$, by projecting 
on $\IR^\times \times SU(2)$ invariant states. Part of this projection
can already be taken care of by restricting to homogeneous functions of
fixed degree $-k$ on $\cS$, or equivalently to sections of $O(-k)$ 
on $\cZ$.

\subsubsection{Quaternionic Penrose transform and exact BPS wave function\label{sec-penrose}}
Remarkably, there is exist a mathematical construction 
valid for any quaternionic-K\"ahler manifold, sometimes known
as the quaternionic Penrose transform \cite{quatman,MR1165872,Neitzke:2007ke}, 
which performs exactly this task, namely takes an element of 
$H^1(\cZ,O(-2))$ to a solution of \eqref{wavebpsqk}.  
This is an analogue of the more familiar Penrose transform 
which maps sections of $H^1(\IC\IP^3,O(-2)$ to massless spin 0
fields on $\IR^4$ \cite{MR506229}. 
Using the complex coordinate system introduced in Section \ref{twisec}, 
it is easy
to provide an explicit integral representation of this transform,
where the element of $H^1(\cZ,O(-2))$ is represented by a holomorphic
function $g(\xi^I,\txi_I,\alpha)$ in the trivialization 
$\lambda=1$ \cite{Neitzke:2007ke}:
\begin{equation} \label{penrosetrans}
\Phi(U, z^a, \bar z^{\bar a}, \zeta^I, \tzeta_I, \sigma) =
e^{2 U}\  \oint \frac{dz}{z}\,g(\xi^I, \txi_I, \alpha)\,,
\end{equation}
In this formula, $\xi^I,\txi_I,\alpha$ are to be expressed as functions
of the coordinates on $\cM_3$ and $z$ via the twistor map \eqref{gentwi}. 
The integral runs over a closed contour which separates $z=0$ from 
$z=\infty$. In 
\cite{Neitzke:2007ke}, it was shown that the left-hand side of
\eqref{penrosetrans} is indeed a solution of the system
of second order differential equations \eqref{wavebpsqk} with
with a fixed value for $\kappa=-1$. Moreover, the Klein-Gordon
inner product on $\cM_3$ \eqref{kg2} may be rewritten in terms of the 
holomorphic function $g$ as
\begin{equation}
\label{inner}
\inprod{\Phi \vert \Phi'} = \int 
d\xi^I d\txi_I d\alpha \, d\bar{\xi}^I d\bar{\txi}_I 
d\bar{\alpha}\ e^{-2(n_V+1) K_\cZ}\,\overline{g(\xi^I, \txi_I, \alpha)}\, g'(\xi^I, \txi_I, \alpha)
\end{equation}
where the integral runs over values of $\xi^I, \txi_I, \alpha,
\bar\xi^I, \bar\txi_I, \bar\alpha$ such that the bracket in 
\eqref{kz-hesse} is strictly positive. In particular, the inner product 
\eqref{inner} is positive definite, as announced at the end of Section
\ref{sphgeoq}.

There also exist versions of
\eqref{penrosetrans},\eqref{inner} appropriate to sections of $H^1(\cZ,O(-k))$
for any $k>0$, which are mapped to sections of $\Lambda^{k-2}(H)$ satisfying
first order differential equations \cite{Neitzke:2007ke}.

Thus the problem of determining the radial wave function of BPS black
holes is reduced to that of finding the appropriate section of 
$H^1(\cZ, O(-2))$. For a black hole with fixed electric and magnetic
charges $q_I, p_I$ and zero NUT charge, the only eigenmode of 
the generators \eqref{cxkillingh} is, up to normalization, 
the ``coherent state''
\begin{equation}
\label{cohst}
g_{p,q}(\xi^I, \txi_I, \alpha) =  e^{\I( p^I \txi_I
-q_I \xi^I)}\,.
\end{equation}
These states are delta-normalizable under inner product
\eqref{inner} (possibly regulated by analytic continuation in $k$), 
and become normalizable after modding out by the discrete Heisenberg
group\footnote{Scaling arguments show that the norm grows as a power of 
$p,q$, rather than exponentially.}.

Applying the Penrose transform \eqref{penrosetrans} to the state
\eqref{cohst}, we find
\begin{equation}
\Phi_{p,q}(U, z^a, \bar z^{\bar a},\zeta^I, \tzeta_I, \sigma) = 
e^{\I p^I \tzeta_I - \I q_I \zeta^I} 
\, e^{2U} \oint \frac{d z}{z} \exp \left[ e^{U} (z \bar{Z} + z^{-1} Z) \right],
\end{equation}
where $Z$ is the central charge \eqref{centchar} of the black hole.
After analytic continuation $(\zeta^I,\tzeta_I)$ to 
$i(\zeta^I,\tzeta_I)$ and $(p^I,q_I)$ 
to $-i(p^I,q_I)$, 
the integral may be evaluated in terms of a Bessel function,
\begin{equation}
\label{exactbpswf}
\Phi(U, z^a, \bar z^{\bar a}, \zeta^I, \tzeta_I, \sigma) 
= 2\pi\, e^{\I p^I \tzeta_I - \I q_I \zeta^I}
\ e^{2U}\ J_0(2  e^U \abs{Z})
\end{equation}
This is the exact radial wave function for a black hole with 
fixed charges $(p^I,q_I)$, at least in the supergravity 
approximation\footnote{In the presence of $R^2$-type corrections,
the geodesic motion receives higher-derivative corrections, and
it is no longer clear how to quantize it.}.

Since the Bessel function $J_0$ 
decays like $\cos(w)/\sqrt{w}$ at large values of $|w|$,
we see that the phase of 
the BPS black hole wave function is stationary at the classical attractor 
point $z^i_{p,q}$, and becomes flatter and flatter in the near-horizon 
limit $U\to -\infty$, while the modulus decays away from these 
points as a power law. The occurrence of large quantum fluctuations
in the near horizon limit may seem at odds with the attractor behavior
for BPS black holes, but is in fact perfectly consistent with the 
picture of a particle moving in an inverted potential $V=-e^{2U}V_{BH}$,
as discussed in Section \ref{bhpot}.
It is a reflection of the infinite fine-tuning of the asymptotic 
conditions which is necessary for obtaining an extremal black hole. 

Returning to the original motivation explained in Section \ref{osvqatt},
we observe that the wave function \eqref{exactbpswf} bears no obvious relation 
to the topological string amplitude. One may however try to rescue
the suggestion in \cite{Ooguri:2005vr} by noting that there is in principle 
an even smaller subspace of the Hilbert space $L^2(\cS)$, corresponding
to ``tri-holomorphic'' on $\cS$; we shall remain deliberately vague about 
the concept of ``tri-holomorphy'' here, referring the 
reader to \cite{Anselmi:1993wm} for some background on this subject,
but merely assume that it divides the functional dimension
by a factor of four. If so, this ``super-BPS'' Hilbert space
of triholomorphic functions on $\cS$ would have functional dimension 
$n_V+2$, and be the natural habitat of a one-parameter generalization
of the topological wave function \cite{Gunaydin:2006bz}.  
One would also expect some quaternionic analogue of the 
Cauchy integral in \eqref{penrosetrans}, which would map the space of 
tri-holomorphic functions on $\cS$ to functions on $\cM_3$ annihilated
by certain differential operators. 
In the symmetric cases studied in the next Section, we
shall indeed be able to construct a ``super-BPS'' Hilbert space,
of functional dimension $n_V+2$,
which carries the smallest possible
unitary representation of the duality group.

\subsection{Very Special Quantum Attractors\label{quasiref}}
We now specialize the construction of Section \ref{bpstwi}
to the case of very special $\CN=2$ supergravities which we introduced 
in Section \ref{veryspe}. Our goal is to produce a framework
for constructing duality-invariant black hole partition functions,
applicable both for these
$\CN=2$ theories and their $\cN=4,8$ variants.

\subsubsection{Quasiconformal Action and Twistors\label{quasitwi}}

Recall that the vector-multiplet moduli space
of very special supergravities are hermitean symmetric tube domains
\eqref{verymod}, built out of the 
invariance groups of Jordan algebras $J$ with a cubic norm $N$.
The result of the $c$-map \cite{Cecotti:1988ad} 
and $c^*$ map \cite{Breitenlohner:1987dg} constructions 
are still symmetric spaces, of the form
\be
\label{m3sym}
{\cal M}_3 = \frac{\QConf(J)}{\widetilde{\Conf}(J) \times SU(2)}\ ,\quad
{\cal M}_3^* = \frac{\QConf(J)}{\Conf(J) \times Sl(2)}
\ee
Here, $\QConf(J)$ is the ``quasi-conformal group'' associated to the
Jordan algebra $J$ (in its quaternionic, rank 4 real form), and 
$\widetilde{\Conf}(J)$ is the compact real form of $\Conf(J)$;
these spaces can read off from Table \ref{tabmod} on page \pageref{tabmod}.

The terminology of ``quasi-conformal group'' refers to the realization
found in \cite{Gunaydin:2000xr}
of $G=\QConf(J)$ as the invariance group of the zero locus 
$\cN_4(\Xi,\bar \Xi)=0$ of 
a homogeneous, degree 4 
polynomial $\cN_4$ in the variables $\Xi=(\xi^I,\txi_I,\alpha)$ (of respective
degree 1,1,2) and  $\bar\Xi=(\bar\xi^I,\bar\txi_I,\bar\alpha)$:
\be
\label{quarticlc}
{\cal N}_4(\Xi;\bar\Xi) = 
\frac12 I_4\left(\xi^I-\bar\xi^{I},\txi_I-\bar\txi_{I}\right) 
+ \left( \alpha-\bar\alpha + 
\bar\xi^{I} \txi_I - \xi^I \bar\txi_I \right)^2
\ee
More precisely, there exists an holomorphic action of $G$ on $\Xi$
such that ${\cal N}_4(\Xi,\bar\Xi)$ gets multiplied by a product 
$f(\Xi)\bar f(\bar\Xi)$\label{footxi}. In \eqref{quarticlc}, $I_4$
is the quartic invariant \eqref{i4jor} of the group $\Conf(J)\subset
\QConf(J)$
associated to the Jordan algebra $J$, acting linearly on 
the symplectic vectors $(\xi^I,\txi_I)$ and $(\bar\xi^I,\bar\txi_I)$.
By analogy with the ``conformal realization'' of $\Conf(J)$, leaving the cubic
light-cone $N(z^i-\bar z^i)$ invariant, this is called the
quasi-conformal realization of $G$.

\begin{figure}
\centerline{\hfill\includegraphics[height=7cm]{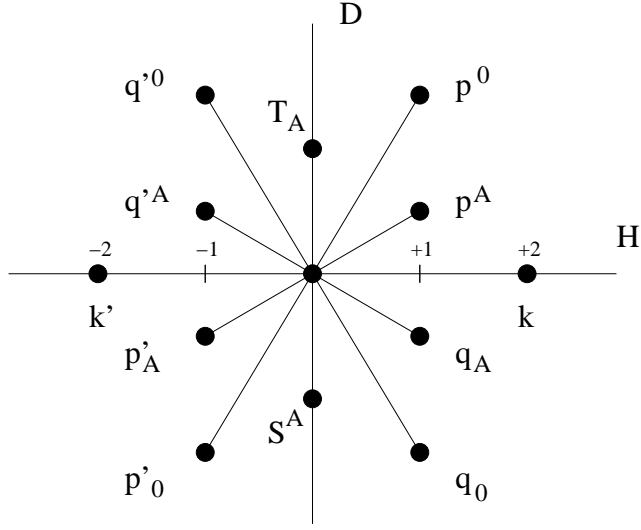}\hfill}
\caption{Two-dimensional projection of the root diagram of the 
quasi-conformal group associated to a cubic
Jordan algebra $J$ (when $J=\IR$, this is the root diagram of $G_2$). 
The five-grading corresponds to the horizontal axis.
The long roots are singlets, generating a $SU(2,1)$ universal subgroup, 
while the short roots are valued in the
Jordan algebra $J$.\label{qconfroot}}
\end{figure}

Group theoretically, the origin of this action is clear: the group 
$G$ admits a 5-graded decomposition, corresponding to 
the horizontal axis in the two-dimensional projection
of the root diagram of $G$ shown in Figure \ref{qconfroot}),
\be
\label{5grad}
\begin{array}{ccccccccccc}
G & = &  G_{-2} & \oplus & G_{-1} & \oplus & G_0 & \oplus & 
G_{+1} &\oplus & G_{+2}  \\
  & \equiv &\{ k' \} &\oplus& \{ p'_I,q^{'I} \}& 
\oplus& \{T_A,S^A,D_A^B\}& 
\oplus& \{ p^I, q_I\} &\oplus& \{ k \}
\end{array}
\ee
In particular, the top space $G_{+2}$ is one-dimensional, 
therefore $G_{+1}\oplus G_{+2}$ form an Heisenberg algebra 
with center $G_{+2}$, which we identify with the Heisenberg algebra 
$[p^I,q_J]=2k\delta^I_J$ of electric, magnetic and NUT isometries
\eqref{heis}. The grade 0 space is $G_0=\Conf(J)\times U(1)
=\{T_A,S^A,D_A^B\}$. Symmetrically,  $G_{-1}\oplus
G_{-2}$ form an Heisenberg algebra $[p'_I,q^{'J}]=2k'\delta^I_J$
with one-dimensional center $G_{-2}$.
Together with $G_{-2}=\{k'\}$ and $G_{+2}=\{k\}$, 
the center $H=D_A^A=[k,k']$ of $G_0$ generates  an $SU(2)$ subgroup
which commutes with $\Conf(J)$, and yields the above 5-grading above.
Finally, $\Conf(J)$ acts linearly on $G_{+1}\sim \{p^I,q_I\}$ 
in the usual way, leaving $G_{+2}\sim \{k\}$ 
invariant. Since the $H$ charge is additive, the sum
$P=G_{-2} \oplus G_{-1} \oplus G_0$ closes under commutation, and
is known as the Heisenberg parabolic subgroup $P$ of $G$. 
The quasi-conformal realization of $G$ is then just the action on 
$P \backslash G$ by right multiplication;   it may be twisted
by a unitary character $\chi$ of $P$, i.e. by considering functions on 
$P\backslash G$
which transform by $\chi$ under the right action of $G$ (mathematically,
this is the induced representation from the parabolic $P$ to $G$
with character $\chi$, see e.g. \cite{Pioline:2003bk} for an
introduction to this concept). 

To be completely explicit, the generators in  $G_{+1} \oplus G_{+2}$ act on 
functions of $\Xi$ as
\be
E_{p^I} = \pa_{\txi_I} - \xi^I \pa_\alpha\  ,
\quad E_{q_I} = -\pa_{\xi^I} - \tilde\xi_I \pa_\alpha\ ,\quad
E_k = \pa_\alpha\ ,
\ee
while the generator $k'$ in $G_{-2}$ acts as
\be
\label{gm2}
E_{k'}=
\left( - \frac{1}{4} \frac{\pa I_4}{\pa \txi_I} - \alpha \xi^I\right)
\pa_{\xi^I} 
+\left( \frac{1}{4} \frac{\pa I_4}{\pa \xi^I} -  \alpha \txi_I\right)
\pa_{\txi_I} +
\frac12( I_4 -  2\alpha^2) \pa_\alpha - k \alpha 
\ee
where $I_4= I_4(\xi^I,\txi_I)$ and $k$ is a complex number parametrizing
the character $\chi$. The rest of the generators can be obtained by 
commutation and $\Conf(J)$ rotations.

Comparing \eqref{quarticlc} and  \eqref{kz-hesse}, and
recalling that the Hesse potential $\Sigma$ for symmetric spaces is 
the square root of the quartic invariant $I_4$, it is manifest that
the log of the ``quartic light-cone'' \eqref{quarticlc} is just the K\"ahler 
potential \eqref{kz-hesse} of the twistor space  
$\cZ=G_\IC\backslash P_\IC$ of the quaternionic-K\"ahler space ${\cal M}_3$;
therefore, the quasi-conformal realization is nothing but the 
holomorphic action of $\QConf(J)$ on the twistor space $\cZ$.
For integer values of the parameter $k$, this representation 
belongs to the ``quaternionic discrete series'' 
representation of $G$ \cite{MR1421947}, a quaternionic analogue
of the usual discrete series for $Sl(2)$. 

We conclude that for very special supergravities,
the BPS Hilbert space carries a unitary representation of 
the three-dimensional U-duality group $G=\QConf(J)$,
given by the ``quaternionic discrete series'' or ``quasi-conformal
realization'' of $G$. 

\subsubsection{Penrose transform and spherical vector}

In Section \ref{sec-penrose}, we have seen that there is a 
Penrose transform which takes holomorphic functions on $\cZ$ to
a function on $\cM_3$ annihilated by some second order differential 
operator. In the present symmetric context, there is an a
priori different way of producing a function $\Phi$ on $G/K$ from a 
vector $f\in {\cal H}$ in a unitary representation of $G$: 
for any $e\in G$, take 
\be
\label{psisph}
\Phi( e ) = \langle f | \rho(e) | f_K \rangle
\ee
where $f_{K}$ is a fixed $K$-invariant vector in ${\cal H}$. 
Since $f_K$ is invariant under $K$, $\Phi$ descends to the quotient $G/K$.
This construction is standard in representation theory, where $f_K$ is
referred to as a spherical vector (see again \cite{Pioline:2003bk}).

Not surprisingly, the geometric and algebraic constructions are in fact 
equivalent, as we illustrate in the simplest case of 
the universal sector $G=SU(2,1)/SU(2)\times U(1)$. The quartic invariant
in this case is the square of a quadric, $I_4=\frac12(\xi^2+\txi^2)^2$.
Using \eqref{gm2} one may check that 
\be
f_K(\xi,\txi,\alpha) = \left( 1 + \xi^2 + \txi^2 + 
\alpha^2 + \frac12 I_4 \right)^{-k/2}
\ee
is the unique vector invariant under $SU(2)\times U(1)$. 
Acting with $\rho(e)$
where $e\in G$ is parameterized by $U,\zeta,\tzeta,\sigma$, one
obtains
\be
[\rho(e)f_K] (\xi,\txi,\alpha) =
\left[e^{2U} + (\txi-\tzeta)^2 + (\xi-\zeta)^2 
+  e^{-2U} {\cal N}_4(\xi,\txi,\alpha;\zeta,\tzeta,\sigma) 
\right]^{-k/2}
\ee 
Thus, we conclude that BPS wave functions, in the unconstrained Hilbert 
space ${\cal H}$, are given by
\be
\Phi(U,\zeta,\tzeta,\sigma) = 
\int \frac{ \bar f(\bar\xi,\bar\txi,\bar\alpha)~ 
d\bar\xi~d\bar\txi~d\bar\alpha}
{\left[e^{2U} + (\bar\txi-\tzeta)^2 + (\bar\xi-\zeta)^2 
+ e^{-2U} {\cal N}_4(\zeta,\tzeta,\sigma,
\bar\xi,\bar\txi,\bar\alpha;) 
\right]^{k/2}}
\ee
By evaluating the contour integral by residues,
one may easily show that, for $k=2$, the function $\Phi$ in \eqref{psisph}
agrees with the Penrose transform  \eqref{penrosetrans} of the function
\be
g(\xi,\txi,\alpha) 
= \int 
\frac{{\bar f}( \bar\xi, \bar\txi, \bar\alpha)\ 
d\bar\xi d\bar\txi d\bar\alpha}
{ (\alpha-\bar\alpha+\bar\txi \xi - \txi \bar\xi)^2 
+ \frac14 \left[(\xi-\bar\xi)^2+(\txi-\bar\txi)^2\right]^2}
\ee
This operator which intertwines between the space of functions 
$g(\xi,\txi,\alpha)$ and ${\bar f}( \bar\xi, \bar\txi, \bar\alpha)$
is an example of the ``twistor transform'' (not to be confused
with the Penrose transform), which maps sections of $H^1(\cZ,O(-k))$
to sections of $H^1(\cZ,O(-4-k)$ \cite{MR610183}.

\subsubsection{The Minimal Representation vs. the 
Topological Amplitude \label{minrep}}
At the end of section \ref{sec-penrose}, we pointed out that the functional
dimension of the BPS Hilbert space $H_1(\cZ,O(-2))$, $2n_V+3$, was 
too large to accommodate the topological string amplitude, which
depends on $n_V+1$ variables. In the symmetric case, it is natural
to ask whether there are smaller representations than the quasi-conformal
realization, which could provide the natural habitat for the 
topological string amplitude. 

In fact, it is known in the mathematics literature 
that the quasiconformal representation,
for low values of the parameter $k$, is no longer 
irreducible \cite{MR1421947}. In particular, the symplectic
space $V=\{ \xi^I, \txi_I \}$ admits a sequence of subspaces 
$V\supset X \supset Y \supset Z$, defined 
by homogeneous polynomial equations of degree 4,3 and 2, respectively
such that each of them is preserved
by the quasi-conformal action of $G$. Here, $X$ is the locus where the
quartic invariant $I_4(\xi,\txi)$ vanishes, $Y\subset X$ is the locus where
the differential $dI_4$ vanishes; finally, $Z \subset Y$ is the locus where
the irreducible component of the Hessian of $I_4$ (viewed as an element 
of the symmetric tensor product $V\otimes_S V$) transforming in the adjoint 
representation of $\Conf(J)$ vanishes, a condition which we'll denote
$d^2 I_4=0$. As shown in  \cite{MR1421947},
each of the subspaces $X,Y,Z$, supplemented with the variable $\alpha$
and for the appropriate choice of $k$, 
furnishes an irreducible unitary representation of $G$, of functional 
dimension $2n_V+3, 2n_V+2,(5n_v+1)/3$ and $n_V+2$ variables, respectively. 
By the ``orbit philosophy'', these are 
associated by to nilpotent co-adjoint orbits
of nilpotency order 5,4,3,2, respectively. 

The smallest of those, known as the  minimal representation
of $G$, is of particular importance to us, as its dimension $n_V+2$ is just 
one more than the number of variables appearing in the topological amplitude.
This representation plays a distinguished role in mathematics, being the 
smallest unitary representation of $G$ and an analogue
of the metaplectic representation of the symplectic group.
In physics, the minimal representation of $Sl(3)$ was used
in the early days for strong interactions \cite{Biedenharn:1972ns},
and more recently in an attempt at quantizing 
BPS membranes\cite{Pioline:2001jn,Pioline:2004xq}.
Its relevance to black hole physics was suggested in \cite{Gunaydin:2001bt}
and expounded in \cite{Pioline:2005vi,Gunaydin:2005mx}.

As first observed in the case of $E_{8(8)}$ in \cite{Gunaydin:2001bt}, 
the minimal representation
can be obtained by quantizing the symplectic space\footnote{This is
sometimes referred to as ``quantizing the quasi-conformal action'',
which may cause some confusion since the quasi-conformal 
realization is quantum mechanical already.} $V$ of the
quasi-conformal realization was acting, namely replace 
$\txi_I \to i \pa_{\xi^I}$ and fixing the ordering ambiguities so
that the algebra of $\QConf(J)$ is preserved. An independent
construction, valid for all simply-laced cases, was given 
in \cite{MR1159103,Kazhdan:2001nx}; a recent unified approach using the
language of Jordan algebra and Freudenthal triple systems 
can be found in \cite{Gunaydin:2004md,Gunaydin:2006vz}.

In order to extract physical information from 
wave functions in the minimal representation, just as in 
the quasiconformal case it is necessary
to embed them in the non-BPS Hilbert space, i.e. map them into functions 
on $\cM_3$ by some analogue of the Penrose transform.  
As explained in the previous subsection, this may be done once
a spherical vector $f_K$ is found. A slight complication is that 
the minimal representation for non-compact
groups in the quaternionic real form (as opposed to the split real
form) do not admit a spherical vector; rather, the decomposition
of the minimal representation under the maximal compact group 
$K=\widetilde{\Conf}(J) \times SU(2)$ has a ``ladder'' structure,
whose lowest component (or ``lowest $K$-type'')
transforms in a spin\footnote{For $n_V<3$, the lowest $K$-type is
a singlet of $SU(2)$, but non-singlet of $\Conf(J)$.}
$(n_V-3)/6$ representation of $SU(2)$.
Replacing $f_K$ in \eqref{psisph} by this lowest $K$-type, one obtains a
section of a symmetric power of $H$ on $\cM_3$. 
The wave function of the lowest $K$-type can be computed
explicitely in a mixed real-holomorphic polarization \cite{gnopw-in-progress}; 
in the semi-classical
approximations, all components of the $K$-type behave as 
\be
f_K(a^A,b^\dagger,x) \sim \exp\left[
-\frac{x^2}{2}  + \frac{I_3( a^A)}{b^\dagger} + 2 i x \sqrt{
\frac{I_3( a^A)}{b^\dagger}} \right]
\ee
where $f_K(a^A,b^\dagger,x)$ is related to $f_K(\xi^0,\xi^A,\alpha)$
by a certain Bogolioubov operator \cite{gnopw-in-progress}. 
We take the fact that
$f_K$ reduces to the classical topological amplitude 
$\exp( I_3( a^A)/b^\dagger)$ in the limit $x\to 0$ as a strong
indication that the minimal representation is the habitat
of a one-parameter generalization of the standard 
topological amplitude. 

Further evidence for this claim 
comes from the fact the holomorphic anomaly 
equations \eqref{Ver1b} obeyed by the usual topological amplitude
follow from the quadratic identities in the 
universal enveloping algebra of the minimal representation
of $G$, upon restriction to the ``Fourier-Jacobi group'' $P/U(1)$, 
where $U(1)$ is the subgroup generated by the Cartan generator $H \in G_0$
\cite{Gunaydin:2006bz}. 
This is in precise analogy with the heat equation satisfied by 
the classical Jacobi theta series,
\be
\label{heat}
\left[i\partial_\tau - \partial_z^2\right]
\theta_1(\tau,z) = 0
\ee
which follows from quadratic relations in the minimal (i.e. metaplectic)
representation of $Sp(4)\supset Sl(2)$.
In this restriction, the generator $k\in G_{+2}$ becomes
central and can be fixed to an arbitrary non-zero value, reducing
the total number of variables from $n_V+2$ down to $n_V+1$.
The usual topological amplitude 
$\Psi_{\IR}(p^I)$ should then arise as a ``Fourier-Jacobi'' coefficient 
of a ``generalized topological amplitude'' $\Psi_{\rm gen}(p^I,k)$ at
$k=1$. The extension of these considerations to realistic cases 
without symmetry, possibly along the lines explained at the end of 
section \ref{sec-penrose}, would clearly have far-reaching consequences for
the enumerative geometry of Calabi-Yau spaces.
.

\subsection{Automorphic Partition Functions\label{autopar}}
We now return to our original motivation for investigating the
radial quantization of BPS black holes, namely the construction
of partition functions for black hole micro-states consistent
with the symmetries of the problem. We shall mainly consider the 
toy model case of very special $\cN=2$ supergravities, but will briefly
discuss the applications to $\cN=4$ and $\cN=8$ supergravity at the
end of this section.

In the previous sections, we discussed how the mini-superspace 
radial quantization of BPS black holes gives rise to Hilbert spaces
of finite functional dimension, furnishing a unitary representation of
the three-dimensional duality group $G_3=\QConf(J)$. It is natural to
expect that $G_3$ should serve as a spectrum-generating symmetry for
black hole micro-states \cite{Ferrara:1997uz,Gunaydin:2000xr,
Gunaydin:2005gd,Pioline:2005vi}. Indeed, it already 
serves as solution-generating
symmetry at the classical level, although it mixes bona-fide black holes 
with solutions with non-zero NUT charge $k$. Thus, we propose that the
black hole indexed degeneracies $\Omega(p,q)$ be given by Fourier 
coefficients of an automorphic form $Z$ on the three-dimensional moduli
space $\cM_3=G(\IZ)\backslash G / K$. More specifically, consider 
\be
\label{fourz}
\Omega(p^I,q_I; U, z^i,\bar z^{\bar i}) 
= \int d\zeta^I\, d\tzeta_I\, d\sigma\ 
e^{-i p^I \tzeta_I + i q_I \zeta^I}\,
Z(U,z^i,\bar z^{\bar i}; \zeta^I,\tzeta_I,\sigma)
\ee
where the integral runs over a fundamental domain
$0\leq (\zeta^I,\tzeta_I,\sigma)\leq 2\pi$ of the discrete Heisenberg
group.
The left-hand side is in principle a function of $U$,
$z^i,\bar z^{\bar i}$: one should view $Z$ as the 
partition function in a thermodynamical ensemble 
with electric and magnetic potentials $\zeta^I$ and $\tzeta_I$,
temperature $T=e^{-U} m_P$ and values $(z^i,\bar z^{\bar i})$
for the vector-multiplet moduli at infinity.
Provided $Z$ is annihilated by appropriate differential operators,
the dependence on $U,z^i,\bar z^{\bar i}$ will be entirely fixed by 
the charges $p^I,q_I$, and leave an overall factor identified as the
actual black hole degeneracy:
\be
\label{omphi}
\Omega(p^I,q_I; U, z^i,\bar z^{\bar i}) = 
\Omega(p^I,q_I)\ 
\Phi_{p,q}(U, z^a, \bar z^{\bar a})
\ee
Now, there is a natural way to construct an automorphic form which
satisfies these requirements: for $e\in G$, take
\be
\label{psisphz}
Z( e ) = \langle f_\IZ | \rho(e) | f_K \rangle
\ee
where $\rho$ is a unitary representation of $G$, $K$ a
spherical vector and $f_{\IZ}$ a $G(\IZ)$-invariant 
vector in this representation. This last condition guarantees that
$Z(g)$ so defined is a function on $G(\IZ)\backslash G/H$. 
We comment on ways to compute $f_{\IZ}$ below.
In particular, one may take for $\rho$ the quasi-conformal representation
described in Section \ref{quasitwi}: the function $\Phi_{p,q}$
in \eqref{omphi} is then just the
black hole wave function \eqref{exactbpswf} (with the dependence on
$\zeta^I$ and $\tzeta_I$ stripped off), while the integer degeneracies 
$\Omega(p,q)$ are encoded in the $G(\IZ)$-invariant vector $f_\IZ$.
In this case, it is known that the Fourier coefficients  have support only 
on charges with $I_4(p^I,q_I)\geq 0$ \cite{MR1988198}. 
One could also consider smaller representations associated to 
the subspaces 
$X,Y$ or $Z$ of $V$: the coefficients $\Omega(p,q)$ would then have support
on charges with $I_4(p,q)=0$, $dI_4=0$ or $d^2I_4=0$, and would presumably
be relevant for `small'' black holes with 3, 2 and 1 charges, respectively.

Thus, we have reduced the problem of computing the black hole 
partition function to that of constructing a $G(\IZ)$-invariant
vector in a unitary representation $\rho$ of the three-dimensional
duality group $G(\IZ)$ \cite{Pioline:2005vi}. 
This is a difficult problem, but there
is a powerful mathematical method, known as the Strong Approximation
Theorem, which allows to address this question
(see \cite{Pioline:2003bk} for a pedestrian introduction
to these techniques): this theorem
states that functions on $G(\IZ)\backslash G(\IR)$ are equivalent 
to functions on $G(\mathbb{A})/G(\IQ)$, where $\IA$ is the field
of adeles, i.e. the (restricted) product of $\IR$ times the 
$p$-adic number fields $\IQ_p$ for all prime $p$, with $\IQ$ 
being diagonally embedded in this product. Since $G(\IQ)$
is the maximal compact subgroup of $G(\IA)$,  the problem of finding 
$f_\IZ$ is reduced to that of finding the spherical vector 
over each $p$-adic field. This point of view has been applied to find
the $G(\IZ)$-invariant vector of the minimal representation
for simply-laced groups in the real form in \cite{MR2094111}.
It would be very interesting to construct the automorphic forms 
attached to quasi-conformal representation, and see if their Fourier
coefficients have the required exponential growth.

We close this section by noting that the construction of automorphic
partition functions outlined in this section can also be applied,
after suitable analytic continuation, to the case of $\cN=4$ and $\cN=8$
supergravity, which have a clear string theory realization. While
the three-dimensional moduli space is no longer quaternionic-K\"ahler,
there are still unitary representations associated to the symplectic
space $V$ and its subspaces $X,Y,Z$, and one can still define 
Fourier coefficients of the type \eqref{fourz}. For $\CN=8$ supergravity,
we expect that exact degeneracies of 1/8-BPS, 1/4-BPS and 
1/2-BPS black holes to be given by automorphic forms of $E_{8(8)}$
based on $V,Y,Z$, respectively (since the 1/4 and 1/2 BPS conditions
are $dI_4(p,q)=0$ and $d^2I_4(p,q)=0$, respectively \cite{Ferrara:1997uz}).
For $\cN=4$ supergravity, we expect 1/4-BPS states to be
counted by an automorphic form of $SO(8,n_v+2)$ (where $n_v$ is the
number of $\cN=4$ vector multiplets in 4 dimensions). This proposal
is distinct from the genus 2 partition function outlined in Section
\ref{dvvform}, and would have to be consistent with it at least in the
large charge regime. In this respect, it is interesting to remark
(see Exercise \ref{tictac} below) 
that $Sp(4)$ can be viewed as a ``degeneration'' of the 
three-dimensional U-duality group $\QConf(J)$ (for any $J$), upon collapsing 
all electric and magnetic charges $p^I$ and $q_I$ to just two charges $p,q$.
Thus, our proposal has the potential to resolve differences
between black holes which have the same continuous U-duality invariant,
but sit in different orbits of the discrete U-duality group. 

\begin{exo} 
\label{tictac}
Show that the root diagram of $Sp(4)$ is ``tic-tac-toe''-shaped.
Compare to the root diagram of $\QConf(J)$ in Figure \ref{qconfroot}
on page \pageref{qconfroot}.
\end{exo}

\section{Conclusion}

In these lectures, we have reviewed some recent attempts at
generalizing the microscopic counting of BPS black holes beyond leading
order. Our main emphasis was on the conjecture by Ooguri, Strominger and
Vafa, which relates the microscopic degeneracies of 
four-dimensional BPS black holes to the topological string amplitude, 
which captures an infinite series of higher-derivative corrections 
in the macroscopic, low energy theory.

By analyzing the case of ``small'' black holes, which can be easily counted
in the heterotic description, we have found that the topological
amplitude captures the microscopic degeneracies with impressive
precision. At the same time, it is clear that some kind of non-perturbative
generalization of the topological string is required, if one wants to
obtain exact agreement for finite charges.

Motivated by the ``holographic'' interpretation of the OSV conjecture as
a channel duality between radial and time-like quantization, we studied
the quantization of the attractor flow for stationary, spherically
symmetric black holes; this was achieved by reformulating the attractor flow 
as a BPS geodesic flow on the moduli space in three dimensions. 
Using the Penrose transform, we were able to compute the  
exact radial wave function for BPS black holes with fixed electric and
magnetic charges, in the supergravity approximation. 
It would be interesting to try and include the effect of higher derivative
corrections, as well as relax the assumption of spherical symmetry.

Contrary to the suggestion in \cite{Ooguri:2005vr}, the BPS wave function 
bears little ressemblance to the 
topological string amplitude. There is however evidence from the
symmetric space case that there exists a ``super-BPS'' Hilbert 
space which can host the topological string wave function, 
or rather a one-parameter generalization thereof. In the general
non-symmetric case, this generalized topological amplitude should
be viewed as a tri-holomorphic function over 
the quaternionic-K\"ahler moduli space (or rather, the Swann
space thereof). Using T-duality between the vector-multiplet
and hyper-multiplet branches in 3 dimensions, it is natural to 
expect that it should encode instanton corrections to the
hypermultiplet geometry in 4 dimensions \cite{npv-to-appear}.

These considerations lend support to the idea that the
three-dimensional duality group should play a role as a 
spectrum-generating symmetry for 4-dimensional 
black holes . Our framework suggests that the black
hole degeneracies should be indeed be related to Fourier coefficients 
of automorphic forms for the three-dimensional U-duality group $G$, 
attached to the representations of $G$ which appear in the radial
quantization of stationary, spherically symmetric BPS black holes.
It would be interesting to construct these automorphic forms 
explicitly, and have a handle on the growth of their Fourier
coefficients, similar to the Rademacher formula for modular
forms of $Sl(2,\IZ)$.

The most direct application of our framework 
is to BPS black holes in the FHSV model, since this is
a quantum realization of the very special $\cN=2$ supergravity
with $J=\IR\oplus \Gamma_{9,1}$; in this case, we expect that the 
black hole partition function is an automorphic form of 
$SO(4,12,\IZ)$, which it would be very interesting to construct.
With some minor amendments, our framework also applies to 
$\CN=4$ and $\CN=8$ backgrounds in string theory, whose three-dimensional 
U-duality groups are $SO(8,24,\IZ)$ and $E_{8(8)}(\IZ)$. In the $\CN=4$
case, our proposal differs from the DVV formula, which relies
on an automorphic form of $Sp(4,\IZ)$, but has the potential 
to distinguish black holes which have the same continuous U-duality invariant,
but sit in different orbits of the discrete U-duality group. 
For $\cN=8$, the entropy of 1/8-BPS BPS black holes in a certain orbit
was computed using the 4D/5D lift in \cite{Shih:2005qf,Pioline:2005vi}.
It would be
interesting to see if an agreement with these formulae can be reached at
least for certain orbits. 

The extension of these ideas to general $\CN=2$ string theories,
possibly using the monodromy group of $X$ as a replacement for the
U-duality group, is of course the most challenging
and potentially rewarding problem, as it is bound to unravel new relations
between number theory, algebraic geometry and physics.

\subsection*{Acknowledgements}

These notes represent the content of lectures delivered at the
RTN Winter School on Strings, Supergravity and Gauge theories,
(CERN, January 16-20, 2006), the 11-th APCTP/KIAS String Winter
School (Pohang, Feb 8-15 2005) and the Winter School on the 
Attractor Mechanism (Frascati, March 20-24, 2006). They have
been updated (most notably Section \ref{qatt}) in February 2007 
to incorporate recent progress. I am very grateful to the respective 
organizers for inviting me to give these lectures, and to the participants 
for their interest and questions. 

I wish to thank A.~Dabholkar, F.~Denef, G.~Moore for an enjoyable 
collaboration leading to the results presented in Sections \ref{smallbh},
and M.~G\"unaydin, A.~Neitzke, S.~Vandoren, and A.~Waldron 
for an on-going collaboration on the subject of Section \ref{qatt}.
I am also grateful to I.~Bena, S.~Bellucci, B.~de Wit, 
R.~Dijkgraaf, E.~Gimon, R.~Kallosh, A.~Keurentjes,
P.~Kraus, M.~Marino, S.~Miller,
M.~Rocek, S.~Vandoren, E.~Verlinde for useful discussions
on the some of the material covered in these lectures.

This research
is supported in part by  the EU under contracts
MTRN--CT--2004--005104, MTRN--CT--2004--512194, and by
ANR (CNRS--USAR) contract No 05--BLAN--0079--01.


\begin{thebibliography}{10%
0}

\bibitem{Hughes:2005wj}
S.~A. Hughes, ``Trust but verify: The case for astrophysical black holes,''
  {\em ECONF} {\bf C0507252} (2005) L006,
  \href{http://xxx.lanl.gov/abs/hep-ph/0511217}{{\tt hep-ph/0511217}}.

\bibitem{Banks:2003vp}
T.~Banks, ``A critique of pure string theory: Heterodox opinions of diverse
  dimensions,'' \href{http://xxx.lanl.gov/abs/hep-th/0306074}{{\tt
  hep-th/0306074}}.

\bibitem{Strominger:1996sh}
A.~Strominger and C.~Vafa, ``Microscopic origin of the {B}ekenstein-{H}awking
  entropy,'' {\em Phys. Lett.} {\bf B379} (1996) 99--104,
  \href{http://xxx.lanl.gov/abs/hep-th/9601029}{{\tt hep-th/9601029}}.

\bibitem{Ooguri:2004zv}
H.~Ooguri, A.~Strominger, and C.~Vafa, ``Black hole attractors and the
  topological string,'' {\em Phys. Rev.} {\bf D70} (2004) 106007,
  \href{http://xxx.lanl.gov/abs/hep-th/0405146}{{\tt hep-th/0405146}}.

\bibitem{Dabholkar:2005by}
A.~Dabholkar, F.~Denef, G.~W. Moore, and B.~Pioline, ``Exact and asymptotic
  degeneracies of small black holes,'' {\em JHEP} {\bf 08} (2005) 021,
  \href{http://xxx.lanl.gov/abs/hep-th/0502157}{{\tt hep-th/0502157}}.

\bibitem{Dabholkar:2005dt}
A.~Dabholkar, F.~Denef, G.~W. Moore, and B.~Pioline, ``Precision counting of
  small black holes,'' {\em JHEP} {\bf 10} (2005) 096,
  \href{http://xxx.lanl.gov/abs/hep-th/0507014}{{\tt hep-th/0507014}}.

\bibitem{Ooguri:2005vr}
H.~Ooguri, C.~Vafa, and E.~P. Verlinde, ``{H}artle-{H}awking wave-function for
  flux compactifications,'' \href{http://xxx.lanl.gov/abs/hep-th/0502211}{{\tt
  hep-th/0502211}}.

\bibitem{Pioline:2005vi}
B.~Pioline, ``{BPS} black hole degeneracies and minimal automorphic
  representations,'' {\em JHEP} {\bf 0508} (2005) 071,
  \href{http://xxx.lanl.gov/abs/hep-th/0506228}{{\tt hep-th/0506228}}.

\bibitem{Gunaydin:2006bz}
M.~Gunaydin, A.~Neitzke, and B.~Pioline, ``Topological wave functions and heat
  equations,'' \href{http://xxx.lanl.gov/abs/hep-th/0607200}{{\tt
  hep-th/0607200}}.

\bibitem{Gunaydin:2005mx}
M.~Gunaydin, A.~Neitzke, B.~Pioline, and A.~Waldron, ``Bps black holes, quantum
  attractor flows and automorphic forms,'' {\em Phys. Rev.} {\bf D73} (2006)
  084019, \href{http://xxx.lanl.gov/abs/hep-th/0512296}{{\tt hep-th/0512296}}.

\bibitem{Neitzke:2007ke}
A.~Neitzke, B.~Pioline, and S.~Vandoren, ``Twistors and black holes,''
  \href{http://xxx.lanl.gov/abs/hep-th/0701214}{{\tt hep-th/0701214}}.

\bibitem{gnpw-in-progress}
M.~Gunaydin, A.~Neitzke, B.~Pioline, and A.~Waldron, ``Quantum attractor
  flows.''
\newblock To appear.

\bibitem{gnopw-in-progress}
M.~Gunaydin, A.~Neitzke, O.~Pavlyk, B.~Pioline, and A.~Waldron,
  ``Quasiconformal, minimal representations and twistors.''
\newblock To appear.

\bibitem{Townsend:1997ku}
P.~K. Townsend, ``Black holes,''
  \href{http://xxx.lanl.gov/abs/gr-qc/9707012}{{\tt gr-qc/9707012}}.

\bibitem{Damour:2004kw}
T.~Damour, ``The entropy of black holes: A primer,''
  \href{http://xxx.lanl.gov/abs/hep-th/0401160}{{\tt hep-th/0401160}}.

\bibitem{Hawking:1974sw}
S.~W. Hawking, ``Particle creation by black holes,'' {\em Commun. Math. Phys.}
  {\bf 43} (1975) 199--220.

\bibitem{Unruh:1976db}
W.~G. Unruh, ``Notes on black hole evaporation,'' {\em Phys. Rev.} {\bf D14}
  (1976) 870.

\bibitem{Wald:2001}
R.~Wald, ``Black hole thermodynamics,'' {\em Living Rev. Relativity} {\bf 4}
  (2001)
  \href{http://xxx.lanl.gov/abs/http://www.livingreviews.org/lrr-2001-6}{{\tt
  http://www.livingreviews.org/lrr-2001-6}}.

\bibitem{Strominger:1998yg}
A.~Strominger, ``Ads(2) quantum gravity and string theory,'' {\em JHEP} {\bf
  01} (1999) 007, \href{http://xxx.lanl.gov/abs/hep-th/9809027}{{\tt
  hep-th/9809027}}.

\bibitem{Cvetic:1995uj}
M.~Cvetic and D.~Youm, ``Dyonic bps saturated black holes of heterotic string
  on a six torus,'' {\em Phys. Rev.} {\bf D53} (1996) 584--588,
  \href{http://xxx.lanl.gov/abs/hep-th/9507090}{{\tt hep-th/9507090}}.

\bibitem{Cvetic:1995kv}
M.~Cvetic and D.~Youm, ``All the static spherically symmetric black holes of
  heterotic string on a six torus,'' {\em Nucl. Phys.} {\bf B472} (1996)
  249--267, \href{http://xxx.lanl.gov/abs/hep-th/9512127}{{\tt
  hep-th/9512127}}.

\bibitem{Witten:1978mh}
E.~Witten and D.~I. Olive, ``Supersymmetry algebras that include topological
  charges,'' {\em Phys. Lett.} {\bf B78} (1978) 97.

\bibitem{Larsen:1995ss}
F.~Larsen and F.~Wilczek, ``Internal structure of black holes,'' {\em Phys.
  Lett.} {\bf B375} (1996) 37--42,
  \href{http://xxx.lanl.gov/abs/hep-th/9511064}{{\tt hep-th/9511064}}.

\bibitem{Tripathy:2005qp}
P.~K. Tripathy and S.~P. Trivedi, ``Non-supersymmetric attractors in string
  theory,'' {\em JHEP} {\bf 03} (2006) 022,
  \href{http://xxx.lanl.gov/abs/hep-th/0511117}{{\tt hep-th/0511117}}.

\bibitem{Goldstein:2005hq}
K.~Goldstein, N.~Iizuka, R.~P. Jena, and S.~P. Trivedi, ``Non-supersymmetric
  attractors,'' {\em Phys. Rev.} {\bf D72} (2005) 124021,
  \href{http://xxx.lanl.gov/abs/hep-th/0507096}{{\tt hep-th/0507096}}.

\bibitem{Kallosh:2005ax}
R.~Kallosh, ``New attractors,'' {\em JHEP} {\bf 12} (2005) 022,
  \href{http://xxx.lanl.gov/abs/hep-th/0510024}{{\tt hep-th/0510024}}.

\bibitem{Kallosh:2006bt}
R.~Kallosh, N.~Sivanandam, and M.~Soroush, ``The non-bps black hole attractor
  equation,'' {\em JHEP} {\bf 03} (2006) 060,
  \href{http://xxx.lanl.gov/abs/hep-th/0602005}{{\tt hep-th/0602005}}.

\bibitem{Sahoo:2006rp}
B.~Sahoo and A.~Sen, ``Higher derivative corrections to non-supersymmetric
  extremal black holes in n = 2 supergravity,''
  \href{http://xxx.lanl.gov/abs/hep-th/0603149}{{\tt hep-th/0603149}}.

\bibitem{Arkani-Hamed:2006dz}
N.~Arkani-Hamed, L.~Motl, A.~Nicolis, and C.~Vafa, ``The string landscape,
  black holes and gravity as the weakest force,''
  \href{http://xxx.lanl.gov/abs/hep-th/0601001}{{\tt hep-th/0601001}}.

\bibitem{Kaura:2006mv}
P.~Kaura and A.~Misra, ``On the existence of non-supersymmetric black hole
  attractors for two-parameter calabi-yau's and attractor equations,''
  \href{http://xxx.lanl.gov/abs/hep-th/0607132}{{\tt hep-th/0607132}}.

\bibitem{Kats:2006xp}
Y.~Kats, L.~Motl, and M.~Padi, ``Higher-order corrections to mass-charge
  relation of extremal black holes,''
  \href{http://xxx.lanl.gov/abs/hep-th/0606100}{{\tt hep-th/0606100}}.

\bibitem{Maldacena:1996ky}
J.~M. Maldacena, ``Black holes in string theory,''
  \href{http://xxx.lanl.gov/abs/hep-th/9607235}{{\tt hep-th/9607235}}.

\bibitem{Peet:2000hn}
A.~W. Peet, ``Tasi lectures on black holes in string theory,''
  \href{http://xxx.lanl.gov/abs/hep-th/0008241}{{\tt hep-th/0008241}}.

\bibitem{David:2002wn}
J.~R. David, G.~Mandal, and S.~R. Wadia, ``Microscopic formulation of black
  holes in string theory,'' {\em Phys. Rept.} {\bf 369} (2002) 549--686,
  \href{http://xxx.lanl.gov/abs/hep-th/0203048}{{\tt hep-th/0203048}}.

\bibitem{Mathur:2005ai}
S.~D. Mathur, ``The quantum structure of black holes,'' {\em Class. Quant.
  Grav.} {\bf 23} (2006) R115,
  \href{http://xxx.lanl.gov/abs/hep-th/0510180}{{\tt hep-th/0510180}}.

\bibitem{Maldacena:1996gb}
J.~M. Maldacena and A.~Strominger, ``Statistical entropy of four-dimensional
  extremal black holes,'' {\em Phys. Rev. Lett.} {\bf 77} (1996) 428--429,
  \href{http://xxx.lanl.gov/abs/hep-th/9603060}{{\tt hep-th/9603060}}.

\bibitem{Johnson:1996ga}
C.~V. Johnson, R.~R. Khuri, and R.~C. Myers, ``Entropy of 4d extremal black
  holes,'' {\em Phys. Lett.} {\bf B378} (1996) 78--86,
  \href{http://xxx.lanl.gov/abs/hep-th/9603061}{{\tt hep-th/9603061}}.

\bibitem{Maldacena:1997de}
J.~M. Maldacena, A.~Strominger, and E.~Witten, ``Black hole entropy in
  {M-theory},'' {\em JHEP} {\bf 12} (1997) 002,
  \href{http://xxx.lanl.gov/abs/hep-th/9711053}{{\tt hep-th/9711053}}.

\bibitem{Dijkgraaf:1996it}
R.~Dijkgraaf, H.~L. Verlinde, and E.~P. Verlinde, ``Counting dyons in {$\CN =
  4$} string theory,'' {\em Nucl. Phys.} {\bf B484} (1997) 543--561,
  \href{http://xxx.lanl.gov/abs/hep-th/9607026}{{\tt hep-th/9607026}}.

\bibitem{Cvetic:1995bj}
M.~Cvetic and A.~A. Tseytlin, ``Solitonic strings and bps saturated dyonic
  black holes,'' {\em Phys. Rev.} {\bf D53} (1996) 5619--5633,
  \href{http://xxx.lanl.gov/abs/hep-th/9512031}{{\tt hep-th/9512031}}.

\bibitem{Cardoso:2004xf}
G.~L. Cardoso, B.~de~Wit, J.~Kappeli, and T.~Mohaupt, ``{Asymptotic degeneracy
  of dyonic N = 4 string states and black hole entropy},''
  \href{http://xxx.lanl.gov/abs/hep-th/0412287}{{\tt hep-th/0412287}}.

\bibitem{Shih:2005uc}
D.~Shih, A.~Strominger, and X.~Yin, ``Recounting dyons in {$\CN = 4$} string
  theory,'' \href{http://xxx.lanl.gov/abs/hep-th/0505094}{{\tt
  hep-th/0505094}}.

\bibitem{Jatkar:2005bh}
D.~P. Jatkar and A.~Sen, ``Dyon spectrum in chl models,'' {\em JHEP} {\bf 04}
  (2006) 018, \href{http://xxx.lanl.gov/abs/hep-th/0510147}{{\tt
  hep-th/0510147}}.

\bibitem{David:2006ji}
J.~R. David, D.~P. Jatkar, and A.~Sen, ``Product representation of dyon
  partition function in chl models,''
  \href{http://xxx.lanl.gov/abs/hep-th/0602254}{{\tt hep-th/0602254}}.

\bibitem{Dabholkar:2006xa}
A.~Dabholkar and S.~Nampuri, ``Spectrum of dyons and black holes in chl
  orbifolds using borcherds lift,''
  \href{http://xxx.lanl.gov/abs/hep-th/0603066}{{\tt hep-th/0603066}}.

\bibitem{Gaiotto:2005hc}
D.~Gaiotto, ``Re-recounting dyons in n = 4 string theory,''
  \href{http://xxx.lanl.gov/abs/hep-th/0506249}{{\tt hep-th/0506249}}.

\bibitem{Ceresole:1995ca}
A.~Ceresole, R.~D'Auria, and S.~Ferrara, ``The symplectic structure of n=2
  supergravity and its central extension,'' {\em Nucl. Phys. Proc. Suppl.} {\bf
  46} (1996) 67--74, \href{http://xxx.lanl.gov/abs/hep-th/9509160}{{\tt
  hep-th/9509160}}.

\bibitem{Fre:1997jk}
P.~Fre, ``Supersymmetry and first order equations for extremal states:
  Monopoles, hyperinstantons, black holes and p- branes,'' {\em Nucl. Phys.
  Proc. Suppl.} {\bf 57} (1997) 52--64,
  \href{http://xxx.lanl.gov/abs/hep-th/9701054}{{\tt hep-th/9701054}}.

\bibitem{Moore:1998pn}
G.~W. Moore, ``Arithmetic and attractors,''
  \href{http://xxx.lanl.gov/abs/hep-th/9807087}{{\tt hep-th/9807087}}.

\bibitem{Mohaupt:2000mj}
T.~Mohaupt, ``{Black hole entropy, special geometry and strings},'' {\em
  Fortsch. Phys.} {\bf 49} (2001) 3--161,
  \href{http://xxx.lanl.gov/abs/hep-th/0007195}{{\tt hep-th/0007195}}.

\bibitem{MR2003030}
K.~Hori, S.~Katz, A.~Klemm, R.~Pandharipande, R.~Thomas, C.~Vafa, R.~Vakil, and
  E.~Zaslow, {\em Mirror symmetry}, vol.~1 of {\em Clay Mathematics
  Monographs}.
\newblock American Mathematical Society, Providence, RI, 2003.
\newblock With a preface by Vafa.

\bibitem{Huebscher:2006mr}
M.~Huebscher, P.~Meessen, and T.~Ortin, ``Supersymmetric solutions of n=2 d=4
  sugra: the whole ungauged shebang,''
  \href{http://xxx.lanl.gov/abs/hep-th/0606281}{{\tt hep-th/0606281}}.

\bibitem{Ferrara:1995ih}
S.~Ferrara, R.~Kallosh, and A.~Strominger, ``{$\CN=2$} extremal black holes,''
  {\em Phys. Rev.} {\bf D52} (1995) 5412--5416,
  \href{http://xxx.lanl.gov/abs/hep-th/9508072}{{\tt hep-th/9508072}}.

\bibitem{Ferrara:1996um}
S.~Ferrara and R.~Kallosh, ``Universality of supersymmetric attractors,'' {\em
  Phys. Rev.} {\bf D54} (1996) 1525--1534,
  \href{http://xxx.lanl.gov/abs/hep-th/9603090}{{\tt hep-th/9603090}}.

\bibitem{Ferrara:1997tw}
S.~Ferrara, G.~W. Gibbons, and R.~Kallosh, ``Black holes and critical points in
  moduli space,'' {\em Nucl. Phys.} {\bf B500} (1997) 75--93,
  \href{http://xxx.lanl.gov/abs/hep-th/9702103}{{\tt hep-th/9702103}}.

\bibitem{Denef:2000nb}
F.~Denef, ``Supergravity flows and d-brane stability,'' {\em JHEP} {\bf 08}
  (2000) 050, \href{http://xxx.lanl.gov/abs/hep-th/0005049}{{\tt
  hep-th/0005049}}.

\bibitem{Behrndt:1996jn}
K.~Behrndt {\em et.~al.}, ``Classical and quantum n = 2 supersymmetric black
  holes,'' {\em Nucl. Phys.} {\bf B488} (1997) 236--260,
  \href{http://xxx.lanl.gov/abs/hep-th/9610105}{{\tt hep-th/9610105}}.

\bibitem{LopesCardoso:2006bg}
G.~Lopes~Cardoso, B.~de~Wit, J.~Kappeli, and T.~Mohaupt, ``Black hole partition
  functions and duality,'' {\em JHEP} {\bf 03} (2006) 074,
  \href{http://xxx.lanl.gov/abs/hep-th/0601108}{{\tt hep-th/0601108}}.

\bibitem{Gunaydin:1983bi}
M.~Gunaydin, G.~Sierra, and P.~K. Townsend, ``{T}he geometry of {$\CN=2$}
  {M}axwell-{E}instein supergravity and {J}ordan algebras,'' {\em Nucl. Phys.}
  {\bf B242} (1984) 244.

\bibitem{MR0466235}
K.~McCrimmon, ``Jordan algebras and their applications,'' {\em Bull. Amer.
  Math. Soc.} {\bf 84} (1978), no.~4, 612--627.

\bibitem{Jordan:1933vh}
P.~Jordan, J.~von Neumann, and E.~P. Wigner, ``On an algebraic generalization
  of the quantum mechanical formalism,'' {\em Annals Math.} {\bf 35} (1934)
  29--64.

\bibitem{Gunaydin:1983rk}
M.~Gunaydin, G.~Sierra, and P.~K. Townsend, ``Exceptional supergravity theories
  and the magic square,'' {\em Phys. Lett.} {\bf B133} (1983) 72.

\bibitem{Ferrara:2006yb}
S.~Ferrara, E.~G. Gimon, and R.~Kallosh, ``Magic supergravities, n = 8 and
  black hole composites,'' \href{http://xxx.lanl.gov/abs/hep-th/0606211}{{\tt
  hep-th/0606211}}.

\bibitem{Ferrara:1995yx}
S.~Ferrara, J.~A. Harvey, A.~Strominger, and C.~Vafa, ``Second quantized mirror
  symmetry,'' {\em Phys. Lett.} {\bf B361} (1995) 59--65,
  \href{http://xxx.lanl.gov/abs/hep-th/9505162}{{\tt hep-th/9505162}}.

\bibitem{Ferrara:1988fr}
S.~Ferrara and M.~Porrati, ``The manifolds of scalar background fields in
  {$\Z_n$} orbifolds,'' {\em Phys. Lett.} {\bf B216} (1989) 289.

\bibitem{etingof}
D.~Etingof~P., Kazhdan and A.~Polishchuk, ``When is the {Fourier} transform of
  an elementary function elementary ?,''
  \href{http://xxx.lanl.gov/abs/math.AG/0003009}{{\tt math.AG/0003009}}.

\bibitem{Pioline:2003uk}
B.~Pioline, ``Cubic free field theory,''
  \href{http://xxx.lanl.gov/abs/hep-th/0302043}{{\tt hep-th/0302043}}.

\bibitem{Duff:2006uz}
M.~J. Duff, ``String triality, black hole entropy and cayley's
  hyperdeterminant,'' \href{http://xxx.lanl.gov/abs/hep-th/0601134}{{\tt
  hep-th/0601134}}.

\bibitem{Kallosh:2006zs}
R.~Kallosh and A.~Linde, ``Strings, black holes, and quantum information,''
  {\em Phys. Rev.} {\bf D73} (2006) 104033,
  \href{http://xxx.lanl.gov/abs/hep-th/0602061}{{\tt hep-th/0602061}}.

\bibitem{Levay:2006kf}
P.~Levay, ``Stringy black holes and the geometry of entanglement,'' {\em Phys.
  Rev.} {\bf D74} (2006) 024030,
  \href{http://xxx.lanl.gov/abs/hep-th/0603136}{{\tt hep-th/0603136}}.

\bibitem{Kallosh:1996uy}
R.~Kallosh and B.~Kol, ``$e_7$ symmetric area of the black hole horizon,'' {\em
  Phys. Rev.} {\bf D53} (1996) 5344--5348,
  \href{http://xxx.lanl.gov/abs/hep-th/9602014}{{\tt hep-th/9602014}}.

\bibitem{Gaiotto:2005gf}
D.~Gaiotto, A.~Strominger, and X.~Yin, ``New connections between 4d and 5d
  black holes,'' {\em JHEP} {\bf 02} (2006) 024,
  \href{http://xxx.lanl.gov/abs/hep-th/0503217}{{\tt hep-th/0503217}}.

\bibitem{Witten:1991zz}
E.~Witten, ``Mirror manifolds and topological field theory,''
  \href{http://xxx.lanl.gov/abs/hep-th/9112056}{{\tt hep-th/9112056}}.

\bibitem{Marino:2004uf}
M.~Marino, ``Chern-simons theory and topological strings,'' {\em Rev. Mod.
  Phys.} {\bf 77} (2005) 675--720,
  \href{http://xxx.lanl.gov/abs/hep-th/0406005}{{\tt hep-th/0406005}}.

\bibitem{Marino:2004eq}
M.~Marino, ``Les houches lectures on matrix models and topological strings,''
  \href{http://xxx.lanl.gov/abs/hep-th/0410165}{{\tt hep-th/0410165}}.

\bibitem{Neitzke:2004ni}
A.~Neitzke and C.~Vafa, ``Topological strings and their physical
  applications,'' \href{http://xxx.lanl.gov/abs/hep-th/0410178}{{\tt
  hep-th/0410178}}.

\bibitem{Vonk:2005yv}
M.~Vonk, ``A mini-course on topological strings,''
  \href{http://xxx.lanl.gov/abs/hep-th/0504147}{{\tt hep-th/0504147}}.

\bibitem{Cordes:1994fc}
S.~Cordes, G.~W. Moore, and S.~Ramgoolam, ``Lectures on 2-d yang-mills theory,
  equivariant cohomology and topological field theories,'' {\em Nucl. Phys.
  Proc. Suppl.} {\bf 41} (1995) 184--244,
  \href{http://xxx.lanl.gov/abs/hep-th/9411210}{{\tt hep-th/9411210}}.

\bibitem{Bershadsky:1993cx}
M.~Bershadsky, S.~Cecotti, H.~Ooguri, and C.~Vafa, ``{Kodaira-Spencer theory of
  gravity and exact results for quantum string amplitudes},'' {\em Commun.
  Math. Phys.} {\bf 165} (1994) 311--428,
  \href{http://xxx.lanl.gov/abs/hep-th/9309140}{{\tt hep-th/9309140}}.

\bibitem{Gopakumar:1998ii}
R.~Gopakumar and C.~Vafa, ``{M-theory and topological strings. I},''
  \href{http://xxx.lanl.gov/abs/hep-th/9809187}{{\tt hep-th/9809187}}.

\bibitem{Gopakumar:1998jq}
R.~Gopakumar and C.~Vafa, ``{M-theory and topological strings. II},''
  \href{http://xxx.lanl.gov/abs/hep-th/9812127}{{\tt hep-th/9812127}}.

\bibitem{Grisaru:1986kw}
M.~T. Grisaru, A.~E.~M. van~de Ven, and D.~Zanon, ``Four loop divergences for
  the n=1 supersymmetric nonlinear sigma model in two-dimensions,'' {\em Nucl.
  Phys.} {\bf B277} (1986) 409.

\bibitem{Candelas:1990rm}
P.~Candelas, X.~C. De~La~Ossa, P.~S. Green, and L.~Parkes, ``{A pair of
  Calabi-Yau manifolds as an exactly soluble superconformal theory},'' {\em
  Nucl. Phys.} {\bf B359} (1991) 21--74.

\bibitem{Antoniadis:1997eg}
I.~Antoniadis, S.~Ferrara, R.~Minasian, and K.~S. Narain, ``{$R^4$ couplings in
  M- and type II theories on Calabi-Yau spaces},'' {\em Nucl. Phys.} {\bf B507}
  (1997) 571--588, \href{http://xxx.lanl.gov/abs/hep-th/9707013}{{\tt
  hep-th/9707013}}.

\bibitem{Marino:1998pg}
M.~Marino and G.~W. Moore, ``{Counting higher genus curves in a Calabi-Yau
  manifold},'' {\em Nucl. Phys.} {\bf B543} (1999) 592--614,
  \href{http://xxx.lanl.gov/abs/hep-th/9808131}{{\tt hep-th/9808131}}.

\bibitem{faber}
C.~Faber and R.~Pandharipande, ``{Hodge integrals and Gromov-Witten theory},''
  \href{http://xxx.lanl.gov/abs/math.AG/9810173}{{\tt math.AG/9810173}}.

\bibitem{thomas-thesis}
R.~P. Thomas, ``{G}auge theories on {C}alabi-{Y}au manifolds,'' 1997.
\newblock Available as of July 2006 at {\tt
  \verb+http://www.ma.ic.ac.uk/~rpwt/thesis.pdf+}.

\bibitem{Nekrasov:2004js}
N.~A. Nekrasov, H.~Ooguri, and C.~Vafa, ``S-duality and topological strings,''
  \href{http://xxx.lanl.gov/abs/hep-th/0403167}{{\tt hep-th/0403167}}.

\bibitem{Kapustin:2004jm}
A.~Kapustin, ``{G}auge theory, topological strings, and {S}-duality,'' {\em
  JHEP} {\bf 09} (2004) 034, \href{http://xxx.lanl.gov/abs/hep-th/0404041}{{\tt
  hep-th/0404041}}.

\bibitem{gw-dt}
D.~Maulik, N.~A. Nekrasov, A.~Okounkov, and R.~Pandharipande,
  ``{G}romov-{W}itten theory and {D}onaldson-{T}homas theory,''
  \href{http://xxx.lanl.gov/abs/math.AG/0312059}{{\tt math.AG/0312059}}.

\bibitem{gw-dt2}
D.~Maulik, N.~A. Nekrasov, A.~Okounkov, and R.~Pandharipande,
  ``{G}romov-{W}itten theory and {D}onaldson-{T}homas theory, {II},''
  \href{http://xxx.lanl.gov/abs/math.AG/0406092}{{\tt math.AG/0406092}}.

\bibitem{okounkov}
A.~Okounkov and R.~Pandharipande, ``{The local Donaldson-Thomas theory of
  curves},'' \href{http://xxx.lanl.gov/abs/math.AG/0512573}{{\tt
  math.AG/0512573}}.

\bibitem{Verlinde:2004ck}
E.~Verlinde, ``Attractors and the holomorphic anomaly,''
  \href{http://xxx.lanl.gov/abs/hep-th/0412139}{{\tt hep-th/0412139}}.

\bibitem{Witten:1993ed}
E.~Witten, ``Quantum background independence in string theory,''
  \href{http://xxx.lanl.gov/abs/hep-th/9306122}{{\tt hep-th/9306122}}.

\bibitem{Gerasimov:2004yx}
A.~A. Gerasimov and S.~L. Shatashvili, ``Towards integrability of topological
  strings. i: Three- forms on calabi-yau manifolds,'' {\em JHEP} {\bf 11}
  (2004) 074, \href{http://xxx.lanl.gov/abs/hep-th/0409238}{{\tt
  hep-th/0409238}}.

\bibitem{D'Hoker:2005ia}
E.~D'Hoker and D.~H. Phong, ``Complex geometry and supergeometry,''
  \href{http://xxx.lanl.gov/abs/hep-th/0512197}{{\tt hep-th/0512197}}.

\bibitem{Berkovits:2002zk}
N.~Berkovits, ``Ictp lectures on covariant quantization of the superstring,''
  \href{http://xxx.lanl.gov/abs/hep-th/0209059}{{\tt hep-th/0209059}}.

\bibitem{Antoniadis:1993ze}
I.~Antoniadis, E.~Gava, K.~S. Narain, and T.~R. Taylor, ``{Topological
  amplitudes in string theory},'' {\em Nucl. Phys.} {\bf B413} (1994) 162--184,
  \href{http://xxx.lanl.gov/abs/hep-th/9307158}{{\tt hep-th/9307158}}.

\bibitem{Antoniadis:1995zn}
I.~Antoniadis, E.~Gava, K.~S. Narain, and T.~R. Taylor, ``{N=2 type II
  heterotic duality and higher derivative F terms},'' {\em Nucl. Phys.} {\bf
  B455} (1995) 109--130, \href{http://xxx.lanl.gov/abs/hep-th/9507115}{{\tt
  hep-th/9507115}}.

\bibitem{LopesCardoso:1998wt}
G.~Lopes~Cardoso, B.~de~Wit, and T.~Mohaupt, ``{Corrections to macroscopic
  supersymmetric black-hole entropy},'' {\em Phys. Lett.} {\bf B451} (1999)
  309--316, \href{http://xxx.lanl.gov/abs/hep-th/9812082}{{\tt
  hep-th/9812082}}.

\bibitem{LopesCardoso:1999cv}
G.~Lopes~Cardoso, B.~de~Wit, and T.~Mohaupt, ``{Deviations from the area law
  for supersymmetric black holes},'' {\em Fortsch. Phys.} {\bf 48} (2000)
  49--64, \href{http://xxx.lanl.gov/abs/hep-th/9904005}{{\tt hep-th/9904005}}.

\bibitem{LopesCardoso:1999ur}
G.~Lopes~Cardoso, B.~de~Wit, and T.~Mohaupt, ``{Macroscopic entropy formulae
  and non-holomorphic corrections for supersymmetric black holes},'' {\em Nucl.
  Phys.} {\bf B567} (2000) 87--110,
  \href{http://xxx.lanl.gov/abs/hep-th/9906094}{{\tt hep-th/9906094}}.

\bibitem{LopesCardoso:1999xn}
G.~Lopes~Cardoso, B.~de~Wit, and T.~Mohaupt, ``Area law corrections from state
  counting and supergravity,'' {\em Class. Quant. Grav.} {\bf 17} (2000)
  1007--1015, \href{http://xxx.lanl.gov/abs/hep-th/9910179}{{\tt
  hep-th/9910179}}.

\bibitem{Wald:1993nt}
R.~M. Wald, ``{Black hole entropy in the Noether charge},'' {\em Phys. Rev.}
  {\bf D48} (1993) 3427--3431,
  \href{http://xxx.lanl.gov/abs/gr-qc/9307038}{{\tt gr-qc/9307038}}.

\bibitem{Jacobson:1993vj}
T.~Jacobson, G.~Kang, and R.~C. Myers, ``On black hole entropy,'' {\em Phys.
  Rev.} {\bf D49} (1994) 6587--6598,
  \href{http://xxx.lanl.gov/abs/gr-qc/9312023}{{\tt gr-qc/9312023}}.

\bibitem{Jacobson:1995uq}
T.~Jacobson, G.~Kang, and R.~C. Myers, ``Increase of black hole entropy in
  higher curvature gravity,'' {\em Phys. Rev.} {\bf D52} (1995) 3518--3528,
  \href{http://xxx.lanl.gov/abs/gr-qc/9503020}{{\tt gr-qc/9503020}}.

\bibitem{Kraus:2005vz}
P.~Kraus and F.~Larsen, ``{Microscopic black hole entropy in theories with
  higher derivatives},'' \href{http://xxx.lanl.gov/abs/hep-th/0506176}{{\tt
  hep-th/0506176}}.

\bibitem{Sen:2005wa}
A.~Sen, ``Black hole entropy function and the attractor mechanism in higher
  derivative gravity,'' {\em JHEP} {\bf 09} (2005) 038,
  \href{http://xxx.lanl.gov/abs/hep-th/0506177}{{\tt hep-th/0506177}}.

\bibitem{Vafa:2004qa}
C.~Vafa, ``{T}wo dimensional {Y}ang-{M}ills, black holes and topological
  strings,'' \href{http://xxx.lanl.gov/abs/hep-th/0406058}{{\tt
  hep-th/0406058}}.

\bibitem{Aganagic:2004js}
M.~Aganagic, H.~Ooguri, N.~Saulina, and C.~Vafa, ``{B}lack holes,
  {$q$}-deformed 2d {Y}ang-{M}ills, and non-perturbative topological strings,''
  \href{http://xxx.lanl.gov/abs/hep-th/0411280}{{\tt hep-th/0411280}}.

\bibitem{Gross:1993hu}
D.~J. Gross and I.~Taylor, Washington, ``Two-dimensional qcd is a string
  theory,'' {\em Nucl. Phys.} {\bf B400} (1993) 181--210,
  \href{http://xxx.lanl.gov/abs/hep-th/9301068}{{\tt hep-th/9301068}}.

\bibitem{Dijkgraaf:2005bp}
R.~Dijkgraaf, R.~Gopakumar, H.~Ooguri, and C.~Vafa, ``Baby universes in string
  theory,'' \href{http://xxx.lanl.gov/abs/hep-th/0504221}{{\tt
  hep-th/0504221}}.

\bibitem{Shih:2005he}
D.~Shih and X.~Yin, ``Exact black hole degeneracies and the topological
  string,'' {\em JHEP} {\bf 04} (2006) 034,
  \href{http://xxx.lanl.gov/abs/hep-th/0508174}{{\tt hep-th/0508174}}.

\bibitem{Gaiotto:2006ns}
D.~Gaiotto, A.~Strominger, and X.~Yin, ``From ads(3)/cft(2) to black holes /
  topological strings,'' \href{http://xxx.lanl.gov/abs/hep-th/0602046}{{\tt
  hep-th/0602046}}.

\bibitem{Mooretoappear}
F.~Denef and G.~W. Moore.
\newblock To appear.

\bibitem{Beasley:2006us}
C.~Beasley {\em et.~al.}, ``Why {$Z_{BH} = |Z_{top}|^2$},''
  \href{http://xxx.lanl.gov/abs/hep-th/0608021}{{\tt hep-th/0608021}}.

\bibitem{deBoer:2006vg}
J.~de~Boer, M.~C.~N. Cheng, R.~Dijkgraaf, J.~Manschot, and E.~Verlinde, ``A
  {Farey} tail for attractor black holes,''
  \href{http://xxx.lanl.gov/abs/hep-th/0608059}{{\tt hep-th/0608059}}.

\bibitem{Dabholkar:2004yr}
A.~Dabholkar, ``Exact counting of black hole microstates,''
  \href{http://xxx.lanl.gov/abs/hep-th/0409148}{{\tt hep-th/0409148}}.

\bibitem{Dabholkar:1989jt}
A.~Dabholkar and J.~A. Harvey, ``Nonrenormalization of the superstring
  tension,'' {\em Phys. Rev. Lett.} {\bf 63} (1989) 478.

\bibitem{Dabholkar:1990yf}
A.~Dabholkar, G.~W. Gibbons, J.~A. Harvey, and F.~Ruiz~Ruiz, ``Superstrings and
  solitons,'' {\em Nucl. Phys.} {\bf B340} (1990) 33--55.

\bibitem{Dijkgraaf:2000fq}
R.~Dijkgraaf, J.~M. Maldacena, G.~W. Moore, and E.~Verlinde, ``{A black hole
  Farey tail},'' \href{http://xxx.lanl.gov/abs/hep-th/0005003}{{\tt
  hep-th/0005003}}.

\bibitem{Harvey:1996ir}
J.~A. Harvey and G.~W. Moore, ``{Fivebrane instantons and $R^2$ couplings in N
  = 4 string theory},'' {\em Phys. Rev.} {\bf D57} (1998) 2323--2328,
  \href{http://xxx.lanl.gov/abs/hep-th/9610237}{{\tt hep-th/9610237}}.

\bibitem{Sen:1995in}
A.~Sen, ``Extremal black holes and elementary string states,'' {\em Mod. Phys.
  Lett.} {\bf A10} (1995) 2081--2094,
  \href{http://xxx.lanl.gov/abs/hep-th/9504147}{{\tt hep-th/9504147}}.

\bibitem{Sen:1997is}
A.~Sen, ``{Black holes and elementary string states in N = 2 supersymmetric
  string theories},'' {\em JHEP} {\bf 02} (1998) 011,
  \href{http://xxx.lanl.gov/abs/hep-th/9712150}{{\tt hep-th/9712150}}.

\bibitem{Sen:2004dp}
A.~Sen, ``How does a fundamental string stretch its horizon?,''
  \href{http://xxx.lanl.gov/abs/hep-th/0411255}{{\tt hep-th/0411255}}.

\bibitem{Dabholkar:2004dq}
A.~Dabholkar, R.~Kallosh, and A.~Maloney, ``A stringy cloak for a classical
  singularity,'' {\em JHEP} {\bf 12} (2004) 059,
  \href{http://xxx.lanl.gov/abs/hep-th/0410076}{{\tt hep-th/0410076}}.

\bibitem{Hubeny:2004ji}
V.~Hubeny, A.~Maloney, and M.~Rangamani, ``String-corrected black holes,''
  \href{http://xxx.lanl.gov/abs/hep-th/0411272}{{\tt hep-th/0411272}}.

\bibitem{Sen:2005ch}
A.~Sen, ``Black holes and the spectrum of half-bps states in n = 4
  supersymmetric string theory,'' {\em Adv. Theor. Math. Phys.} {\bf 9} (2005)
  527--558, \href{http://xxx.lanl.gov/abs/hep-th/0504005}{{\tt
  hep-th/0504005}}.

\bibitem{Sen:2005pu}
A.~Sen, ``Black holes, elementary strings and holomorphic anomaly,'' {\em JHEP}
  {\bf 07} (2005) 063, \href{http://xxx.lanl.gov/abs/hep-th/0502126}{{\tt
  hep-th/0502126}}.

\bibitem{Kraus:2005zm}
P.~Kraus and F.~Larsen, ``Holographic gravitational anomalies,'' {\em JHEP}
  {\bf 01} (2006) 022, \href{http://xxx.lanl.gov/abs/hep-th/0508218}{{\tt
  hep-th/0508218}}.

\bibitem{Kraus:2006nb}
P.~Kraus and F.~Larsen, ``Partition functions and elliptic genera from
  supergravity,'' {\em JHEP} {\bf 01} (2007) 002,
  \href{http://xxx.lanl.gov/abs/hep-th/0607138}{{\tt hep-th/0607138}}.

\bibitem{Kraus:2006wn}
P.~Kraus, ``Lectures on black holes and the ads(3)/cft(2) correspondence,''
  \href{http://xxx.lanl.gov/abs/hep-th/0609074}{{\tt hep-th/0609074}}.

\bibitem{Thiemann:1992jj}
H.~A. Kastrup and T.~Thiemann, ``{C}anonical quantization of spherically
  symmetric gravity in {A}shtekar's selfdual representation,'' {\em Nucl.
  Phys.} {\bf B399} (1993) 211--258,
  \href{http://xxx.lanl.gov/abs/gr-qc/9310012}{{\tt gr-qc/9310012}}.

\bibitem{Kuchar:1994zk}
K.~V. Kuchar, ``{G}eometrodynamics of {S}chwarzschild black holes,'' {\em Phys.
  Rev.} {\bf D50} (1994) 3961--3981,
  \href{http://xxx.lanl.gov/abs/gr-qc/9403003}{{\tt gr-qc/9403003}}.

\bibitem{Cavaglia:1994yc}
M.~Cavaglia, V.~de~Alfaro, and A.~T. Filippov, ``Hamiltonian formalism for
  black holes and quantization,'' {\em Int. J. Mod. Phys.} {\bf D4} (1995)
  661--672, \href{http://xxx.lanl.gov/abs/gr-qc/9411070}{{\tt gr-qc/9411070}}.

\bibitem{Hollmann:1996cb}
H.~Hollmann, ``{G}roup theoretical quantization of {S}chwarzschild and
  {T}aub-{NUT},'' {\em Phys. Lett.} {\bf B388} (1996) 702--706,
  \href{http://xxx.lanl.gov/abs/gr-qc/9609053}{{\tt gr-qc/9609053}}.

\bibitem{Hollmann:1996ra}
H.~Hollmann, ``A harmonic space approach to spherically symmetric quantum
  gravity,'' \href{http://xxx.lanl.gov/abs/gr-qc/9610042}{{\tt gr-qc/9610042}}.

\bibitem{Breitenlohner:1998yt}
P.~Breitenlohner, H.~Hollmann, and D.~Maison, ``{Q}uantization of the
  {R}eissner-{N}ordstr{\"o}m black hole,'' {\em Phys. Lett.} {\bf B432} (1998)
  293--297, \href{http://xxx.lanl.gov/abs/gr-qc/9804030}{{\tt gr-qc/9804030}}.

\bibitem{Mandal:2005wv}
G.~Mandal, ``Fermions from half-bps supergravity,'' {\em JHEP} {\bf 08} (2005)
  052, \href{http://xxx.lanl.gov/abs/hep-th/0502104}{{\tt hep-th/0502104}}.

\bibitem{Maoz:2005nk}
L.~Maoz and V.~S. Rychkov, ``Geometry quantization from supergravity: The case
  of 'bubbling ads','' {\em JHEP} {\bf 08} (2005) 096,
  \href{http://xxx.lanl.gov/abs/hep-th/0508059}{{\tt hep-th/0508059}}.

\bibitem{Rychkov:2005ji}
V.~S. Rychkov, ``D1-d5 black hole microstate counting from supergravity,'' {\em
  JHEP} {\bf 01} (2006) 063, \href{http://xxx.lanl.gov/abs/hep-th/0512053}{{\tt
  hep-th/0512053}}.

\bibitem{Grant:2005qc}
L.~Grant, L.~Maoz, J.~Marsano, K.~Papadodimas, and V.~S. Rychkov,
  ``Minisuperspace quantization of 'bubbling ads' and free fermion droplets,''
  {\em JHEP} {\bf 08} (2005) 025,
  \href{http://xxx.lanl.gov/abs/hep-th/0505079}{{\tt hep-th/0505079}}.

\bibitem{Biswas:2006tj}
I.~Biswas, D.~Gaiotto, S.~Lahiri, and S.~Minwalla, ``Supersymmetric states of n
  = 4 yang-mills from giant gravitons,''
  \href{http://xxx.lanl.gov/abs/hep-th/0606087}{{\tt hep-th/0606087}}.

\bibitem{Mandal:2006tk}
G.~Mandal and N.~V. Suryanarayana, ``Counting 1/8-{BPS} dual-giants,''
  \href{http://xxx.lanl.gov/abs/hep-th/0606088}{{\tt hep-th/0606088}}.

\bibitem{Maldacena:1998uz}
J.~M. Maldacena, J.~Michelson, and A.~Strominger, ``{Anti-de Sitter}
  fragmentation,'' {\em JHEP} {\bf 02} (1999) 011,
  \href{http://xxx.lanl.gov/abs/hep-th/9812073}{{\tt hep-th/9812073}}.

\bibitem{Pioline:2005pf}
B.~Pioline and J.~Troost, ``Schwinger pair production in ads(2),'' {\em JHEP}
  {\bf 03} (2005) 043, \href{http://xxx.lanl.gov/abs/hep-th/0501169}{{\tt
  hep-th/0501169}}.

\bibitem{Breitenlohner:1987dg}
P.~Breitenlohner, G.~W. Gibbons, and D.~Maison, ``{F}our-dimensional black
  holes from {K}aluza-{K}lein theories,'' {\em Commun. Math. Phys.} {\bf 120}
  (1988) 295.

\bibitem{Hull:1998br}
C.~M. Hull and B.~L. Julia, ``Duality and moduli spaces for time-like
  reductions,'' {\em Nucl. Phys.} {\bf B534} (1998) 250--260,
  \href{http://xxx.lanl.gov/abs/hep-th/9803239}{{\tt hep-th/9803239}}.

\bibitem{Ferrara:1989ik}
S.~Ferrara and S.~Sabharwal, ``Quaternionic manifolds for type {II} superstring
  vacua of {Calabi-Yau} spaces,'' {\em Nucl. Phys.} {\bf B332} (1990) 317.

\bibitem{Damour:2002et}
T.~Damour, M.~Henneaux, and H.~Nicolai, ``Cosmological billiards,'' {\em Class.
  Quant. Grav.} {\bf 20} (2003) R145--R200,
  \href{http://xxx.lanl.gov/abs/hep-th/0212256}{{\tt hep-th/0212256}}.

\bibitem{Pioline:2002qz}
B.~Pioline and A.~Waldron, ``Quantum cosmology and conformal invariance,'' {\em
  Phys. Rev. Lett.} {\bf 90} (2003) 031302,
  \href{http://xxx.lanl.gov/abs/hep-th/0209044}{{\tt hep-th/0209044}}.

\bibitem{deAlfaro:1976je}
V.~de~Alfaro, S.~Fubini, and G.~Furlan, ``Conformal invariance in quantum
  mechanics,'' {\em Nuovo Cim.} {\bf A34} (1976) 569.

\bibitem{kinnersley}
W.~Kinnersley, ``{G}eneration of stationary {E}instein-{M}axwell fields,'' {\em
  J. Math. Phys.} {\bf 14, no. 5} (1973) 651--653.

\bibitem{MR1701415}
A.~A. Kirillov, ``Merits and demerits of the orbit method,'' {\em Bull. Amer.
  Math. Soc. (N.S.)} {\bf 36} (1999), no.~4, 433--488.

\bibitem{Cecotti:1988qn}
S.~Cecotti, S.~Ferrara, and L.~Girardello, ``Geometry of type {II} superstrings
  and the moduli of superconformal field theories,'' {\em Int. J. Mod. Phys.}
  {\bf A4} (1989) 2475.

\bibitem{Bagger:1983tt}
J.~Bagger and E.~Witten, ``{M}atter couplings in {$\CN=2$} supergravity,'' {\em
  Nucl. Phys.} {\bf B222} (1983) 1.

\bibitem{Gutperle:2000ve}
M.~Gutperle and M.~Spalinski, ``Supergravity instantons for {$N = 2$}
  hypermultiplets,'' {\em Nucl. Phys.} {\bf B598} (2001) 509--529,
  \href{http://xxx.lanl.gov/abs/hep-th/0010192}{{\tt hep-th/0010192}}.

\bibitem{Behrndt:1997ch}
K.~Behrndt, I.~Gaida, D.~Lust, S.~Mahapatra, and T.~Mohaupt, ``From type iia
  black holes to t-dual type iib d-instantons in n = 2, d = 4 supergravity,''
  {\em Nucl. Phys.} {\bf B508} (1997) 659--699,
  \href{http://xxx.lanl.gov/abs/hep-th/9706096}{{\tt hep-th/9706096}}.

\bibitem{deVroome:2006xu}
M.~de~Vroome and S.~Vandoren, ``Supergravity description of spacetime
  instantons,'' \href{http://xxx.lanl.gov/abs/hep-th/0607055}{{\tt
  hep-th/0607055}}.

\bibitem{MR1096180}
A.~Swann, ``Hyper-{K}\"ahler and quaternionic {K}\"ahler geometry,'' {\em Math.
  Ann.} {\bf 289} (1991), no.~3, 421--450.

\bibitem{MR664330}
S.~M. Salamon, ``Quaternionic {K}\"ahler manifolds,'' {\em Invent. Math.} {\bf
  67} (1982), no.~1, 143--171.

\bibitem{Rocek:2005ij}
M.~Rocek, C.~Vafa, and S.~Vandoren, ``Hypermultiplets and topological
  strings,'' {\em JHEP} {\bf 02} (2006) 062,
  \href{http://xxx.lanl.gov/abs/hep-th/0512206}{{\tt hep-th/0512206}}.

\bibitem{Giddings:1988wv}
S.~B. Giddings and A.~Strominger, ``Baby universes, third quantization and the
  cosmological constant,'' {\em Nucl. Phys.} {\bf B321} (1989) 481.

\bibitem{Witten:1982df}
E.~Witten, ``Constraints on supersymmetry breaking,'' {\em Nucl. Phys.} {\bf
  B202} (1982) 253.

\bibitem{Alvarez-Gaume:1983at}
L.~Alvarez-Gaume, ``Supersymmetry and the atiyah-singer index theorem,'' {\em
  Commun. Math. Phys.} {\bf 90} (1983) 161.

\bibitem{Friedan:1983xr}
D.~Friedan and P.~Windey, ``Supersymmetric derivation of the atiyah-singer
  index and the chiral anomaly,'' {\em Nucl. Phys.} {\bf B235} (1984) 395.

\bibitem{Gauntlett:1992yj}
J.~P. Gauntlett, ``Low-energy dynamics of supersymmetric solitons,'' {\em Nucl.
  Phys.} {\bf B400} (1993) 103--125,
  \href{http://xxx.lanl.gov/abs/hep-th/9205008}{{\tt hep-th/9205008}}.

\bibitem{Gauntlett:1993sh}
J.~P. Gauntlett, ``Low-energy dynamics of n=2 supersymmetric monopoles,'' {\em
  Nucl. Phys.} {\bf B411} (1994) 443--460,
  \href{http://xxx.lanl.gov/abs/hep-th/9305068}{{\tt hep-th/9305068}}.

\bibitem{quatman}
S.~M. Salamon, ``{D}ifferential geometry of quaternionic manifolds,'' {\em
  Annales Scientifiques de l'\'Ecole Normale Sup\'erieure} {\bf Sér. 4, 19}
  (1986) 31--55.

\bibitem{MR1165872}
R.~J. Baston, ``Quaternionic complexes,'' {\em J. Geom. Phys.} {\bf 8} (1992),
  no.~1-4, 29--52.

\bibitem{MR506229}
M.~F. Atiyah, N.~J. Hitchin, and I.~M. Singer, ``Self-duality in
  four-dimensional {R}iemannian geometry,'' {\em Proc. Roy. Soc. London Ser. A}
  {\bf 362} (1978), no.~1711, 425--461.

\bibitem{Anselmi:1993wm}
D.~Anselmi and P.~Fre, ``Topological sigma models in four-dimensions and
  triholomorphic maps,'' {\em Nucl. Phys.} {\bf B416} (1994) 255--300,
  \href{http://xxx.lanl.gov/abs/hep-th/9306080}{{\tt hep-th/9306080}}.

\bibitem{Cecotti:1988ad}
S.~Cecotti, ``Homogeneous kahler manifolds and t algebras in n=2 supergravity
  and superstrings,'' {\em Commun. Math. Phys.} {\bf 124} (1989) 23--55.

\bibitem{Gunaydin:2000xr}
M.~Gunaydin, K.~Koepsell, and H.~Nicolai, ``{C}onformal and quasiconformal
  realizations of exceptional {L}ie groups,'' {\em Commun. Math. Phys.} {\bf
  221} (2001) 57--76, \href{http://xxx.lanl.gov/abs/hep-th/0008063}{{\tt
  hep-th/0008063}}.

\bibitem{Pioline:2003bk}
B.~Pioline and A.~Waldron, ``{A}utomorphic forms: {A} physicist's survey,''
  \href{http://xxx.lanl.gov/abs/hep-th/0312068}{{\tt hep-th/0312068}}.

\bibitem{MR1421947}
B.~H. Gross and N.~R. Wallach, ``On quaternionic discrete series
  representations, and their continuations,'' {\em J. Reine Angew. Math.} {\bf
  481} (1996) 73--123.

\bibitem{MR610183}
M.~G. Eastwood and M.~L. Ginsberg, ``Duality in twistor theory,'' {\em Duke
  Math. J.} {\bf 48} (1981), no.~1, 177--196.

\bibitem{Biedenharn:1972ns}
L.~C. Biedenharn, R.~Y. Cusson, M.~Y. Han, and O.~L. Weaver, ``Hadronic regge
  sequences as primitive realizations of sl(3,r) symmetry,'' {\em Phys. Lett.}
  {\bf B42} (1972) 257--260.

\bibitem{Pioline:2001jn}
B.~Pioline, H.~Nicolai, J.~Plefka, and A.~Waldron, ``{$R^4$} couplings, the
  fundamental membrane and exceptional theta correspondences,'' {\em JHEP} {\bf
  03} (2001) 036, \href{http://xxx.lanl.gov/abs/hep-th/0102123}{{\tt
  hep-th/0102123}}.

\bibitem{Pioline:2004xq}
B.~Pioline and A.~Waldron, ``The automorphic membrane,'' {\em JHEP} {\bf 06}
  (2004) 009, \href{http://xxx.lanl.gov/abs/hep-th/0404018}{{\tt
  hep-th/0404018}}.

\bibitem{Gunaydin:2001bt}
M.~Gunaydin, K.~Koepsell, and H.~Nicolai, ``{T}he minimal unitary
  representation of ${E_{8(8)}}$,'' {\em Adv. Theor. Math. Phys.} {\bf 5}
  (2002) 923--946, \href{http://xxx.lanl.gov/abs/hep-th/0109005}{{\tt
  hep-th/0109005}}.

\bibitem{MR1159103}
D.~Kazhdan and G.~Savin, ``The smallest representation of simply laced
  groups,'' in {\em Festschrift in honor of I. I. Piatetski-Shapiro on the
  occasion of his sixtieth birthday, Part I (Ramat Aviv, 1989)}, vol.~2 of {\em
  Israel Math. Conf. Proc.}, pp.~209--223.
\newblock Weizmann, Jerusalem, 1990.

\bibitem{Kazhdan:2001nx}
D.~Kazhdan, B.~Pioline, and A.~Waldron, ``{M}inimal representations, spherical
  vectors, and exceptional theta series. {I},'' {\em Commun. Math. Phys.} {\bf
  226} (2002) 1--40, \href{http://xxx.lanl.gov/abs/hep-th/0107222}{{\tt
  hep-th/0107222}}.

\bibitem{Gunaydin:2004md}
M.~Gunaydin and O.~Pavlyk, ``{M}inimal unitary realizations of exceptional
  {U}-duality groups and their subgroups as quasiconformal groups,'' {\em JHEP}
  {\bf 01} (2005) 019, \href{http://xxx.lanl.gov/abs/hep-th/0409272}{{\tt
  hep-th/0409272}}.

\bibitem{Gunaydin:2006vz}
M.~Gunaydin and O.~Pavlyk, ``A unified approach to the minimal unitary
  realizations of noncompact groups and supergroups,''
  \href{http://xxx.lanl.gov/abs/hep-th/0604077}{{\tt hep-th/0604077}}.

\bibitem{Ferrara:1997uz}
S.~Ferrara and M.~Gunaydin, ``{O}rbits of exceptional groups, duality and {BPS}
  states in string theory,'' {\em Int. J. Mod. Phys.} {\bf A13} (1998)
  2075--2088, \href{http://xxx.lanl.gov/abs/hep-th/9708025}{{\tt
  hep-th/9708025}}.

\bibitem{Gunaydin:2005gd}
M.~Gunaydin, ``{U}nitary realizations of {U}-duality groups as conformal and
  quasiconformal groups and extremal black holes of supergravity theories,''
  {\em AIP Conf. Proc.} {\bf 767} (2005) 268--287,
  \href{http://xxx.lanl.gov/abs/hep-th/0502235}{{\tt hep-th/0502235}}.

\bibitem{MR1988198}
N.~R. Wallach, ``Generalized {W}hittaker vectors for holomorphic and
  quaternionic representations,'' {\em Comment. Math. Helv.} {\bf 78} (2003),
  no.~2, 266--307.

\bibitem{MR2094111}
D.~Kazhdan and A.~Polishchuk, ``Minimal representations: spherical vectors and
  automorphic functionals,'' in {\em Algebraic groups and arithmetic},
  pp.~127--198.
\newblock Tata Inst. Fund. Res., Mumbai, 2004.

\bibitem{npv-to-appear}
A.~Neitzke, B.~Pioline, and S.~Vandoren.
\newblock To appear.

\bibitem{Shih:2005qf}
D.~Shih, A.~Strominger, and X.~Yin, ``Counting dyons in n = 8 string theory,''
  \href{http://xxx.lanl.gov/abs/hep-th/0506151}{{\tt hep-th/0506151}}.

\end{thebibliography}

\providecommand{\href}[2]{#2}\begingroup\raggedright\endgroup

\end{document}